\documentclass[]{interact}
\usepackage{float}
\usepackage{longtable}
\usepackage{epstopdf}% To incorporate .eps illustrations using PDFLaTeX, etc.
\usepackage[caption=false]{subfig}% Support for small, `sub' figures and tables
\usepackage[natbibapa,nodoi]{apacite}% Citation support using apacite.sty. Commands using natbib.sty MUST be deactivated first!
\usepackage{hyperref}

\usepackage{color,soul}%For colors

\theoremstyle{plain}% Theorem-like structures provided by amsthm.sty

\theoremstyle{definition}

\theoremstyle{remark}

\begin{document}

% \articletype{ARTICLE TEMPLATE}% Specify the article type or omit as appropriate

\title{Tailored Real-time AR Captioning Interface for Enhancing Learning Experience of Deaf and Hard-of-Hearing (DHH) Students}

% Enhancing Deaf and Hard-of-Hearing (DHH) Students' Learning Experience: A User-Centered Approach to AR Captioning Interface Development

% Enhancing Deaf and Hard-of-Hearing (DHH) Students' Learning Experience: A User-Centered Design of an AR Captioning Interface for Enhanced Learning in Specialized Settings

% Enhancing Deaf and Hard-of-Hearing (DHH) Students' Learning Experience: A User-Centered Approach to AR Captioning Interface Development

% User-Centered Development of AR Caption Interface to Enhance Educational Experience of DHH Students in Specialized Educational Settings
% Empowering Deaf and Hard-of-Hearing Education: A User-Centered Design of an AR Captioning Interface for Enhanced Learning in Specialized Settings

\author{
\name{Yasith Samaradivakara\textsuperscript{a,}\textsuperscript{b}\thanks{Yasith Samaradivakara. Email: yasith@ahlab.org}, Asela Pathirage \textsuperscript{a}, Thavindu Ushan \textsuperscript{a},  Prasanth Sasikumar \textsuperscript{b}, Kasun Karunanayaka \textsuperscript{a}, Chamath Keppitiyagama \textsuperscript{a} and Suranga Nanayakkara \textsuperscript{b}}
\affil{\textsuperscript{a} Augmented Human Lab, National University of Singapore, Singapore; \textsuperscript{b} University of Colombo School of Computing, Sri Lanka}
}

\maketitle

\begin{abstract}
Deaf and hard-of-hearing (DHH) students face significant challenges in specialized educational settings, such as limited exposure to written and spoken language, a lack of tailored educational tools, and restricted access to resources, impacting their language literacy development and overall educational experience. We, therefore, employed a User-Centered Design (UCD) process, collaborating with 8 DHH students and 2 Teachers of the Deaf (ToDs) from a School of Deaf to effectively develop and utilize a real-time captioning augmented reality (AR) system to their school settings, aiming to enhance their learning experience. User study with 24 DHH participants revealed a strong preference (87.5\%) for our system, underscoring its potential to enhance learning experience. We present a comprehensive needs analysis, the UCD process, system implementation, and user feedback, showcasing the effectiveness of tailored AR caption interfaces for DHH students. We also discuss the implications for future development of educational technologies for DHH students.
\end{abstract}


\section{Introduction}
For most Deaf and Hard-of-Hearing (DHH) individuals language proficiency is a huge barrier causing numerous challenges in their social, professional, and educational life \citep{paul2020current,EvidenceBasedPractices2020,ziadat2020learning}. Studies have shown that there is a language proficiency gap between students with hearing abilities and those who have hearing difficulties \citep{QiMitchell2012, Garberoglio2021}.  Previous work has emphasised that early intervention and exposure to language, and access to rich-language environments enhances language proficiency development of DHH individuals \citep{Lederberg2013}. Some common approaches included solutions such as bilingual education which incorporates spoken words to their traditional educational settings \citep{DevelopingLanguage2014, EffectsOfBilingual2018}. Additionally, studies have also indicated that visual presentation of information such as written words significantly enhances language learning experience \citep{AssessingTheEffectiveness2018, JOURNALOF2021}.

With the recent advancements of speech recognition, caption interfaces have been used as a promising solution to provide real-time access to spoken words for DHH individuals \citep{DynamicSubtitles2015,BetweenText2008,TheProsAndCons2002,CloseToTheAction2017,AProposedSet1998}. However, in the educational context, most of these caption interfaces are displayed either through projectors, students' laptops, or mobile phones.  These have issues as students have to switch their focus between the teacher and the caption interface, ultimately enhancing their cognitive load \citep{CollaborativeGaze2014}. Recently, these caption interfaces have evolved from traditional digital displays to immersive AR interfaces, offering the potential to deliver information seamlessly within the user's field of view without the need for handheld devices. 

Previous studies have demonstrated the use of real-time AR captioning to support educational activities (e.g., classroom discussions, lectures) for DHH students \citep{ExploringAugmentedReality2018, Ioannou2017Augmented}. Building on these findings, we are exploring the effective utilization of real-time AR caption interface to a specialised School of Deaf. We adopted a User-Centered Development (UCD) process, collaborating with 8 DHH students and 2 Specialized Teachers from the Deaf School to design the overall system. Initiating with a comprehensive needs analysis, conducted with both teachers and students, we identified significant challenges in language literacy among DHH students, primarily due to limited exposure to written and spoken language. Need analysis further highlighted the critical gap in the availability of educational tools tailored to the unique requirements of  DHH students in specialized educational settings. Based on these insights, we developed our initial AR prototype  which delivers localized real-time captions directly in users line-of-sight. Through semi-structured interviews, we identified the challenges in the initial prototype that affect the overall learning experience. Employing an iterative co-design process we developed and refined our prototype. We designed caption presentation methods, placement methods, markups for inaccurate transcriptions and wearer voice visualizations addressing the existing design limitations and enhancing the learning experience while supporting the overall learning outcomes.

We conducted set of user studies involving 24 DHH students to evaluate the designed prototypes addressing five research questions detailed in section \ref{sec:rq}. Finally, we evaluated the overall system and results show that 21 students (87.5\%) preferred to use our system with strong (78.18) System Usability Score \citep{SUS1995}. User feedback suggested that the system enhances the overall learning experience and supports educational activities. We discussed the results, limitations of the system, and concluded with potential future work.

\section{Related Work}

\subsection*{Education for DHH Students}
DHH students face many challenges, particularly in the learning of language and communication development during childhood. Unlike their hearing peers, DHH students often encounter challenges in gaining language and communication skills, primarily due to limited exposure to spoken language in their environment. Their interactions mainly occur with family members and individuals who are not DHH, resulting in communication barriers that can interfere with their language and social development \citep{paul2020current,EvidenceBasedPractices2020,ziadat2020learning}. The primary objective in the education of DHH students is to achieve inclusive education. This involves integrating DHH students into mainstream schools, ensuring they receive equitable educational opportunities alongside their hearing peers \citep{jachova2023deaf}. However, this integration demands specialized support services to address the unique challenges faced by DHH students, particularly due to literacy skills \citep{jachova2023deaf, spencer2010evidence}. Due to these reasons DHH students admitted to specialized schools with personalized support including sign-language interpreters, adapted curriculum's and personalized monitoring etc. With the recent advancements of technology innovative tools have been introduced to support and enhance the educational experiences of DHH students. With proper understanding of unique needs and providing effective solutions can potentially enhance the learning experiences and literacy skills. Consequently, this support can facilitate the effective integration of DHH students into mainstream educational settings, substantially improving their overall educational journey.

\subsection*{AR in Education for DHH students}During the past years AR has become a trend in educational research \citep{arici2019research,akcayir2017advantages,garzon2019systematic,izaguirre2020mobile}. Researchers have highlighted the educational benefits of AR, highlighting its potential applications in fields such as science, technology, mathematics, and language learning \citep{akcayir2017advantages,ibanez2018augmented, Al-Megren_Almutairi_2018}. Previous studies have also demonstrated the utility of AR tools to support and enhance the learning experience for DHH students in educational settings using visual-centric information (e.g. sign avatars) \citep{Luo2022Avatar, Adamo-Villani2016Holographic,Kercher2012Improving}. These studies have shown that such approaches are highly effective in terms of educational enhancements, student interest, active engagement, and collaboration \citep{researchtrends,akcayir2017advantages,garzon2019systematic}. Two case studies \citep{Ioannou2017Augmented}  have identified that AR interfaces can be used to enhance communication, feedback, vocabulary acquisition, and reading comprehension for DHH learners. In our work, we design an AR caption interface to provide an enhanced learning experience, with greater student engagement and motivation.

\subsection*{AR Caption Interfaces for DHH Education} The development of caption interfaces tailored for DHH students in classroom environments poses unique challenges. Marshcark et al. \citep{marschark2006benefits} observed that DHH students relying on real-time captions often underperform compared to their hearing peers. This could be mainly due to less glanceability, decreased visual contact with the speaker and need to consistently switch between the speaker and the traditional interface. In response, Jain et al. \citep{ExploringAugmentedReality2018} explored the use of AR headsets to deliver the captions directly in users' line-of-sight. This approach aims to provide a more seamless and immersive experience for DHH students, enabling them to access captions without distracting their attention away from the educational content. We build upon their findings, aiming to adapt and refine these interfaces and address identified limitations.

\subsubsection*{AR Caption Presentation Methods}
The impact of text presentation methods on DHH individuals, who have special needs, is a crucial area of exploration. Studies like those of Duchnicky and Kolers \citep{duchnicky1983readability} and more recent works \citep{ReadingOnSmartGlasses2018} have delved into the reading speed, comprehension, and task load of non-deaf users with different text presentation methods. However, the impact on DHH individuals who have distinct language proficiency, visual attention, and information processing \citep{Dye2007Which, Luckner2013Response, Proksch2002Changes} needs to be further explored. Masson \citep{masson1983conceptual} identified reading longer text could overload users, leading to reduced comprehension. Moreover, Jain et al. \citep{ExploringAugmentedReality2018} identified that presenting captions as single-line at a time affects the comprehension of DHH users, while Peng et al. \citep{SpeechBubbles2018} found that adding more lines increases comprehension in communication. While these methods shown their unique benefits, in learning context for DHH students it's essential to explore how each presentation method affects the learning outcomes, therefore in our work we employ these presentation methods and more effective presentation methods considering the amount and presentation of text optimizing comprehension, reading speed, cognitive load, and engagement.

\subsubsection*{AR Caption Placement}The strategic placement of captions in AR environments is crucial for enhancing the learning experience of DHH students. Dynamic captioning, a technique widely used in video captioning for movies and television \citep{DynamicSubtitles2015, DynamicCaptioning2010}, presents a foundational concept. Following that Jain et al. \citep{ExploringAugmentedReality2018} proposed an approach, that allows users to manipulate captions in a 3D space which found to be useful in interactive sessions 
and also suggested a fixed captioning strategy for more static, non-interactive sessions. Our work build upon these findings, exploring how caption placements influence DHH students' learning experience, and aims to identify optimal solutions while addressing current limitations.

\subsubsection*{Automatic Speech Recognition (ASR) Transcription Errors} Real-time caption interfaces mostly use ASR as a captioning service. However, ASR is imperfect; causing challenges in the understandability of captions \citep{Kafle2016Effect}, especially for DHH individuals. While addressing this issue,  previous studies have used visual markups to represent the transcription errors \citep{confidence,remote}. Berke et al. \citep{confidence} changed the appearance of text to indicate low-confidence words in video captions. Based upon these methods, we incorporated visual markups in our co-design process, seeking input from participants to refine ideal designs that enhance the clarity and accuracy of captions for DHH users.

\section{Needs Analysis}
Our needs analysis was conducted with 8 DHH students (4 females and 4 males, Age: 14-18, Hearing Loss: Moderate - 1, Severe - 4, Extremely Severe/ Profound - 3) and 2 Teachers of the Deaf (ToD) from a School of Deaf. Hearing loss levels were classified based on thresholds defined in clinical audiology \citep{Clark1981-ca}. These students have been enrolled in the school since the age of 6 and have exclusively received instructions in sign language. The objective of the need analysis was to understand the educational challenges faced by DHH students, their adaptation strategies, and the broader impacts of current educational approaches. We conducted focus group discussions and semi-structured interviews to gather data. Interviews were performed in-person spanning within a one week time. Communication with participants was facilitated by a Teacher of Deaf who is an expert in Sri Lankan Sign Language \footnote[1]{\href{https://en.wikipedia.org/wiki/Sri_Lankan_Sign_Language}{\url{https://en.wikipedia.org/wiki/Sri_Lankan_Sign_Language}}(last accessed January 02, 2024)}  which is the sign language used by students and teachers. To analyze the data, we used a thematic approach \citep{BraunClarke2006} that combined both deductive and inductive methods. Deductive themes were identified first, informed by the expertise of experienced Teachers of the Deaf and grounded in existing literature, which highlighted challenges such as difficulties with literacy and the limitations of current teaching methodologies. This was followed by an inductive analysis, allowing themes to emerge directly from the data and uncovering additional challenges, such as limited access to educational materials and the critical role of localized captions in online resources.

The transcriptions were independently coded by two researchers to ensure reliability. Discrepancies in coding were resolved through discussion, and themes were refined collaboratively. To ensure the reliability of the analysis, interrater reliability was assessed using the Intraclass Correlation Coefficient (ICC), with the average fixed raters yielding a score of 0.888, indicating excellent agreement between coders.

The analysis revealed significant challenges in three primary areas: literacy, teaching methodologies, and access to resources. Following provides a qualitative summary.

\subsubsection{Literacy Challenges}
Both teachers emphasized the difficulties DHH students face in reading and comprehending written text. For instance, Teacher 2 (T2) \textit{``Students take a lot of time to read and comprehend exam papers, often struggling to finish within the allocated time.''}. She further emphasized it mentioning, \textit{``Examinations doesn't have additional support or different papers, students are mostly familiar with sign language not written language which slow them down.''}. Several students expressed their struggles with reading speed and comprehension of written text, with Student 4 (S4) noting, \textit{``I can read words but it takes more time for me to read and comprehend.''}. The unfamiliarity with textbook vocabulary was a common issue, as expressed by S6: \textit{``I find it difficult to read and understand textbooks by myself as most words are unfamiliar for me.''}. S6 also mentioned that textbooks don't have explanations for individual words or support with signs, \textit{``If the books have signs I will understand the words.''}. Three students mentioned of requiring additional support such as lip readings, \textit{``Its hard for me to read books alone; I need support with lip reading.''} (S7).

\subsubsection{Educational Methodology Limitations}
The teachers noted the shortcomings of current educational methods in supporting the development of reading and writing skills. One teacher remarked, \textit{``Certain words, particularly those in textbooks, cannot be directly translated into sign language. This makes it difficult for students to read and understand these words.''} (T1). The lack of sign language representation for certain concepts was mentioned as leading to oversimplified content, impacting vocabulary development. Furthermore, the reliance on sign language in education has been observed to hinder students' proficiency in written and spoken language. T2 pointed out, \textit{``Students tend to grasp sign language well, but they face challenges in connecting those signs with their written counterparts.''}. This observation was echoed by a student (S6) who shared their personal experience, saying, \textit{``I try to associate written words with signs when I come across them. However, I frequently find that many written words lack corresponding signs.''}. Teachers also mentioned that teaching advanced subjects with signs takes a lot of time, \textit{``For such subjects, we end up writing everything on the whiteboard and explain, which is incredibly time-consuming.''} (T1), T2 highlighted its impact on maintaining syllabus on-time, \textit{``We are sometimes forced to omit parts of the syllabus, struggling to keep pace with mainstream schools. This is a significant disadvantage for students in completing their education on time.''}.

\subsubsection{Access to Learning Resources}
Students highlighted difficulties in accessing suitable learning materials. One student shared the challenge of using online resources: \textit{“On YouTube, most videos don’t have localized (Sinhala)\footnote[2]{\url{https://en.wikipedia.org/wiki/Sinhala_language} (last accessed January 02, 2024)} captions, I try to understand by looking at the pictures which take weeks to watch a short clip.”} (S1). Additionally, teachers pointed out the challenges in linking students to external educational opportunities such as workshops, informational sessions, and additional resources, making it difficult to enrich their learning experience beyond the classroom.\\

The analysis revealed profound challenges faced by DHH students, particularly in language literacy, attributed to limited exposure to written and spoken language, limited educational methodologies, and lack of supporting tools. This underscores the critical need for advanced educational aids, such as captioning systems, to complement sign language and address the unique requirements of DHH students in specialized settings.

\section{User Centered Design Process}
\begin{figure}[ht]
  \centering
  \includegraphics[width=0.45\linewidth]{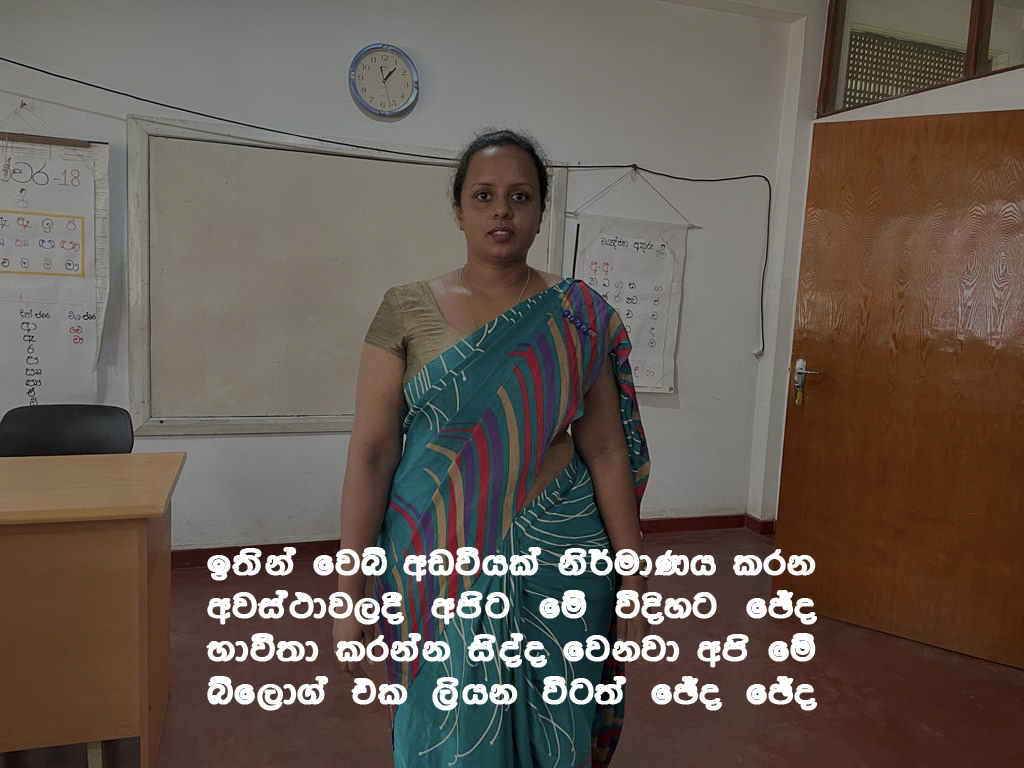}
  \captionsetup{justification=centering}
  \caption{Initial Prototype (Non-optimized AR interface - 4 lines of text with 20 words) [Refer the Appendix \ref{appendix} for English translations]}
  \label{Fig:Initial}
\end{figure}

Considering the insights from the need analysis, we initialized the development of AR captioning interface following a UCD process. Our design process was driven by the goal of creating an optimal AR caption interface for DHH students, effectively utilized in their classroom settings, addressing their unique needs and enhancing their learning experience. To accomplish this, we introduced a preliminary AR interface (Figure \ref{Fig:Initial}), similar to traditional captions, displaying multiple lines of text directly in the users' line-of-sight. We then conducted semi-structured interviews to gather feedback on the challenges in the prototype affecting the learning experience. Subsequently, we engaged in a co-design process, where DHH students actively participated in refining our design by providing sketches of their ideal designs and offering feedback on our prototypes. Finally, we synthesized our findings, leveraging insights from prior research, user preferences, and the co-design phase to create optimized prototypes tailored to overcome the identified challenges.

\subsection{Semi-Structured Interviews and Co-Design}
We conducted semi-structured interviews with the same 8 DHH students from need analysis, aiming to more deeply understand the challenges in preliminary prototype. For each student, we asked for the issues that they encountered while using our prototype in classroom sessions that affects their overall learning experience and outcomes. Following the semi-structured interviews, we asked students to describe and provide their ideal designs to address the challenges.

\subsubsection{Issues}

\begin{enumerate}
    \item \textbf{Low Comprehension:} Students found the amount of text presented at a time overwhelming, making it difficult to grasp the overall meaning and requiring significant focus. S5 highlighted, \textit{``Too many words shown at a time, it's hard to grasp the overall meaning.''}. S4 added, \textit{``It's difficult to follow and understand the captions.''}. S7 mentioned an issue with placement, \textit{``It's difficult to lip-read when text is placed away from the lips of the speaker and I often miss some text trying to switch between lip reading and text.''}
    \item \textbf{Low Reading Speed:} Majority (n=5) found reading with the prototype time-consuming. S3 explained, \textit{``It’s harder for me to focus on which word to read with a lot of text.''}. S6 added that \textit{``Continuing to read further down the caption lines slows me down.''}. Students also mentioned that placement affects their reading speed, \textit{``Consistently shifting my eyes between lip-reading and text reading slows me down.''} (S4).
    
    \item \textbf{High Task Load:} Many students (n=6) felt that the prototype causes excessive mental strain, often leading to discouragement and reluctance to read. For example, S2 mentioned, \textit{``It makes me nervous when there are a lot of words at once.''}. S7 stated, \textit{``Lot of text is overwhelming as I feel like I have a big task to complete.''}. Additionally, students suggested that text should be closer to the lip, \textit{``I need to lip read with the words but in here I have to move my eyes which takes a lot of effort.''} (S8).
    
    \item \textbf{Transcription Errors:} Students noticed issues with the transcriptions, S1 mentioned, \textit{``In the middle of a sentence, it showed just a letter. I'm curious whether it showed the actual word the teacher said. However, it distracted the reading.''}. S3 also suggested, \textit{``It would be better if you could mark any errors, so I won't be confused.''}.

    \item \textbf{Wearer's Voice Transcription:} Students faced difficulties with the transcription feature, which did not differentiate between their personal utterances and external speech, leading to confusion. S2 pointed out, \textit{``I think, What I say also gets displayed; which is annoying that I can't separate out what the teacher said.''} Furthermore, the feature disrupted students who vocalize text for speech practice, with S4 noting, \textit{``When I read, I usually speak it aloud but showing it in the glass is distracting.''}. 

\end{enumerate}

\begin{figure}[H]
  \centering
   % \begin{subfigure}[b]{0.3\textwidth}
  {\includegraphics[width=0.3\textwidth]{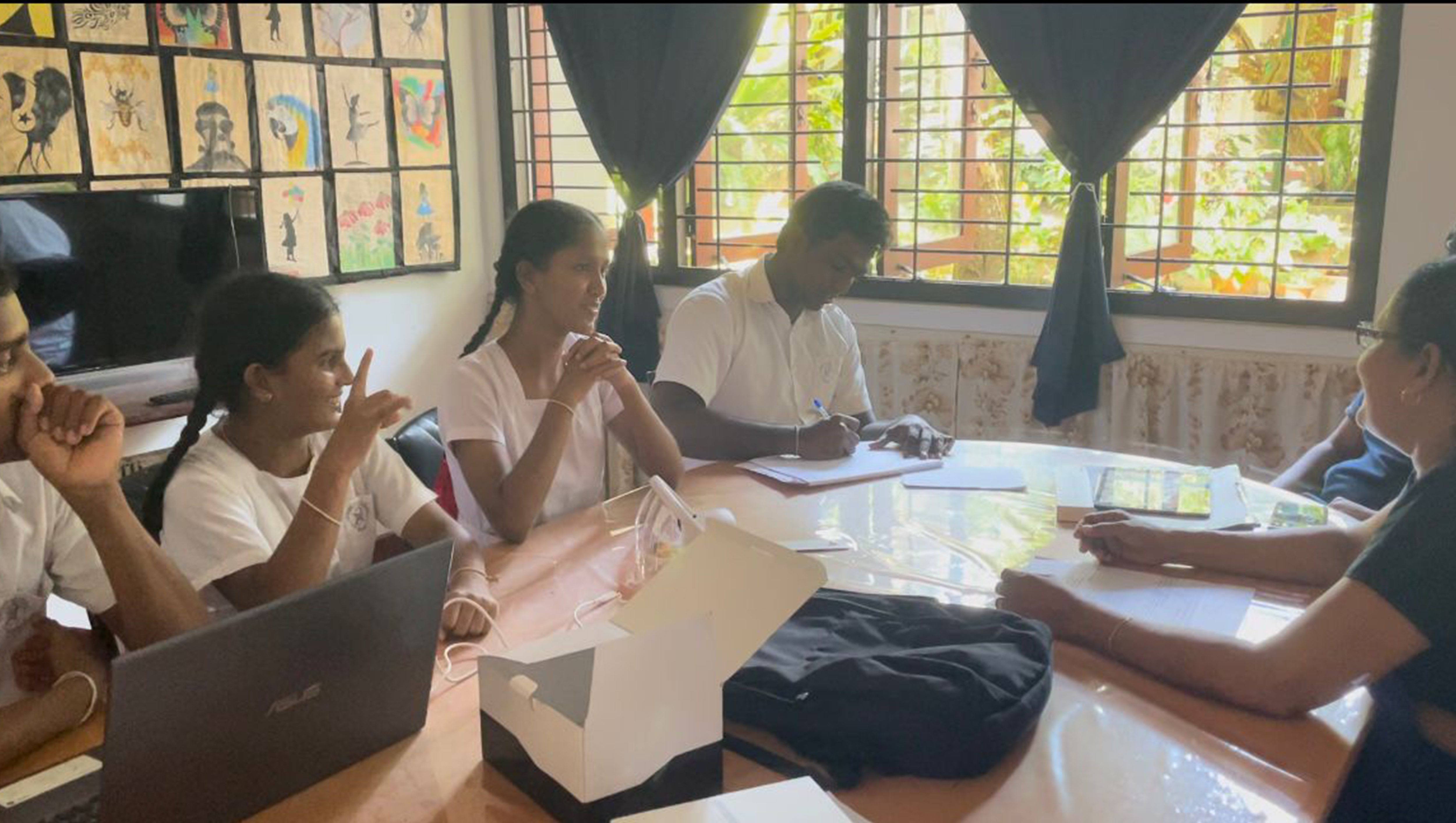}
  % \caption{Caption Presentation and Placement Methods}
  \label{fig:main_sub1}}
  % \end{subfigure}
   % \begin{subfigure}[b]{0.3\textwidth}
   {\includegraphics[width=0.3\textwidth]{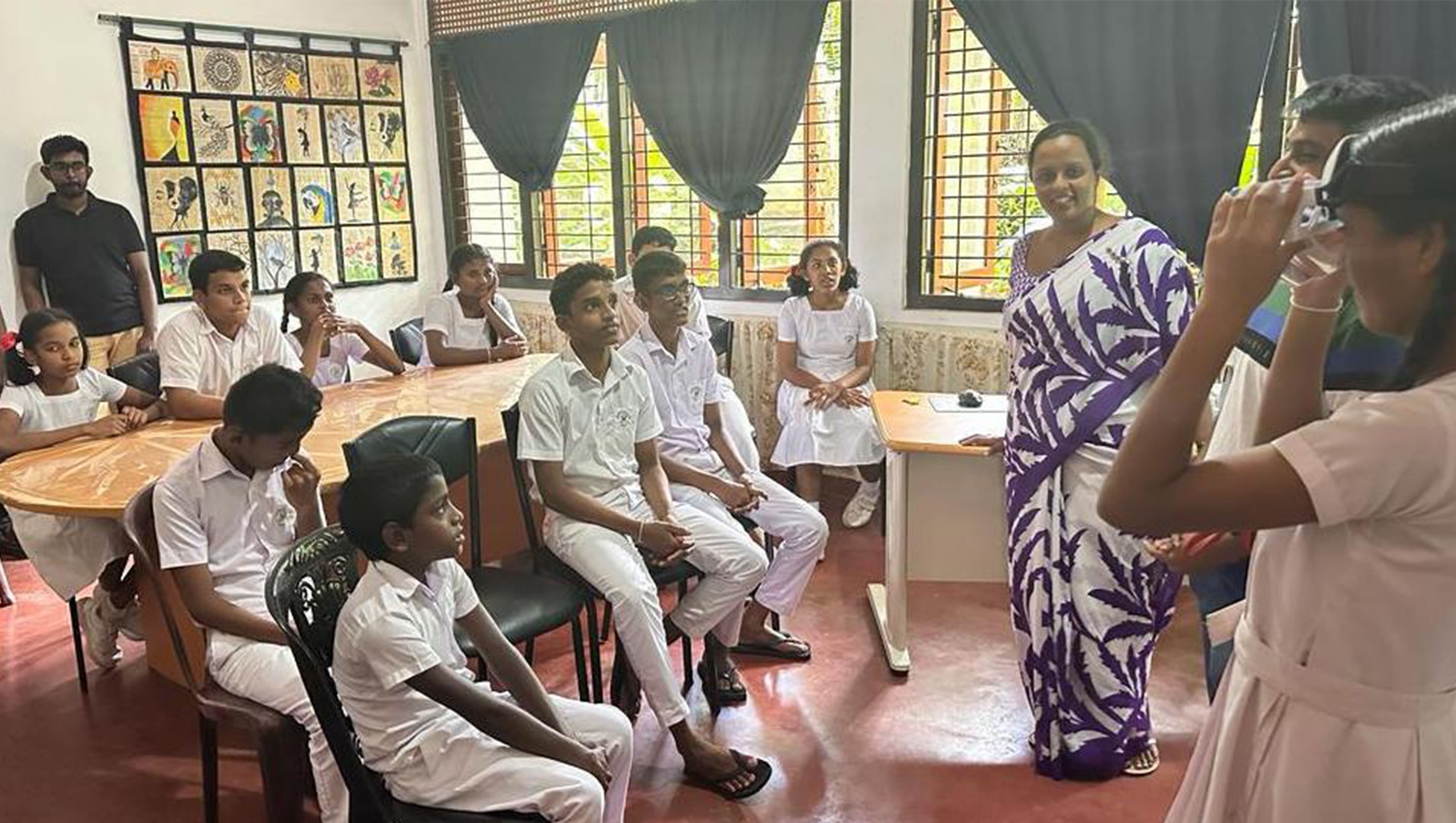}
  % \caption{Caption Presentation and Placement Methods}
  \label{fig:main_sub2}}
  % \end{subfigure}
   % \begin{subfigure}[b]{0.3\textwidth}
   {\includegraphics[width=0.3\textwidth]{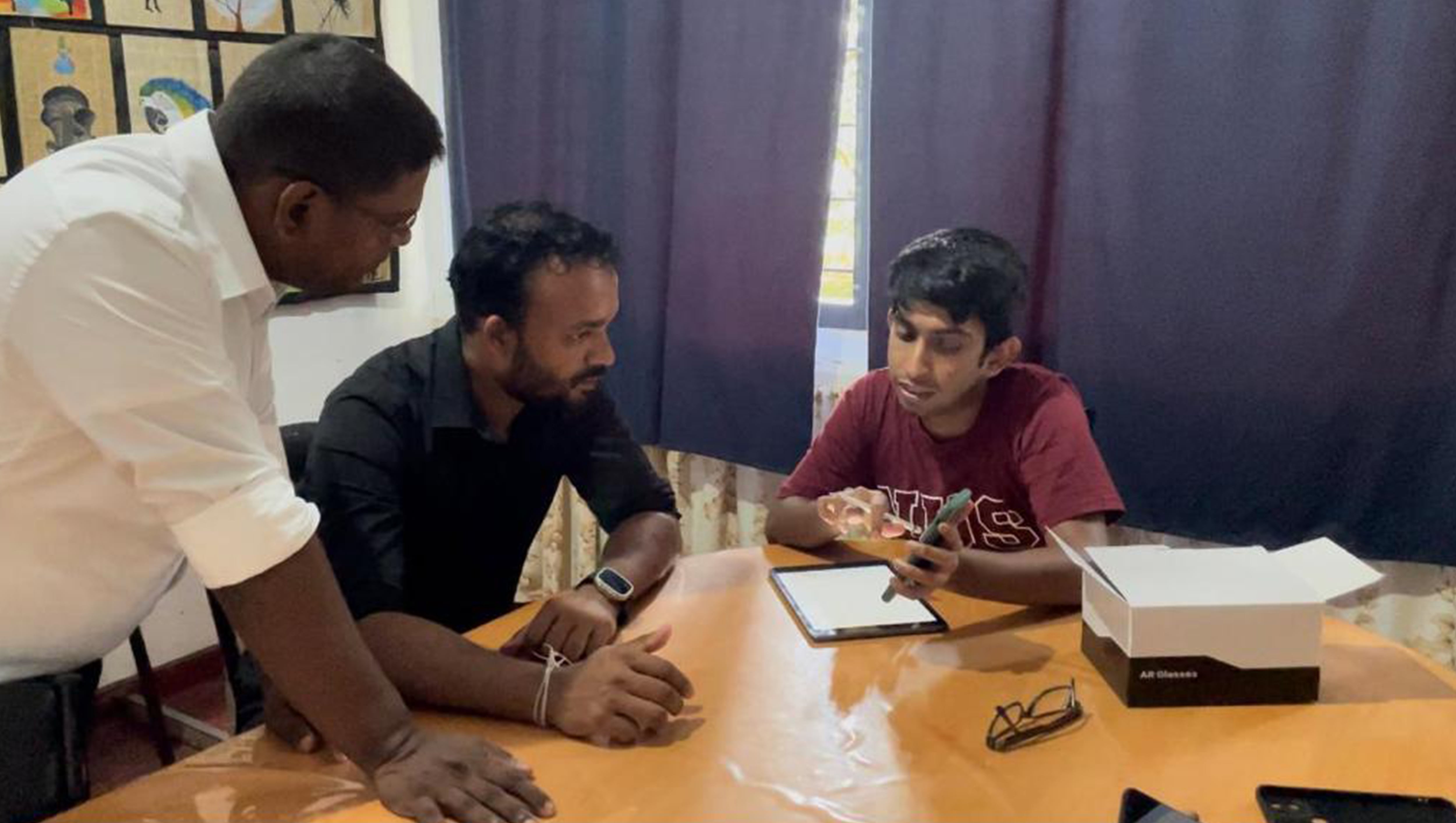}
  % \caption{Caption Presentation and Placement Methods}
  \label{fig:main_sub3}}
  % \end{subfigure}
   % \begin{subfigure}[b]{0.3\textwidth}
   {\includegraphics[width=0.3\textwidth]{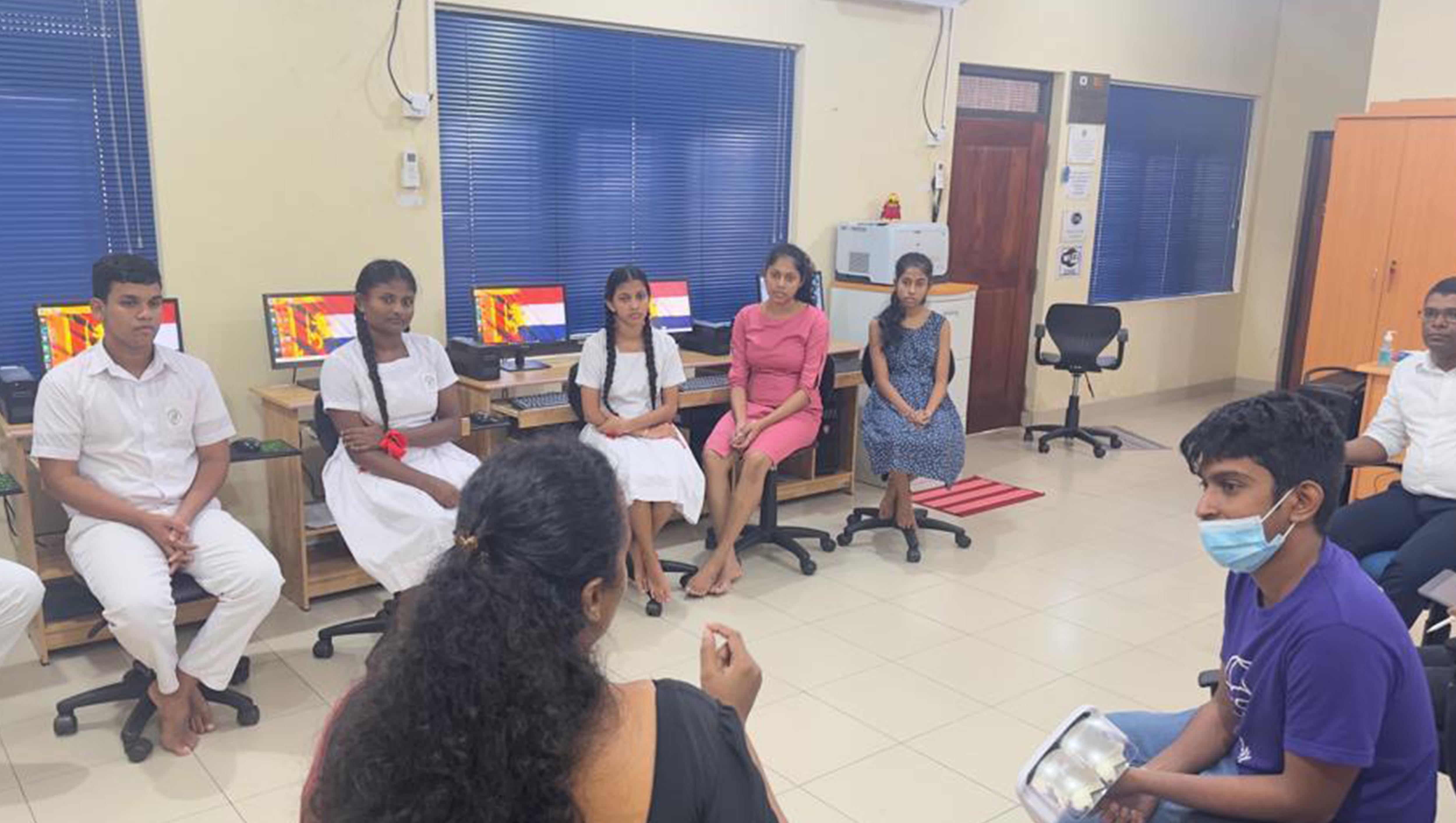}
  % \caption{Caption Presentation and Placement Methods}
  \label{fig:main_sub4}}
  % \end{subfigure}
   % \begin{subfigure}[b]{0.3\textwidth}
   {\includegraphics[width=0.3\textwidth]{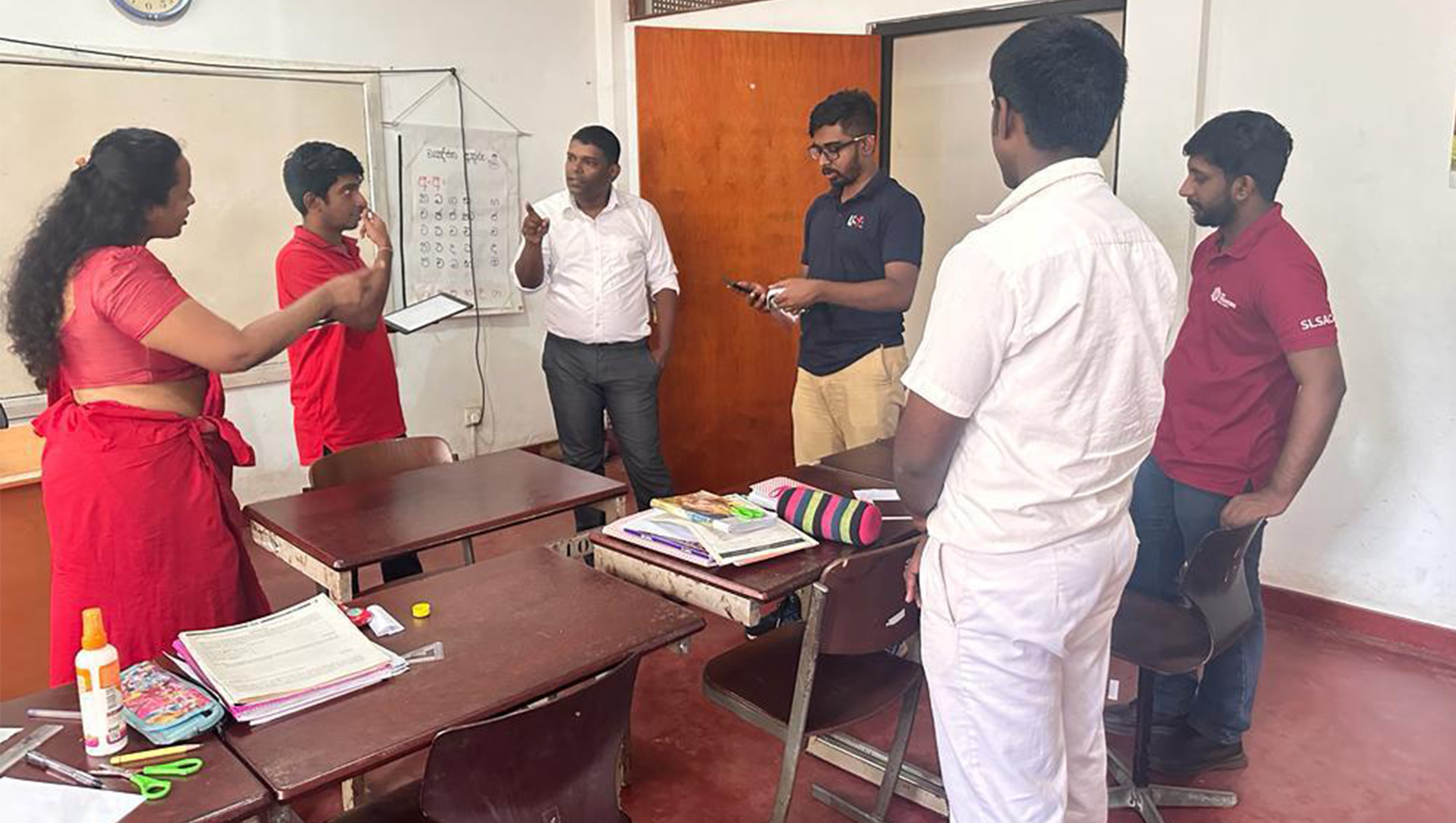}
  % \caption{Caption Presentation and Placement Methods}
  \label{fig:main_sub5}}
  % \end{subfigure}
   % \begin{subfigure}[b]{0.3\textwidth}
   {\includegraphics[width=0.3\textwidth]{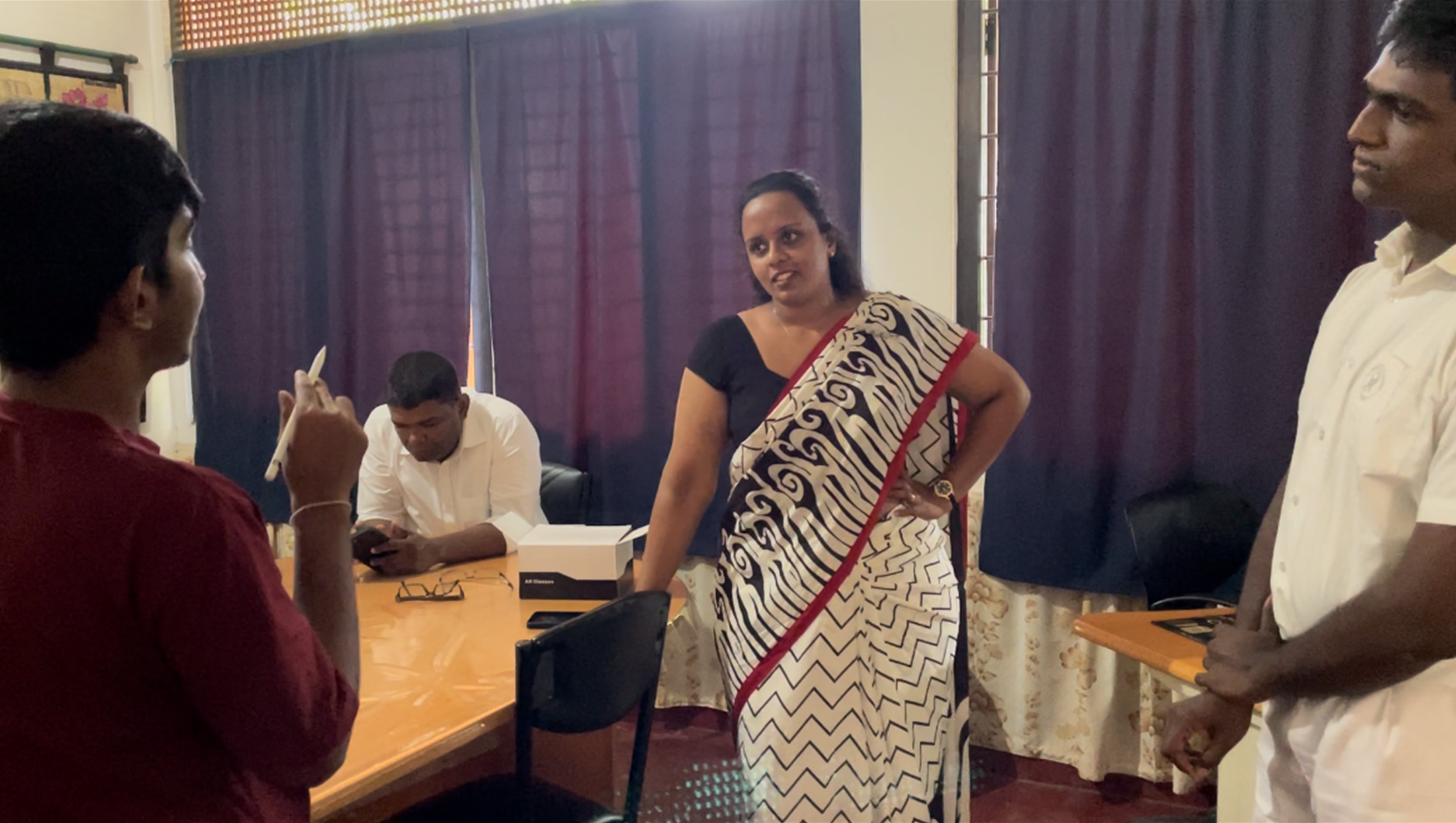}
  % \caption{Caption Presentation and Placement Methods}
  \label{fig:main_sub6}}
  % \end{subfigure}
  \caption{User-Centered Design Process (Semi-Structured Interviews, Focus Groups, Co-Design Sessions)}
  \label{fig:main}
\end{figure}

\subsection{Designs by DHH Students} 
\begin{figure}[ht]
  \centering
  % \begin{subfigure}[b]{0.6\textwidth}
  \subfloat[Caption presentation and placement methods]
   {\includegraphics[width=0.6\textwidth]{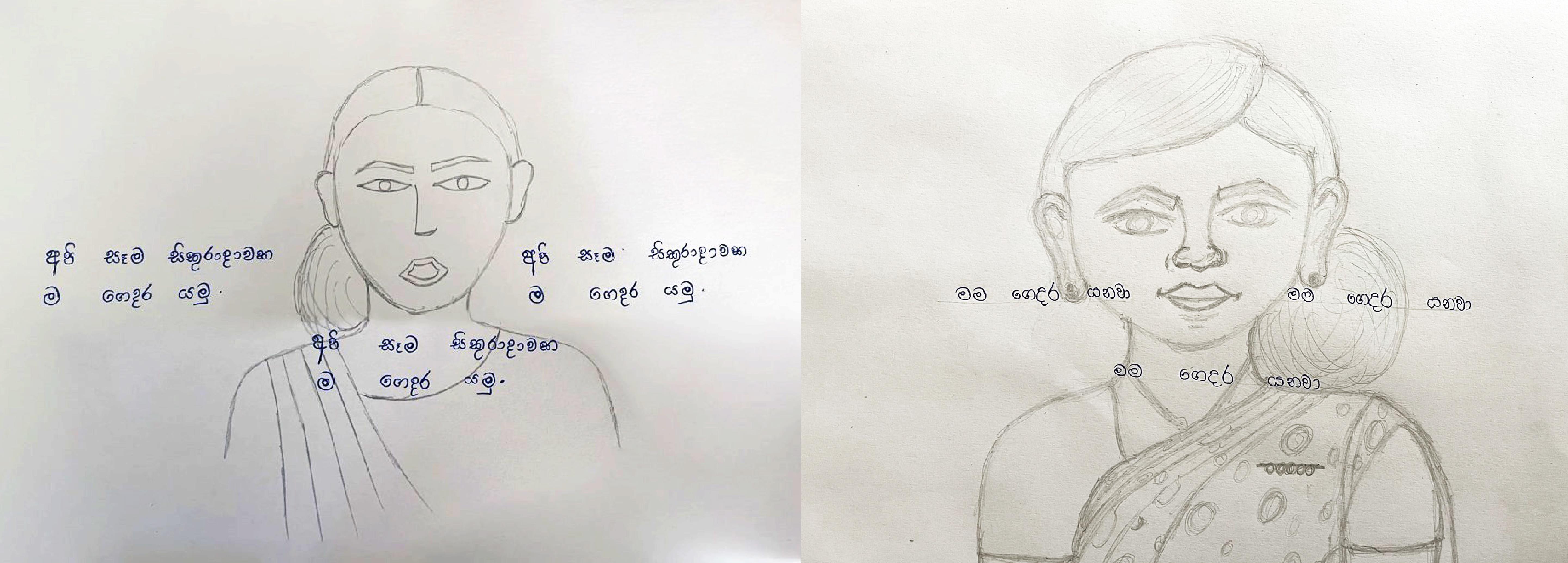}
  \label{fig:idealsub1}}\par
  % \end{subfigure}\\
  % \begin{subfigure}[b]{0.35\textwidth}
  \subfloat[Markups for inaccurate transcribed word]
  { \includegraphics[width=0.35\textwidth]{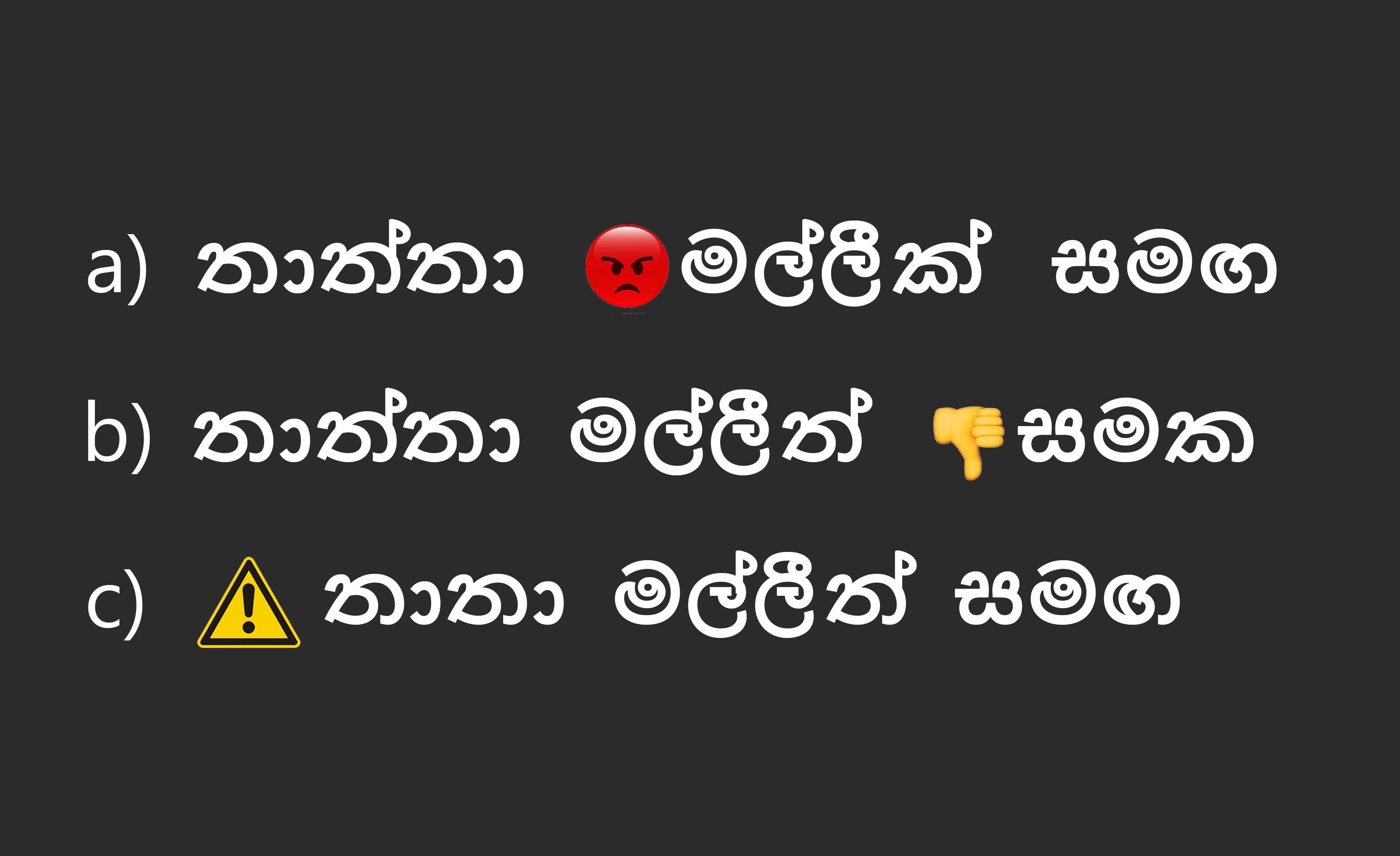}
  \label{fig:idealsub2}}
  % \end{subfigure}
  % \begin{subfigure}[b]{0.35\textwidth}
  \subfloat[Markups for personal utterances]
   {\includegraphics[width=0.35\textwidth]{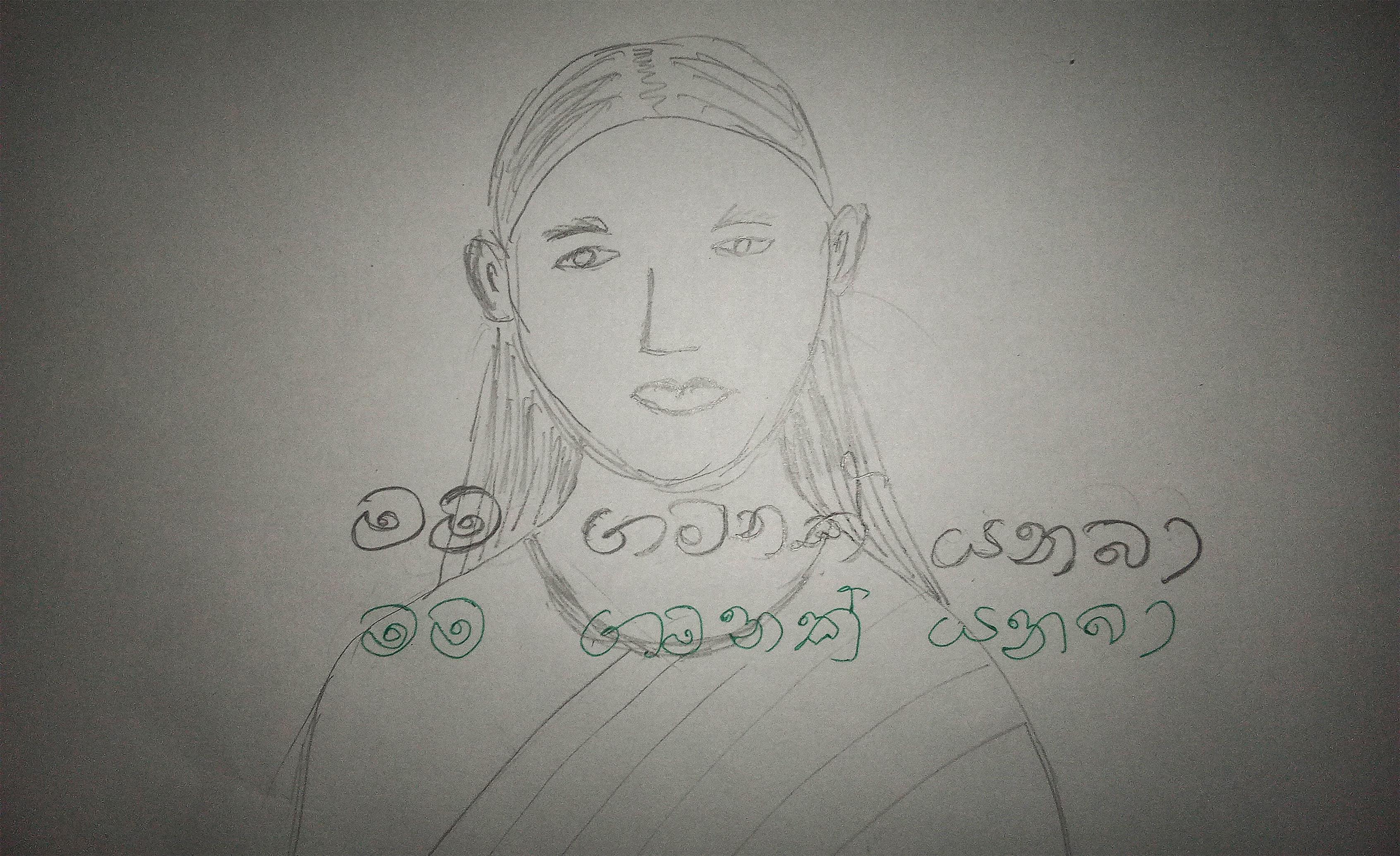}
  \label{fig:idealsub3}}
  % \end{subfigure}
  \caption{Designs envisioned by DHH Students: (a) Illustrating the number of caption lines (one or two), words per line (three), and the placement (left to the speaker, right to the speaker, directly below), (b) Markups for inaccurately transcribed word (a-c), (c) Illustrating using different colors to separate personal utterances [Refer the Appendix \ref{appendix} for English translations]}
  \label{fig:ideal}
\end{figure}

Students provided their designs (Figure \ref{fig:ideal}) and suggestions to address the challenges.

\textbf{Amount of Content:} Most students (n=6) responded with designs (Figure \ref{fig:idealsub1}) maximum of two lines of captions with three words per line. S2 highlighted, \textit{``Three words are the best, If there is more I feel like I can't read.''}. All the students agreed that the amount of content should not exceed that limit, which makes them discouraged and reluctant to read. Students also suggested incorporating finger-guided reading method when displaying captions, drawing from their school experiences where teachers point out word by word in a passage to guide students to read. Participant (S2) remarked, \textit{``Could you highlight each word as I go along? It simplifies the reading process.''}. 

\textbf{Association with the Speaker:} Similar to \citep{ExploringAugmentedReality2018} students liked when captions move with head movement and like to keep captions closer to the speakers' face. However, most students (n=6) found positioning of traditional captions problematic, causing extra effort \textit{``I have to move my head to position captions closer to the speaker, which is quite challenging and uncomfortable over an extended period.''} (S4). Some students (n=3) find it distracting when captions occupy the central view, \textit{``I prefer to focus on either captions or lips at a time such that I can switch if I need.''} (S6). Following that students responded with designs (Figure \ref{fig:idealsub1}) with captions closer to the speakers' face either left, right or directly below.

\textbf{Error Handling:} Students primary concern was clarity in transcription; they preferred to avoid engaging with text if its accuracy was uncertain. As S7 mentioned, \textit{``If I know the words with errors, I can focus on lip-reading rather than reading those words which makes it confusing.''} However, students suggested to represent markups with minimum distraction for uncertain words. Most students (n=6) suggested using an icon next to the word. Proposed designs (Figure \ref{fig:idealsub2}) included an angry emoticon next to the word (S4), a triangle with an exclamation point (S3), and thumbs down icon next to the word (S2). From all the designs most students (n=5) preferred angry emoticon, highlighting it's use in educational activities. Additionally, S1 suggested including a red squiggly line underneath the word, mentioning its familiarity with word processing software.

\textbf{Wearer's Voice:} To visually represent the personal utterances students (n=5) suggested to use a different color code for personal utterances (Figure \ref{fig:idealsub3}). Although, students mentioned separating out personal utterances from speaker's transcriptions could be helpful to minimize the distraction. 

\subsection{Design Goal and Proposal}
Through semi-structured interviews, we identified several challenges with the initial prototype. During subsequent co-design sessions, students proposed design solutions to these issues. Leveraging these insights and methods from prior work, we developed AR prototypes specifically addressing each identified challenge. Notably, considering the linguistic proficiency of our students, all prototypes were designed to support captions in Sinhala, the students' only proficient language.

\subsubsection*{Caption Presentation Method}
\begin{figure}[ht] 
  \centering
  \includegraphics[width=0.95\linewidth]{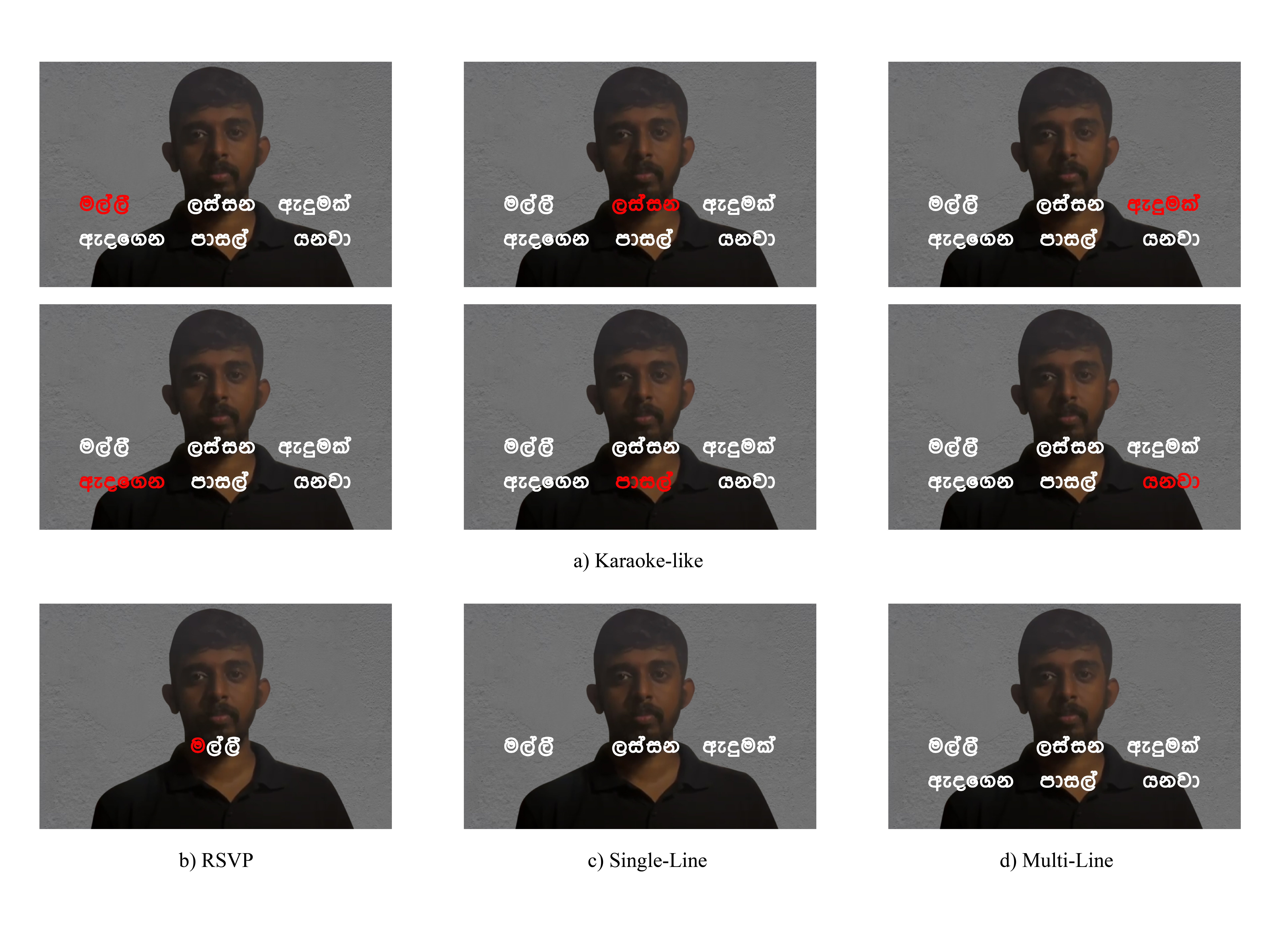} 
  \caption{Caption presentation methods [Refer the Appendix \ref{appendix} for English translations]}
  \label{fig:presentation}
\end{figure}

To meet students' needs for optimal text presentation that enhances reading speed and comprehension while reducing cognitive load, we developed various prototypes showcasing different presentation methods. These designs, illustrated in Figure \ref{fig:presentation}, were informed by prior research and refined through a co-design process with our students.

\begin{itemize}
    \item \textbf{Karaoke-Like Scrolling:} In the co-design phase, students highlighted the need for a method that facilitates easy navigation through text. They proposed a sequential word-by-word highlighting approach for captions, akin to karaoke. We implemented this method as a self-guided technique, allowing students to navigate through words using a Bluetooth mouse, with each word being highlighted in sequence. Captions are displayed two lines at a time, featuring a scrolling function that progresses to the next set of lines as the student reaches the last word of the current display. This approach was favored for its ease of use, with students preferring mouse-click navigation over visual cursors, which they found effort-intensive and distracting.
   \item \textbf{Single-Line Scrolling:} Single-line presentations are advantageous for their minimal spatial requirements, effectively reducing cognitive load and facilitating faster reading speeds \citep{ReadingOnSmartGlasses2018}. In this approach, captions are divided into lines that appear successively.
   \item \textbf{Multi-Line Scrolling:} Building on the single-line concept, this method presents captions in two-line segments. This modification is based on findings from previous studies \citep{SpeechBubbles2018} indicating that two-line presentations can enhance comprehension compared to single-line displays.
   \item \textbf{Rapid Serial Visual Presentation (RSVP):} Introduced by Forster \citep{VisualPerception1970}, RSVP boosts reading efficiency by displaying text sequentially, word by word, at a consistent screen location. A distinctive feature of this method is the use of a visual anchor, often a red letter placed at the one-third mark of each word, aiding in eye stabilization. The compact layout of RSVP is particularly well-suited for speed reading in our context. We adopted the algorithm from \citep{ReadingOnSmartGlasses2018} and employed OpenSpritz\footnote[3]{\href{https://github.com/Miserlou/Glance-Bookmarklet}{https://github.com/Miserlou/Glance-Bookmarklet} (last accessed January 02, 2024)} , an open-source speed reading bookmarklet, in our implementation.
\end{itemize}

% \begin{figure}[ht]
%   \centering
%   \includegraphics[width=\linewidth]{figures/tp_full-02.jpg}
%   \caption{Designed Prototypes: a) Caption Presentation Methods, b) Caption Placements, c) Word-by-Word Highlighting Strategies}
%   \label{Fig:Presentation_Types}
%   \Description{description}
% \end{figure}

\subsubsection*{Highlighting Strategy:}
To implement word-by-word highlighting in Karaoke-like presentation, we selected several well-known text-highlighting techniques from previous studies (Table \ref{tab:highlightprior}).
% % Highlight reference table
\begin{table}[ht]
  \tbl{Highlighting Strategies from Prior Work}
  {\begin{tabular}{ll}
    \toprule
    Highlighting Strategy & Prior Research\\
    \midrule
    \textcolor{red}{Font Color} (color\_r) &  Ponce and Mayer \citep{AnEyeMovement2014}\\
    \textbf{Bold Face} (bold) &  Rello, Saggion, and Baeza-Yates \citep{KeywordHighlighting2014}\\
    \hl{Background} (color\_bg) &  Chi, Hong, Gumbrecht, and Card \citep{ScentHighlights2005}\\
    % UPPERCASE (uc) &  Berger, Niebuhr, and Fischer \citep{ElicitingExtraProminence2018}\\
    \underline{Underlining} (ul) & Vertanen and Kristensson \citep{OnTheBenefits2008}\\
    \textit{Italicized} (it) & Berger et al. \citep{ElicitingExtraProminence2018}\\
    Font {\large Size} (size) & Wang, Nagano, Kashino, and Igarashi \citep{VisualizingVideoSounds2016}\\
    \bottomrule
  \end{tabular}}
  \label{tab:highlightprior}
\end{table}

\begin{figure}[ht]
  \centering
  % \begin{subfigure}[b]{0.3\textwidth}
  \subfloat[Font Color]
   {\includegraphics[width=0.3\textwidth]{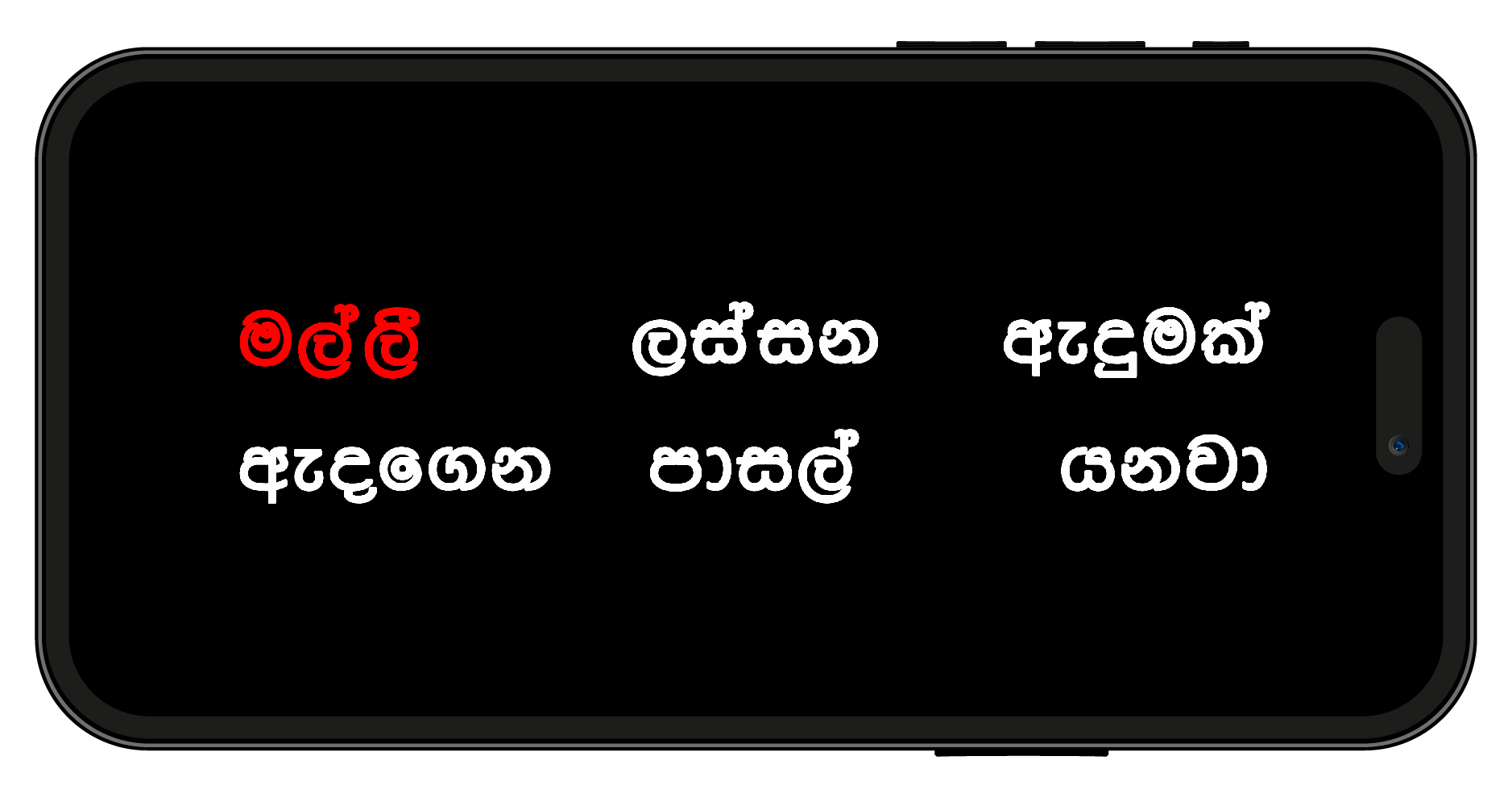}
  \label{fig:text-highlight_sub1}}
  % \end{subfigure}
  % \begin{subfigure}[b]{0.3\textwidth}
  \subfloat[Font Size]
   {\includegraphics[width=0.3\textwidth]{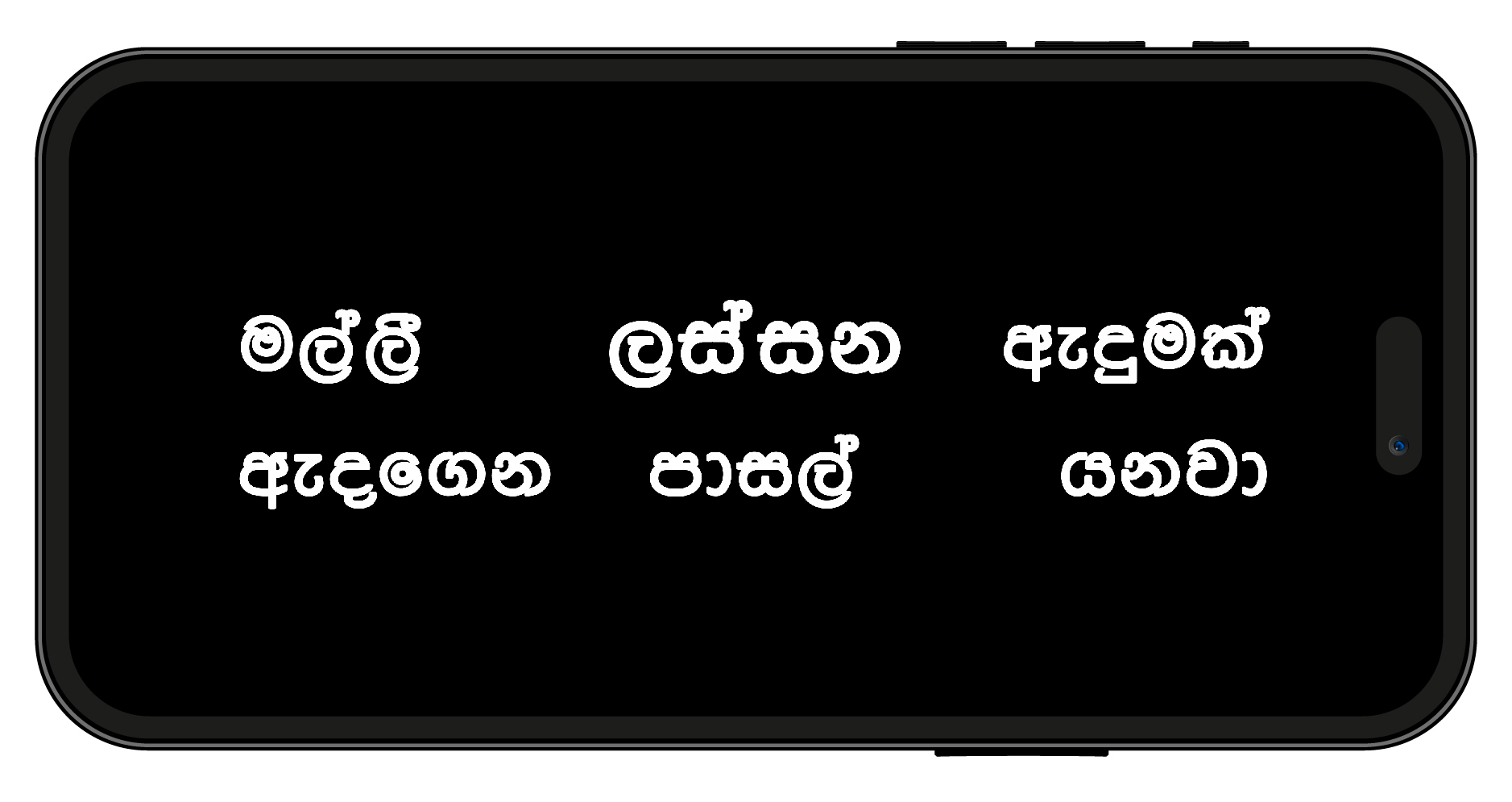}
  \label{fig:text-highlight_sub2}}
  % \end{subfigure}
  % \begin{subfigure}[b]{0.3\textwidth}
  \subfloat[Underlining]
  { \includegraphics[width=0.3\textwidth]{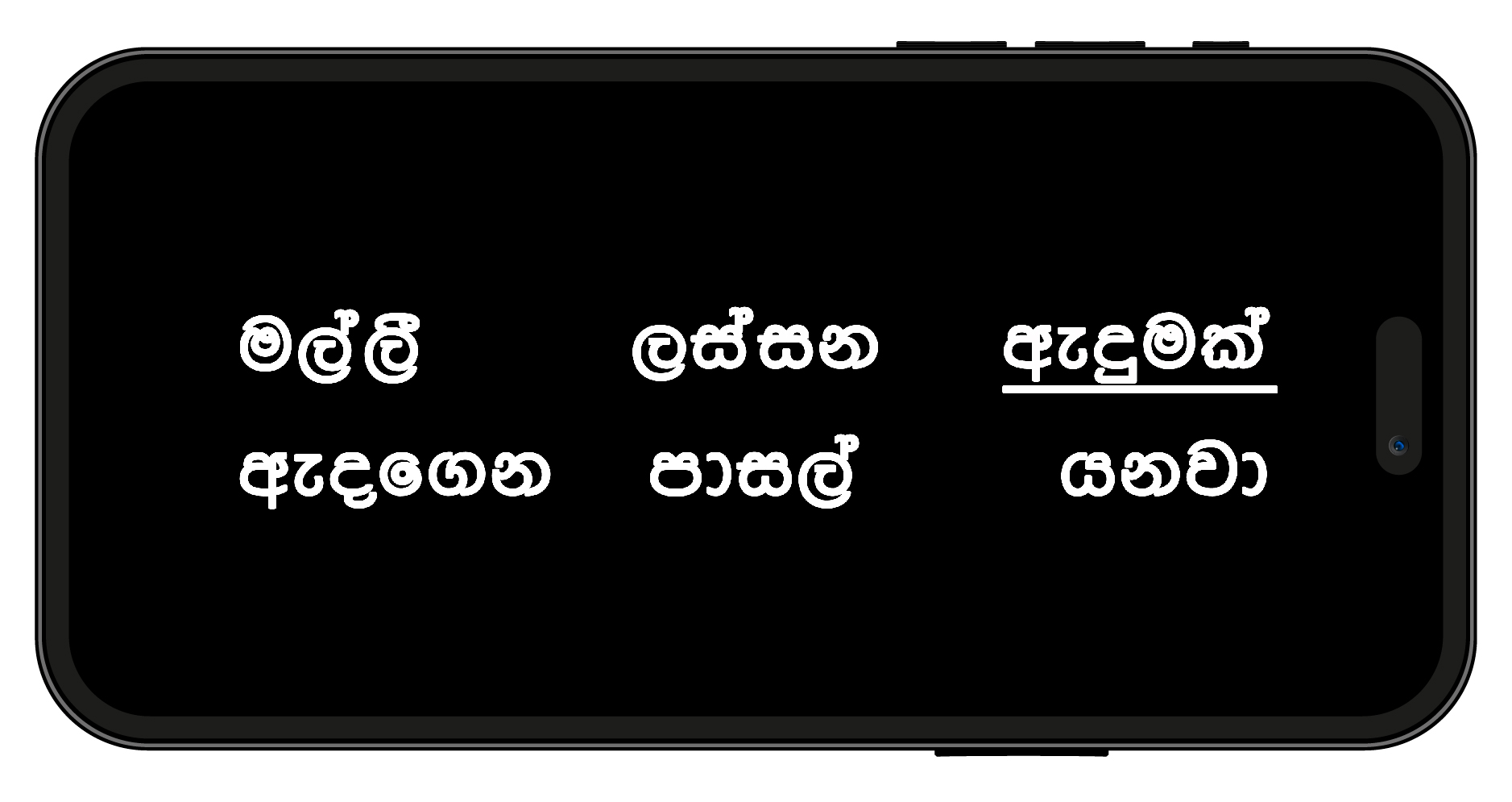}
  \label{fig:text-highlight_sub3}}\par
  % \end{subfigure}
  % \begin{subfigure}[b]{0.3\textwidth}
  \subfloat[Bold]
   {\includegraphics[width=0.3\textwidth]{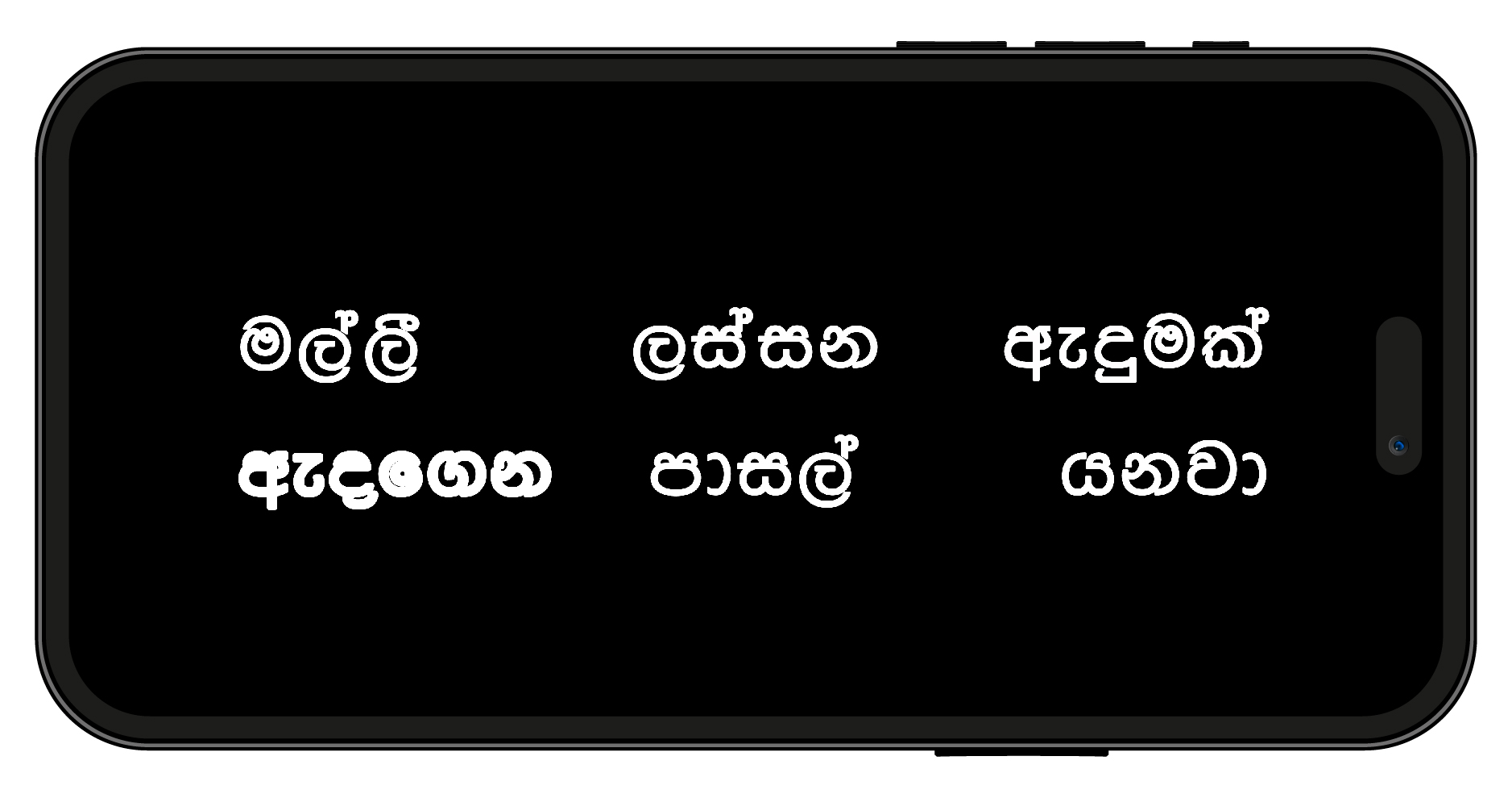}
  \label{fig:text-highlight_sub4}}
  % \end{subfigure}
  % \begin{subfigure}[b]{0.3\textwidth}
  \subfloat[Background Color]
   {\includegraphics[width=0.3\textwidth]{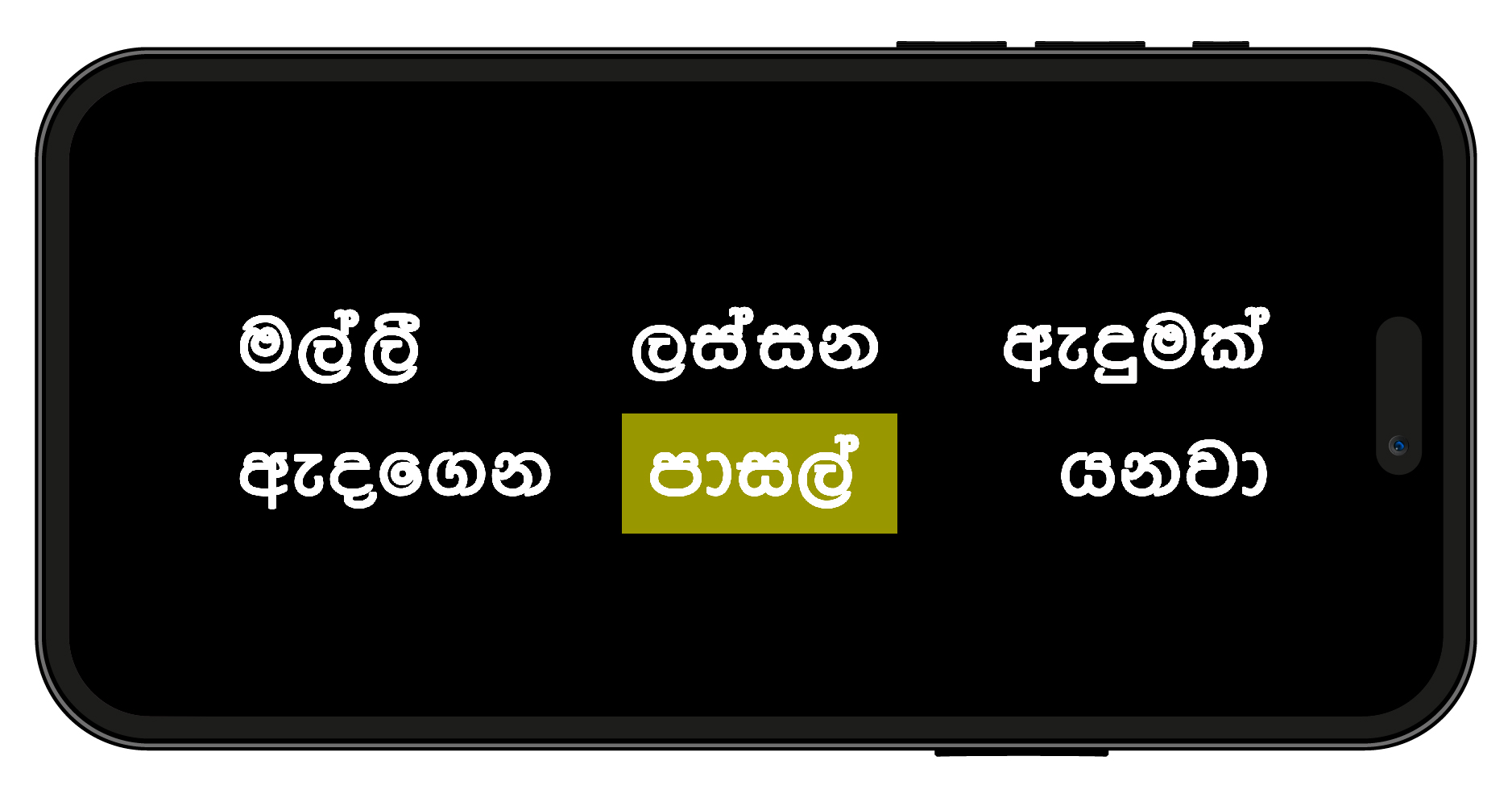}
  \label{fig:text-highlight_sub5}}
  % \end{subfigure}
  % \begin{subfigure}[b]{0.3\textwidth}
  \subfloat[Italicized]
   {\includegraphics[width=0.3\textwidth]{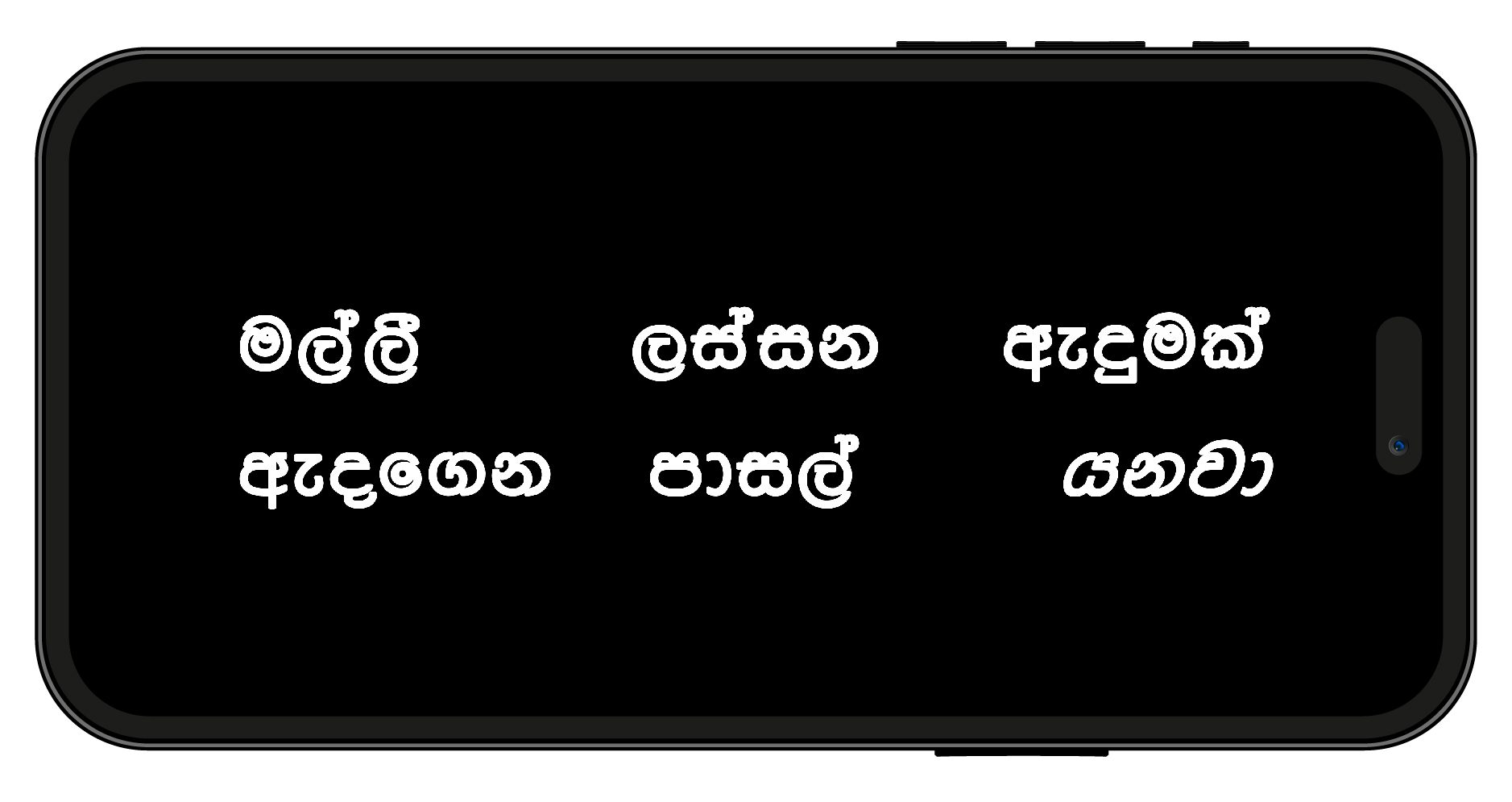}
  \label{fig:text-highlight_sub6}}
  % \end{subfigure}
  \caption{Text highlighting strategies [Refer the Appendix \ref{appendix} for English translations]}
  \label{fig:text-highlight}
\end{figure}

\subsubsection*{Placement of Caption:} 
\begin{figure}[ht]
  \centering
  % \begin{subfigure}[b]{0.4\textwidth}
  \subfloat[Left to the Speaker]
  {\includegraphics[width=0.4\textwidth]{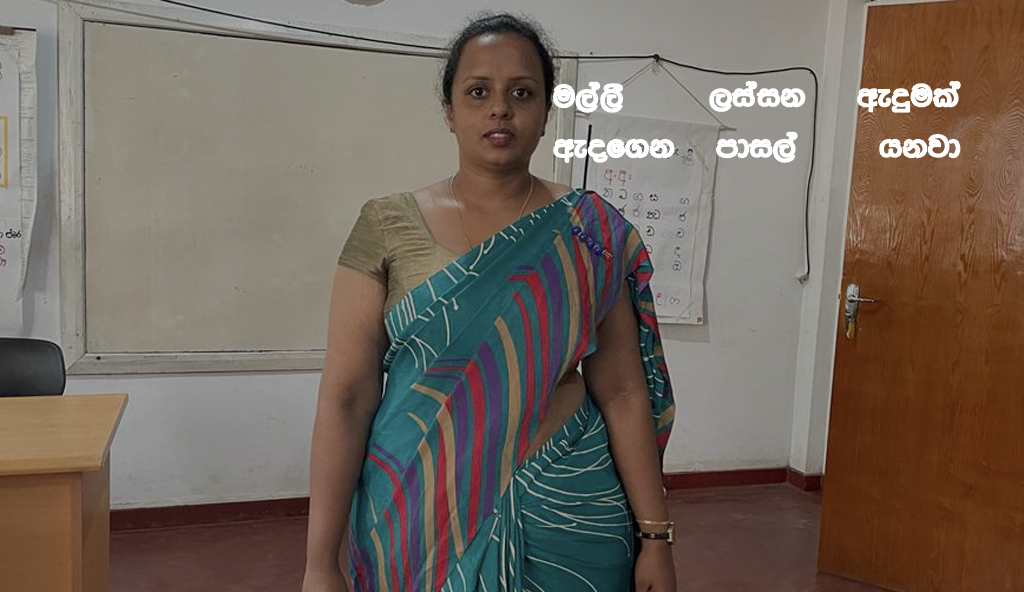}
  \label{fig:placements_sub1}}
  % \end{subfigure}
  % \begin{subfigure}[b]{0.4\textwidth}
  \subfloat[Right to the speaker]
   {\includegraphics[width=0.4\textwidth]{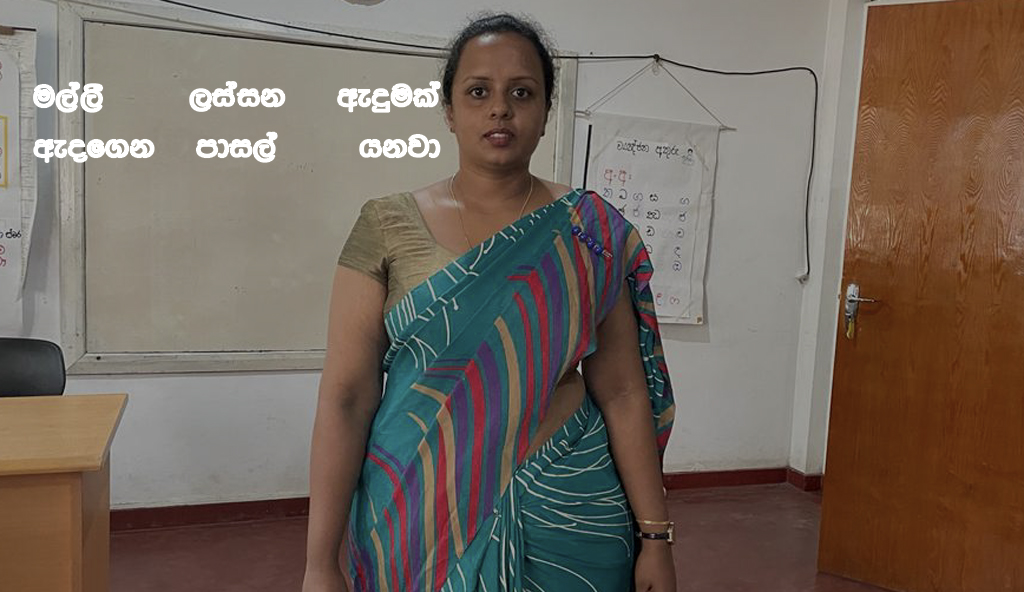}
  \label{fig:placements_sub2}}\par
  % \end{subfigure}
  % \begin{subfigure}[b]{0.4\textwidth}
  \subfloat[Directly Below]
   {\includegraphics[width=0.4\textwidth]{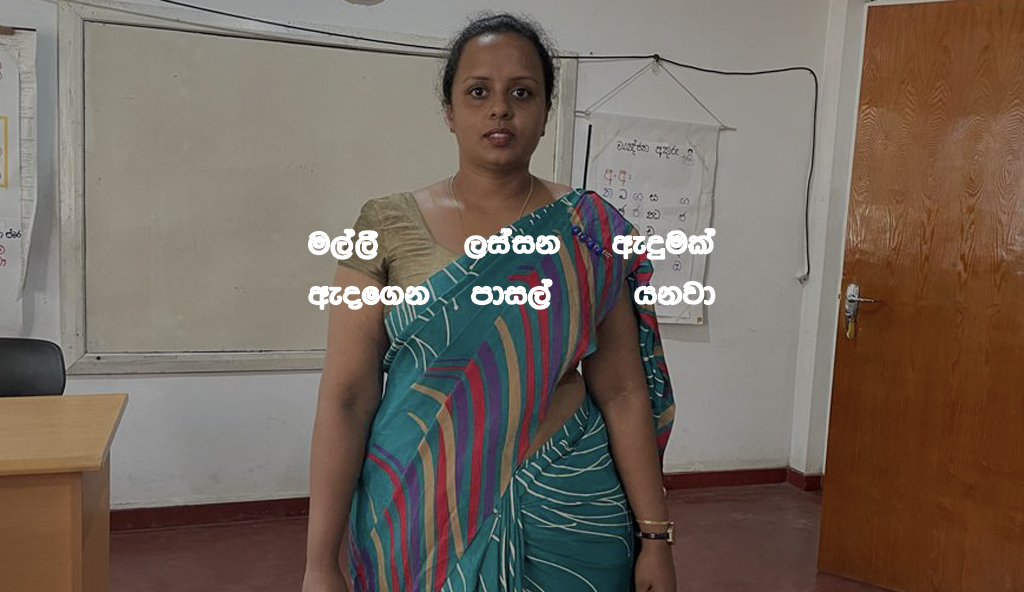}
  \label{fig:placements_sub3}}
  % \end{subfigure}
  % \begin{subfigure}[b]{0.4\textwidth}
  \subfloat[Traditional Caption]
   {\includegraphics[width=0.4\textwidth]{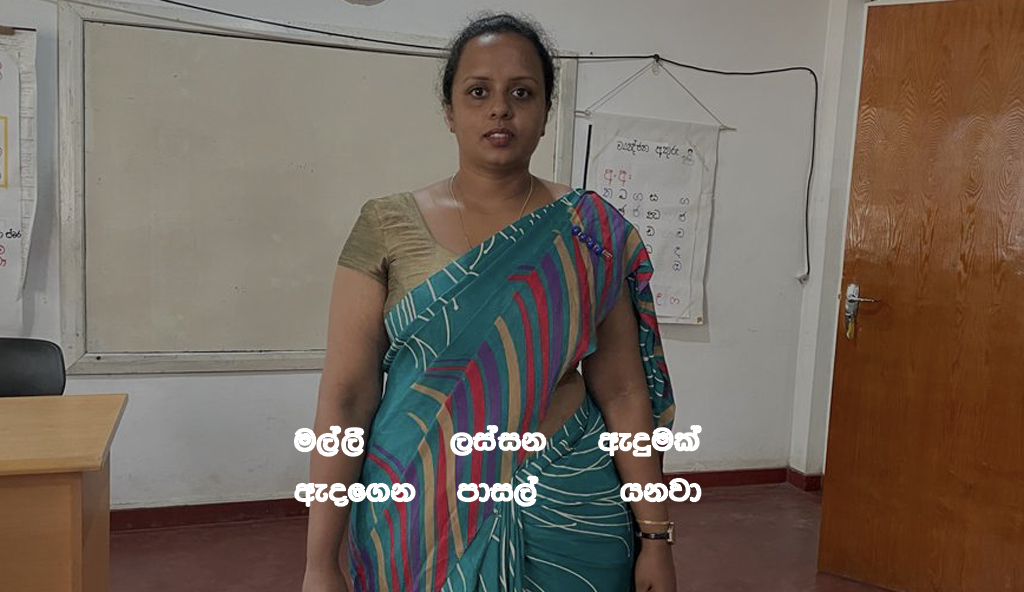}
  \label{fig:placements_sub4}}
  % \end{subfigure}
  \caption{Caption placement methods [Refer the Appendix \ref{appendix} for English translations]}
  \label{fig:placements}
\end{figure}

During the co-design sessions (see Figure \ref{fig:main}), students recommended the option to adjust caption placement relative to the speaker's position. Students' designs suggested that captions should be placed closer to the speaker's face and specifically near the mouth, either left, right, or directly-below. In response, we developed a prototype to dynamically reposition captions within their line of sight, aligning them with the speaker's face. Users are able to drag and reposition the captions either to left, right, directly-below the speakers face as shown in Figure \ref{fig:placements}. Additionally, the prototype enables caption resizing, accommodating changes in the speaker's distance from the user.

% students (n=3) more familiar with book reading wanted positioning captions left on the screen for a natural reading experience, \textit{``I like to keep captions left to the teacher; it gives a more natural reading experience, like reading a book''} (P2). students who do not rely on lip reading preferred to keep captions either left or right to the speaker, P2 noted, \textit{``When captions are positioned in line with the teacher's face, it feels very comfortable to read, rather than displaying it on the teacher's body.''}, P1 \textit{``I don't need to read lips much often, I prefer keeping captions separated from the teacher.''}. students responded with designs \ref{Fig:cap_ideal} with captions closer to the speaker's face either left, right, or directly below. 

% conclusion - captions should be kept closer to the speaker improving learning outcomes and also does not significant on the position either left right or directly

\subsubsection*{Markup for Uncertain Words}
\begin{figure}[ht]
  \centering
  % \begin{subfigure}[b]{0.33\textwidth}
  \subfloat[No Markup]
   {\includegraphics[width=0.33\textwidth]{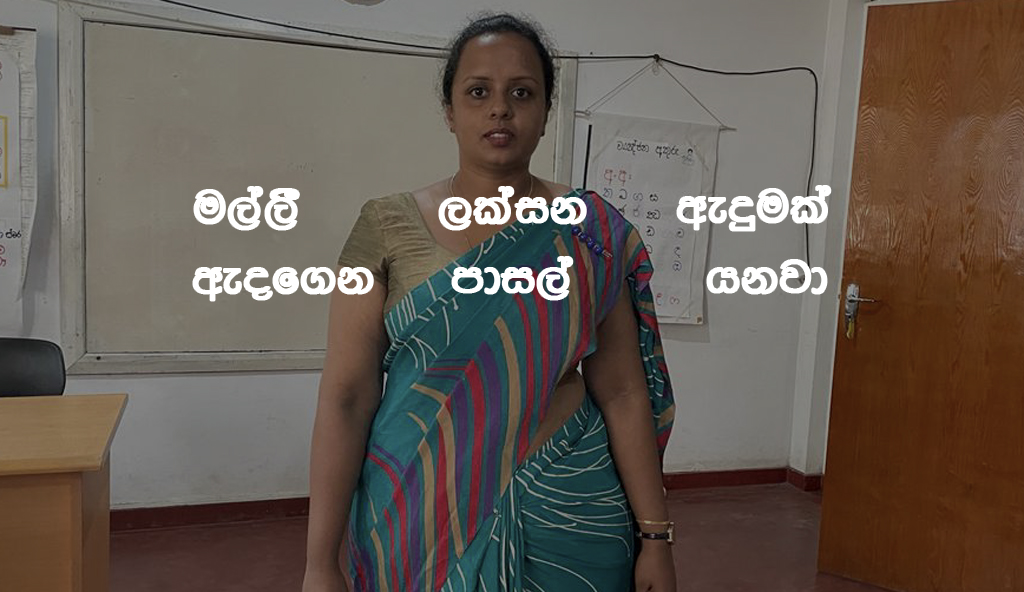}
  \label{fig:m1}}
  % \end{subfigure}
  % \begin{subfigure}[b]{0.33\textwidth}
  \subfloat[Italicized]
   {\includegraphics[width=0.33\textwidth]{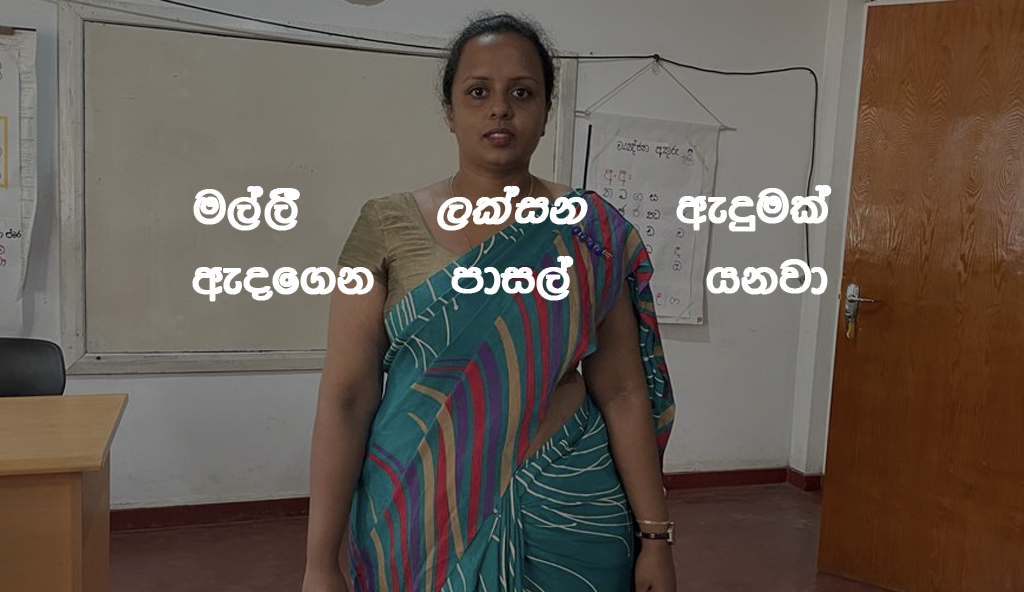}
  \label{fig:m2}}
  % \end{subfigure}
  % \begin{subfigure}[b]{0.33\textwidth}
  \subfloat[Emoticon]
  { \includegraphics[width=0.33\textwidth]{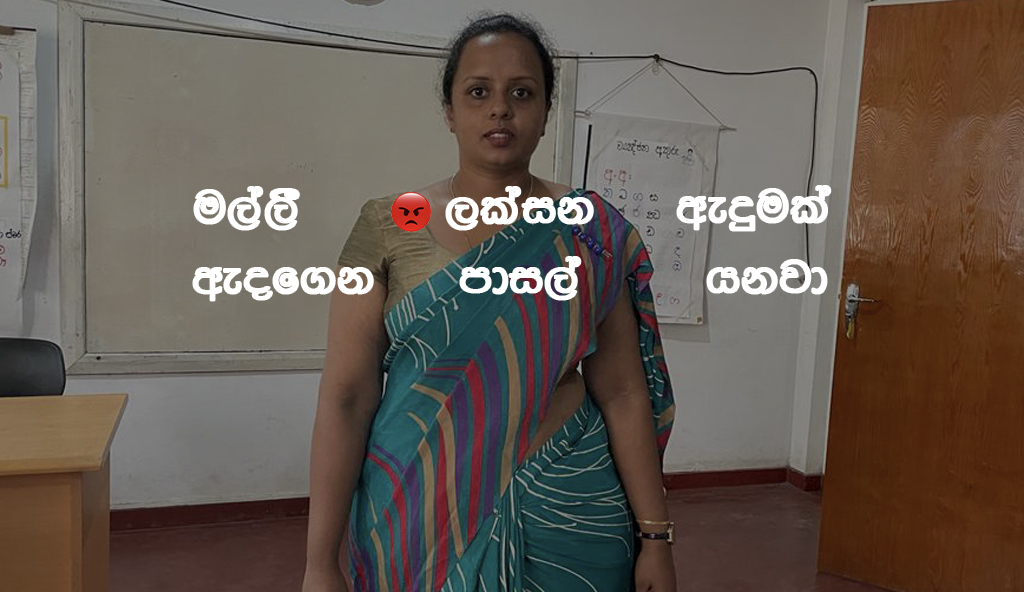}
  \label{fig:m3}}\par
  % \end{subfigure}
  % \begin{subfigure}[b]{0.33\textwidth}
  \subfloat[Bold with Yellow Font]
   {\includegraphics[width=0.33\textwidth]{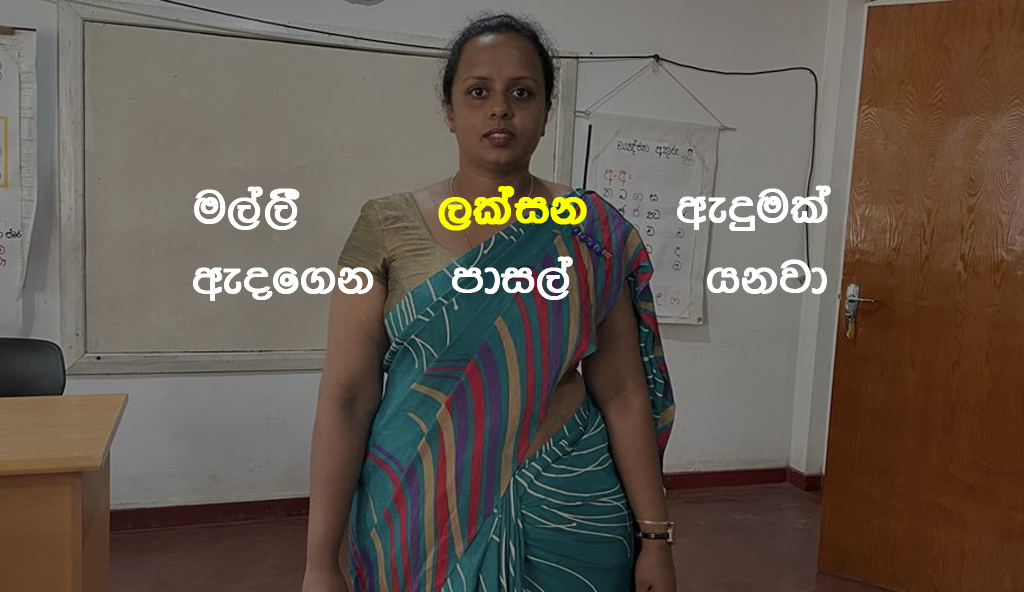}
  \label{fig:m4}}
  % \end{subfigure}
  % \begin{subfigure}[b]{0.33\textwidth}
  \subfloat[Red Squiggly Line]
   {\includegraphics[width=0.33\textwidth]{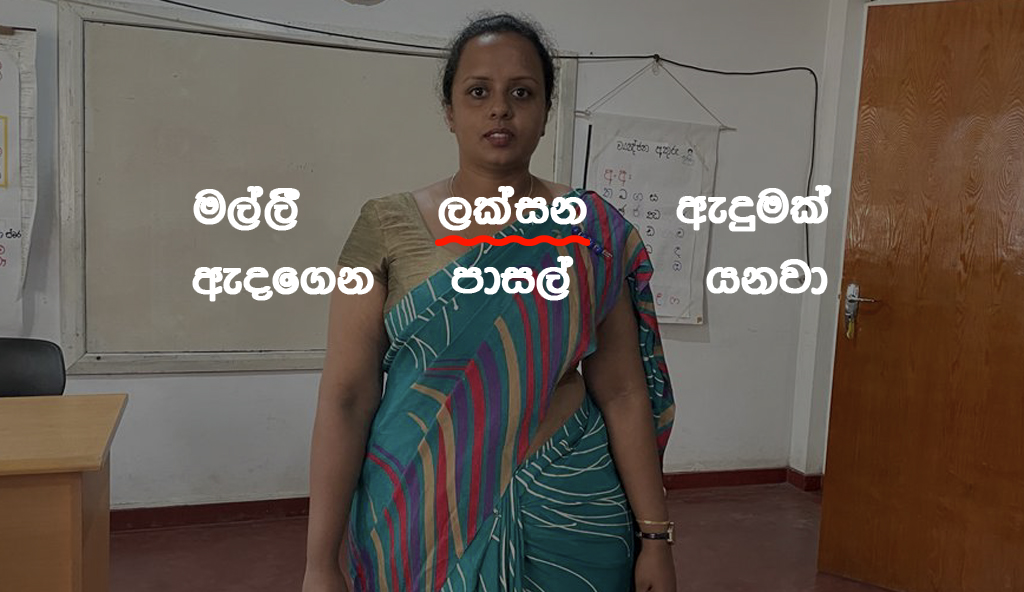}
  \label{fig:m5}}
  % \end{subfigure}
  \caption{Markup methods for uncertain words [Refer the Appendix \ref{appendix} for English translations]}
  \label{fig:uncertain}
\end{figure}

The study by Berke et al. \citep{confidence} found that users showed a preference for employing italics, underlining, and the combination of yellow text with bold formatting to denote uncertain words, with no markup as base condition. During our co-design sessions, students proposed alternative designs that incorporated emoticons. They also expressed a preference for red squiggly lines over underlining, citing their familiarity with this convention from word processing software. Based on these insights, we developed prototypes that incorporated these design elements (Figure \ref{fig:uncertain}).

\subsubsection*{Visualization of Personal Utterances}
\begin{figure}[ht]
  \centering
  % \begin{subfigure}[b]{0.4\textwidth}
  \subfloat[No Change]
   {\includegraphics[width=0.4\textwidth]{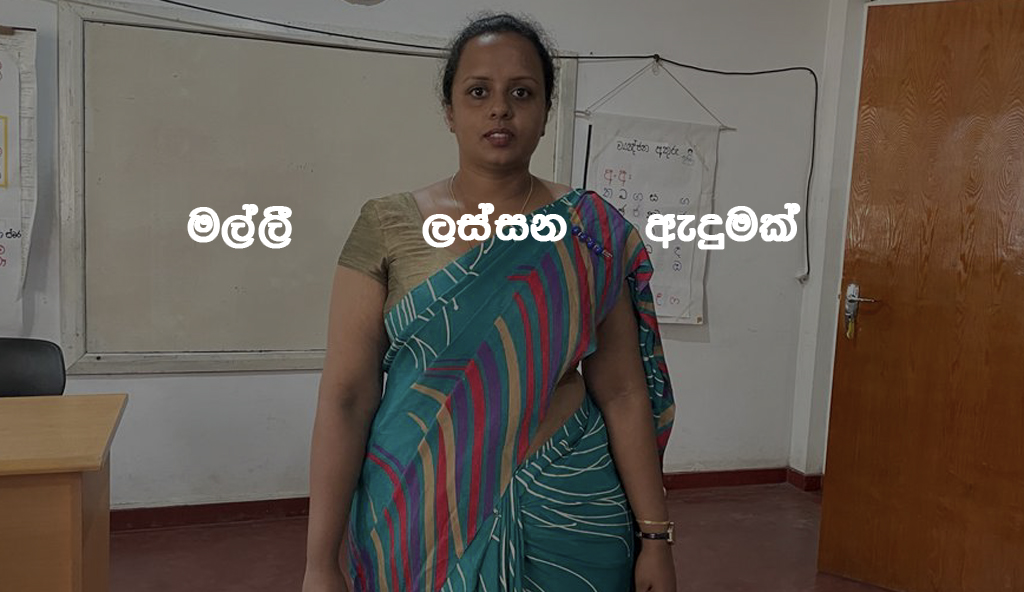}
  \label{fig:v1}}
  % \end{subfigure}
  % \begin{subfigure}[b]{0.4\textwidth}
  \subfloat[No Separation]
   {\includegraphics[width=0.4\textwidth]{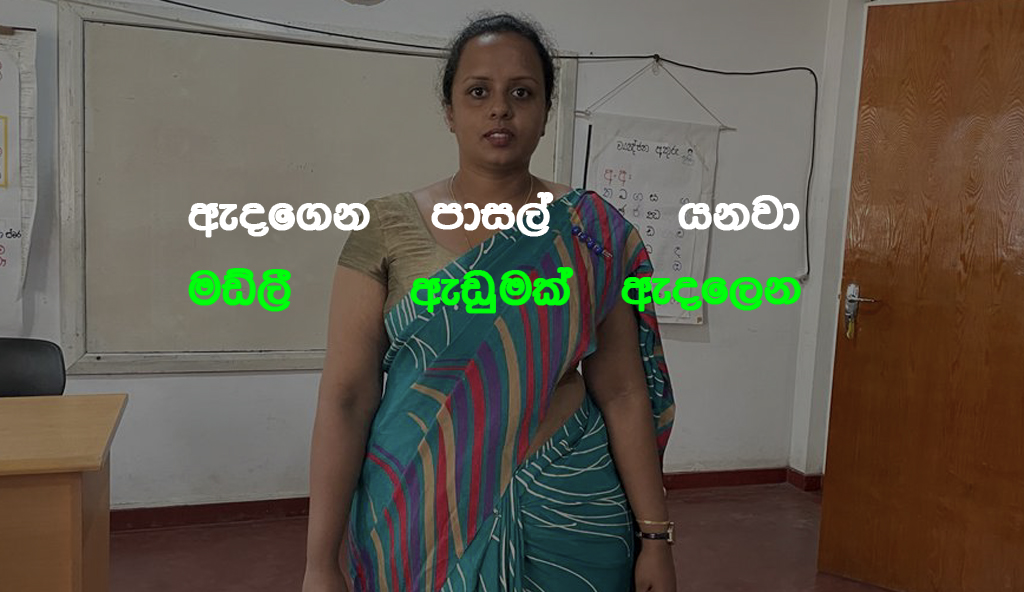}
  \label{fig:v2}}
  % \end{subfigure}
  % \begin{subfigure}[b]{0.4\textwidth}
  \subfloat[Separation without color]
   {\includegraphics[width=0.4\textwidth]{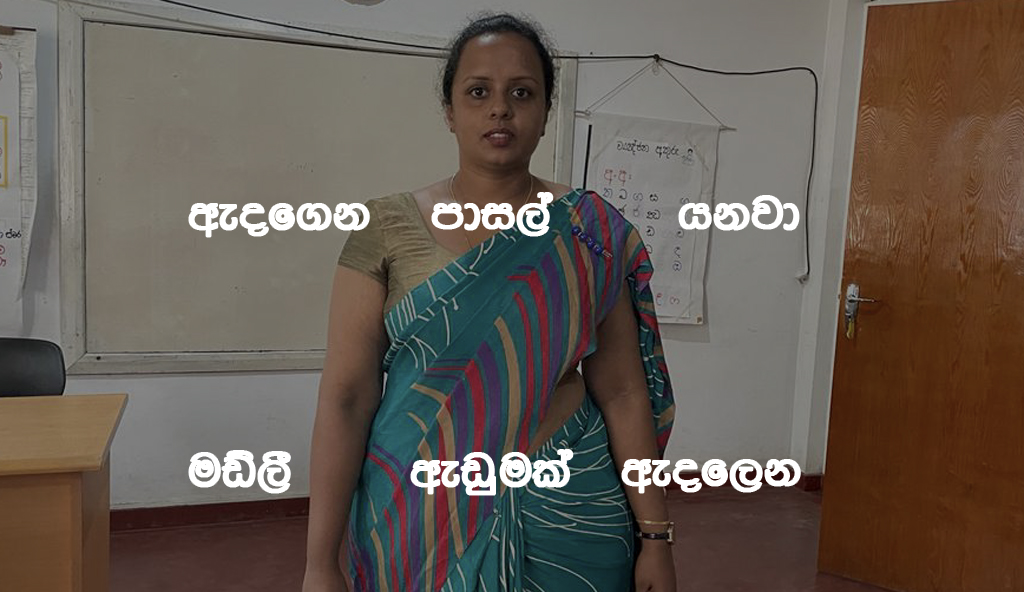}
  \label{fig:v3}}
  % \end{subfigure}
  % \begin{subfigure}[b]{0.4\textwidth}
  \subfloat[Separation with color]
   {\includegraphics[width=0.4\textwidth]{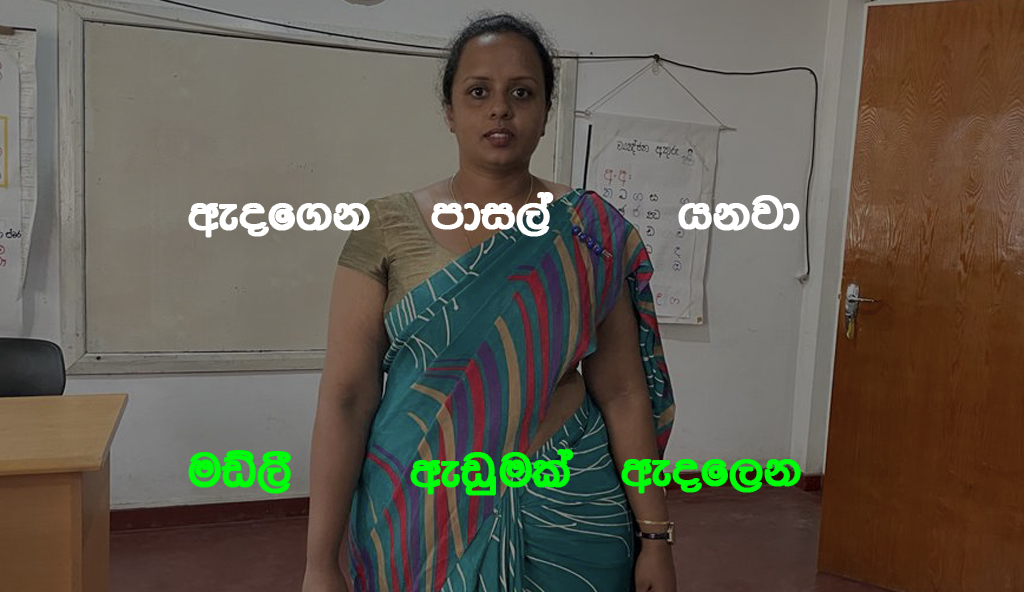}
  \label{fig:v4}}
  % \end{subfigure}
  \caption{Visualization methods for personal utterances [Refer the Appendix \ref{appendix} for English translations]}
  \label{fig:utterance}
\end{figure}

In co-design students suggested distinct color coding for personal utterances, with `Green' emerging as the most preferred choice. Students also highlighted separating personal utterances could be less distracting and easy to distinguish between speaker's and personal utterance, this has also been useful in group conversations to associate the speakers \citep{SpeechBubbles2018}. Subtitles in television, movies are typically placed at bottom-centred \citep{DynamicSubtitles2015,baker1982handbook} making it less distracting, we propose a similar approach for displaying personal utterances to allow users to maintain focus on the primary dialogue without obstructing the visual field. However, some students (n=3) mentioned filtering out personal utterances could be less distracting. Considering those we developed prototypes to evaluate each design affect on reading experience (Figure \ref{fig:utterance}).

\section{System Implementation}
\label{sec:system_implementation}

\begin{figure}[ht]
  \centering
  \includegraphics[width=0.9\linewidth]{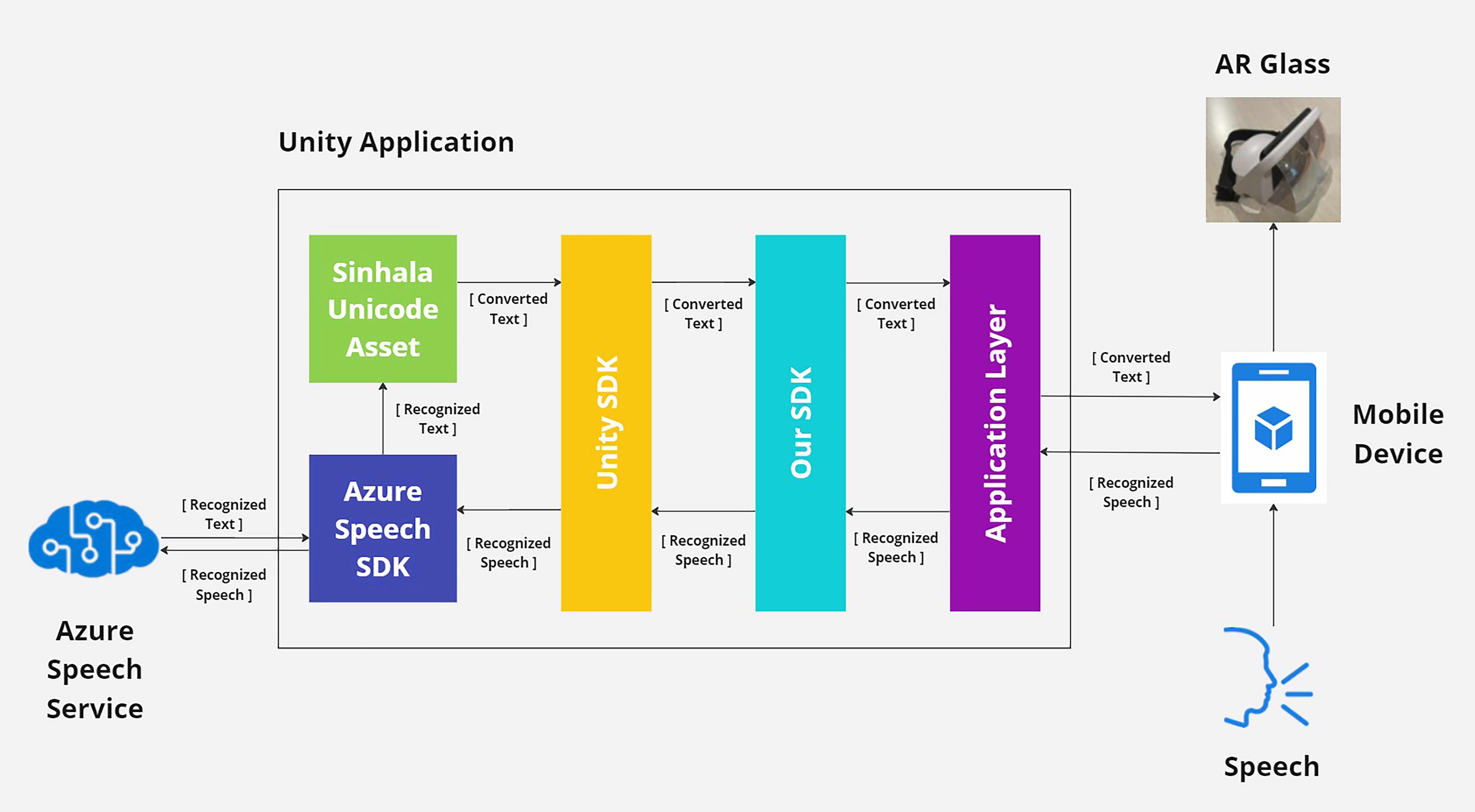}
  \captionsetup{justification=centering}
  \caption{High-level system architecture}
  \label{Fig:system}
\end{figure}

Our system comprises two main components: Augmented Reality Glass\footnote[4]{\href{https://www.alibaba.com/product-detail/2022-AR-Smart-Glasses-3D-Augmented_60816962011.html?spm=a2700.shop_index.scGlobalHeaderSmall.51.6ce94b95LHWVCP}{Alibaba-AR Glass} (last accessed January 02, 2024)} and a Unity Application. The AR glass utilizes widely available smartphones mounted on low-cost transparent headsets, reflecting the smartphone screen into the user’s field of view. We developed our application using our own Software Development Kit (SDK)\footnote[5]{\url{https://github.com/ThavinduUshan/SeEar-SDK} (last accessed March 15, 2024)} leveraging Unity SDK version 2021.3.25F1\footnote[6] {\url{https://unity.com/releases/editor/whats-new/2021.3.25} (last accessed January 04, 2024)}. Our SDK is equipped with capabilities for various caption presentation methods, dynamic caption positioning, markup tools for flagging transcription inaccuracies, word-by-word highlighting features, and personal utterance visualization techniques. This SDK integrates real-time localized captioning capabilities through the Microsoft Azure Speech-to-Text\footnote[7]{\url{https://learn.microsoft.com/en-us/azure/ai-services/speech-service/overview} (last accessed January 02, 2024)}(si-LK) service, facilitating the development of prototype applications for user studies. To address challenges in Unity Sinhala Unicode conversions, we also created a Unity Sinhala Unicode Asset\footnote[8]{\url{https://github.com/ThavinduUshan/SeEar-SDK} (last accessed March 15, 2024)}. Both the SDK and Unicode Asset is developed to support multiple platforms. High-level system architecture diagram is shown in (Figure \ref{Fig:system}), illustrating the interplay between the components.

\section{Research Questions}
\label{sec:rq}
After the design phase, to understand the effect of the designed prototypes to the overall learning experience and outcomes we investigate the following questions.

\textbf{RQ1:}
How does the placement of real-time captions closer to the speaker affect learning outcomes of DHH students in classroom settings?

\textbf{RQ2:}
How do we visually represent the uncertain words in real-time transcriptions?

\textbf{RQ3:}
What sequential word-by-word highlighting strategy is most effective?

\textbf{RQ4:}
How does the real-time caption presentation method impact overall learning outcomes, and to what extent does the implementation of a reading assistant presentation method enhance these learning outcomes?

\textbf{RQ5:}
How do we visually represent the personal utterances of the wearer? 

\textbf{RQ6:}
Do users express a subjective preference for using our system during their classroom sessions?

\section{User Studies}
\begin{figure}[H]
    \centering    
    \includegraphics[width=1\linewidth]{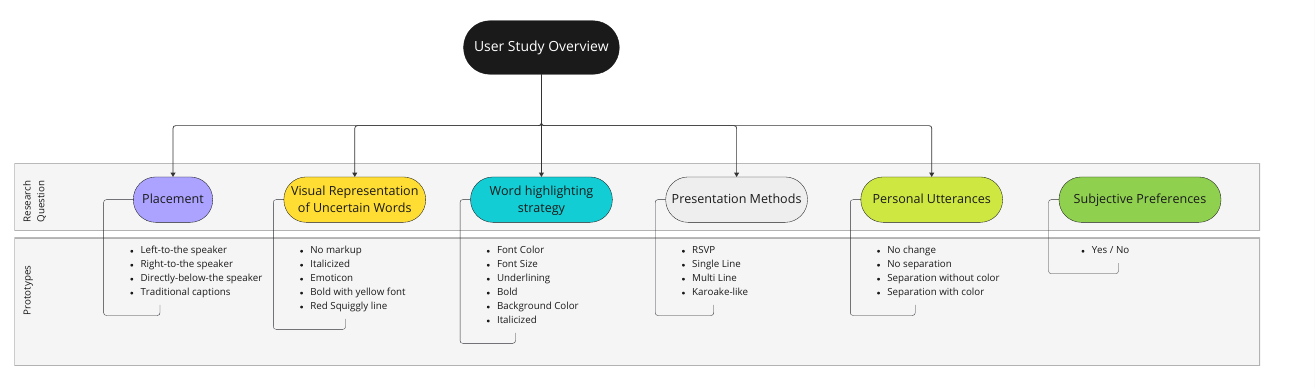}
    \caption{Summary of User Studies}
    \label{fig:user-study-summary}
\end{figure}

We conducted a series of five user studies and a comprehensive system evaluation to address our RQs. All the user studies were conducted with the same group of participants, and each study's design, apparatus, procedure, and results are provided separately. Summary of the studies are described in (Figure \ref{fig:user-study-summary})

% \begin{figure}[ht]
%   \centering
%   \includegraphics[width=0.4\linewidth]{figures/user-study.jpg}
%   \captionsetup{justification=centering}
%   \caption{Participant engaging in a user study}
%   \label{Fig:user-study}
% \end{figure}

\subsection{Participants}
For the studies, we recruited 24 participants (11 females \& 13 males) with ages ranging from 13-19 (\(M = 15.458, SD = 1.911\)) from a School of Deaf. Our participants had varying degrees of hearing moderate (6), severe (8), and extremely severe (10).  All the participants' native language was “Sinhala” and since our study uses reading from an AR glass we confirmed that all the participants had non-disabled vision for daily activities without glasses. Teachers confirmed that all the participants can read and understand the study materials we used in the studies. Communication with participants was facilitated by a Teacher of Deaf who is an expert in Sign Language.

\subsection{Study 1: Placement of Captions}
\subsubsection{Study Design}
This study aimed to explore the effect of placing captions closer to the speaker on learning outcomes, addressing our first research question (RQ1). We developed a prototype using our system (Section \ref{sec:system_implementation}) that allows participants to dynamically reposition captions relative to the speaker in four positions (as illustrated in Figure \ref{fig:placements}). Within the study, participants were asked to evaluate each design in a 5-point Likert scale in comprehension, reading speed, and comfort, with a particular focus on their influence on learning outcomes. 

\subsubsection{Procedure}
Before the study, we explained to the participants the objective of the study and introduced them to the system. A teacher conducted a session with a chosen text from a school textbook\footnote[9]{\url{http://www.edupub.gov.lk/Administrator/Sinhala/9/sin\%20g-9/sinhala\%20g9.pdf} (last accessed January 02, 2024)} and participants read captions in real-time. Each participant read a text of ‘300 words’ from each design (Figure \ref{fig:placements}) and we counterbalanced the order of designs with a Latin-square design. The study took about 5 hours in total. 

After the study, we asked participants to rate each design in a 5-point Likert Scale according to the following variables Table \ref{tab:study1q}.

% Comprehension: How well did you understand the content while using this design? (Very Poorly - 1, Very Well - 5)

% Reading Speed: How would you describe your 
% reading speed when using this design? (Very Slow - 1, Very Fast - 5)

% Comfort: How comfortable was it to read with this design? (Very Uncomfortable - 1, Very Comfortable - 5)

% Finally, participants ranked the designs according to their preference from Most Preferred to Least Preferred (1 to 4).

\begin{table}[ht]
  \tbl{List of Questions Used in Study 1}
  {\begin{tabular}{p{0.5\textwidth}p{0.4\textwidth}}
    \toprule
    Questions & Scale\\
    \midrule
    Comprehension: How well did you understand the content while using this design? &  5-point Likert Scale from Very poorly to Very well \\
    Reading Speed: How would you describe your 
reading speed when using this design? & 5-point Likert Scale from Very slow to Very fast \\
    Comfort: How comfortable was it to read with this design? & 5-point Likert Scale from Very uncomfortable to Very comfortable\\
    \bottomrule
  \end{tabular}}
  \label{tab:study1q}
\end{table}

Finally, participants ranked the designs according to their preference from Most Preferred to Least Preferred (1 to 4).

\subsubsection{Study Results}
We performed an analysis based on the data we collected from all 24 participants. Table \ref{tab:prototype_results} provides a summary of mean values (with standard deviations) of comprehension, reading speed, comfort for each design and the rankings based on participants' preferences.

\begin{table}[ht]
  \tbl{Descriptive results of study 1}
  {\begin{tabular}{cccccccc}
    \toprule
    & \multicolumn{2}{c}{Comprehension} & \multicolumn{2}{c}{Reading Speed} & \multicolumn{2}{c}{Comfort} & Rank\\
 & M& SD& M& SD& M& SD&\\
    \midrule
     left-to-the-speaker (Design A)& 4.042 & 0.751& 3.458 & 0.779 & 4.167& 0.637 &3\\
     right-to-the-speaker (Design B)& 4.167 & 0.482 & 3.417 & 0.584 & 4.208 & 0.721 &2\\
     directly-below-the-speaker (Design C)& 4.375 & 0.495 & 4.042 & 0.690 & 3.958 & 0.550 &1\\
     traditional captions (Design D)& 3.292 & 0.690 & 3.375 & 0.647 & 3.875 & 0.680 & 4\\
    \bottomrule
  \end{tabular}}
  \label{tab:prototype_results}
\end{table}

\textbf{Comprehension:}
Friedman test \citep{Friedman1999} indicated significance among the four prototypes (Friedman's \(\chi^2 = 31.871, df = 3, p < .001\)). Post-hoc pairwise comparisons using Wilcoxon signed-rank test \citep{Friedman1999} revealed that Design C and Design B rated significantly higher for comprehension compared to Design D, with \(p < 0.001\). Design A showed a higher comprehension compared to Design D, \(p < 0.05\).  We could not find any statistically significance between other prototypes, \(p > 0.05\). Results suggested that comprehension is significantly higher when captions are placed closer to the speaker compared to traditional captions.\\
\textbf{Reading Speed:}
Friedman test indicated significance among the prototypes (Friedman's \(\chi^2 = 12.516, df = 3, p = 0.005\)), and Wilcoxon signed-rank test revealed that participants reported a significantly higher reading speed in Design C compared to Design A, Design B, and Design D with \(p < 0.05\). We could not find any significance among other prototypes, \(p > 0.05\). Results suggested that placing captions directly below the speaker significantly increases the reading speed compared to placing captions either left, right to the speaker or traditional bottom-centered. \\
\textbf{Comfort:}
Friedman test did not reveal significant differences among the prototype,  \(p > 0.05\). However, Design A and Design B showed higher mean comfort values compared to Design C and Design D. The results indicate that captions placed separately from the speaker enhances reading comfort compared to captions displayed on the speaker's body, but not significantly. \\
\textbf{Preference Ranking:}
Friedman showed a significant among the prototypes (Friedman's \(\chi^2 = 10.050, df = 3, p < .05\)). Wilcoxon signed-rank test revealed that Design C is significantly preferred over Design A, Design B and Design D, with \(p < 0.05\). However, we could not find any significance among other prototypes, \(p > 0.05\).

\begin{figure}[H]
  \centering
  % \begin{subfigure}[b]{0.48\textwidth}
  \subfloat[Mean Comprehension]
  { \includegraphics[width=0.48\textwidth]{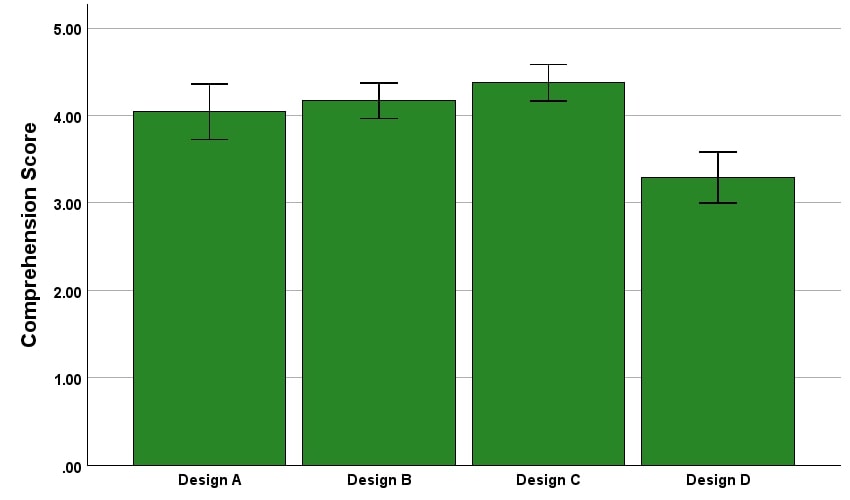}
  \label{fig:place_com}}
  % \end{subfigure}
  % \begin{subfigure}[b]{0.48\textwidth}
  \subfloat[Mean Reading Speed]
   {\includegraphics[width=0.48\textwidth]{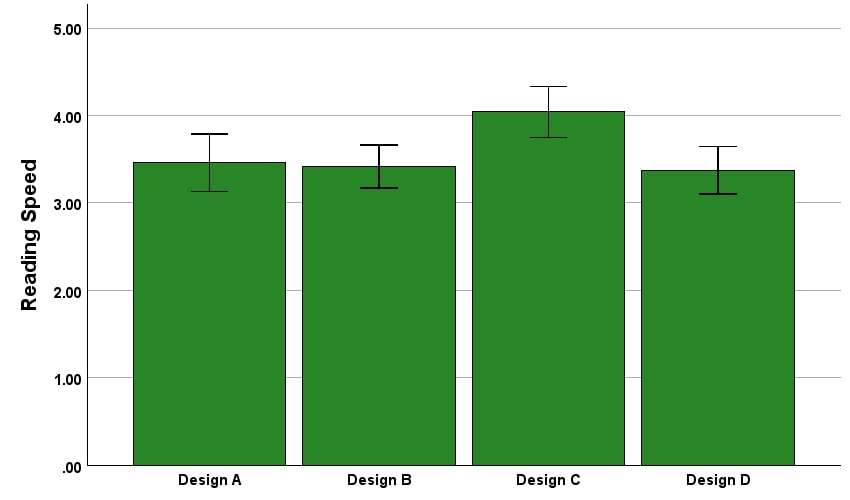}
  \label{fig:place_rs}}\par
  % \end{subfigure}
  % \begin{subfigure}[b]{0.48\textwidth}
  \subfloat[Mean Comfort]
   {\includegraphics[width=0.48\textwidth]{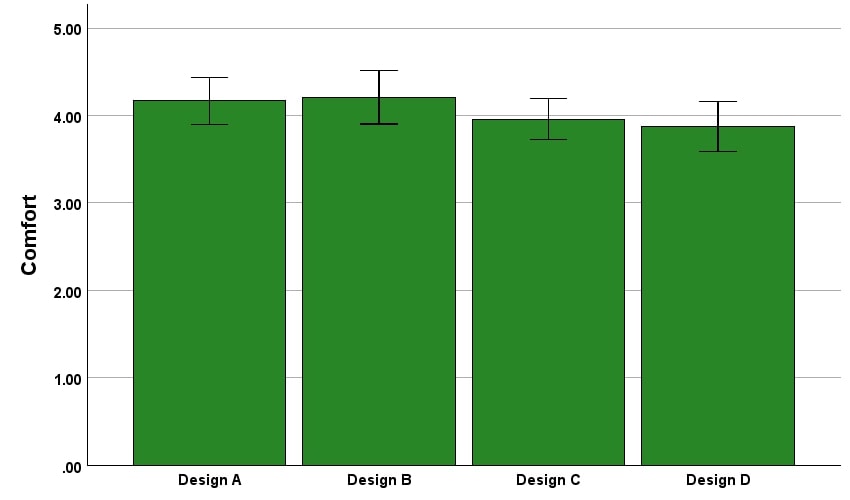}
  \label{fig:place_cm}}
  % \end{subfigure}
  % \begin{subfigure}[b]{0.49\textwidth}
  \subfloat[Preference]
   {\includegraphics[width=0.48\textwidth]{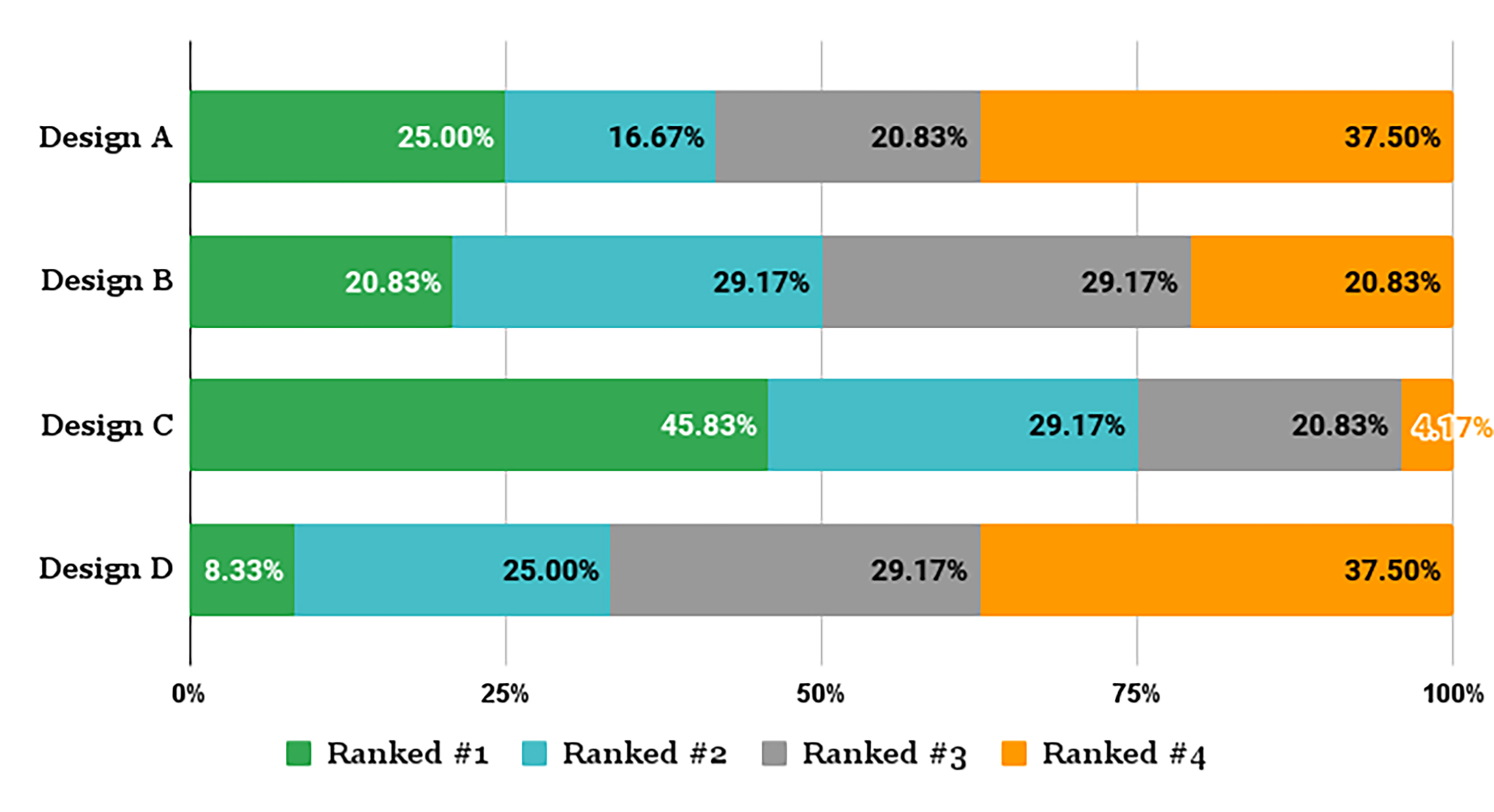}
  \label{fig:place_rank}}
  % \end{subfigure}
  \caption{Comparison of Caption Placements}
  \label{Fig:placement_results}
\end{figure}

The results suggest that Design C is notably preferred over Design A, Design B and Design D (Figure \ref{Fig:placement_results}).

\subsection{Study 2: Markups for Uncertain Words}
\subsubsection{Study Design}
In this study, we performed an in-depth analysis of how various markups for uncertain words affect the reading experiences, addressing our second research question (RQ2). We developed five prototypes (Figure \ref{fig:uncertain}) with design elements chosen in user-centred design process. Within the study, participants were asked to evaluate each design under the list of questions explained in Table \ref{tab:study2q}

\subsubsection{Apparatus and Procedure}
Before the study, we transcribed captions from five classroom sessions conducted by a teacher in real-time and used confidence values of the ASR with a probability threshold of 0.995 used by Berke et al. \citep{confidence} to separate uncertain words. These words are marked before the study with each design element. This is done to maintain the consistency in uncertainties among each design. 

Participants read captions from each design (Figure \ref{fig:uncertain}) in all five sessions and we counterbalanced the order of designs with a Latin-Square design. The study spanned for five days with an average of 6 hours per day.

After completing each reading session, participants answered three questions (Table \ref{tab:study2q} Q1, Q2, Q3), and after completing the study they ranked the designs according to preference. 

\begin{table}[ht]
  \caption{List of Questions Used in Study 2}
  {\begin{tabular}{p{0.5\textwidth}p{0.4\textwidth}}
    \toprule
    Questions & Scale\\
    \midrule
    Q1. I found visual cue clear and understandable. &  5-point Likert Scale from Strongly Disagree to Strongly Agree. \\
    Q2. I found it easy to notice uncertain words using the visual cue without detracting from my focus on the main content. & 5-point Likert Scale from Strongly Disagree to Strongly Agree \\
    Q3. I found the visual cue distracting & Yes=1/No=0\\
    Q4. Rank the visual markups according to preference& Most Preferred to Least Preferred (1 to 5) \\
    \bottomrule
  \end{tabular}}
  \label{tab:study2q}
\end{table}

\subsubsection{Study Results}
\begin{figure}[H]
  \centering
  % \begin{subfigure}[b]{0.49\textwidth}
  \subfloat[Clear and Understandable]
   {\includegraphics[width=0.49\textwidth]{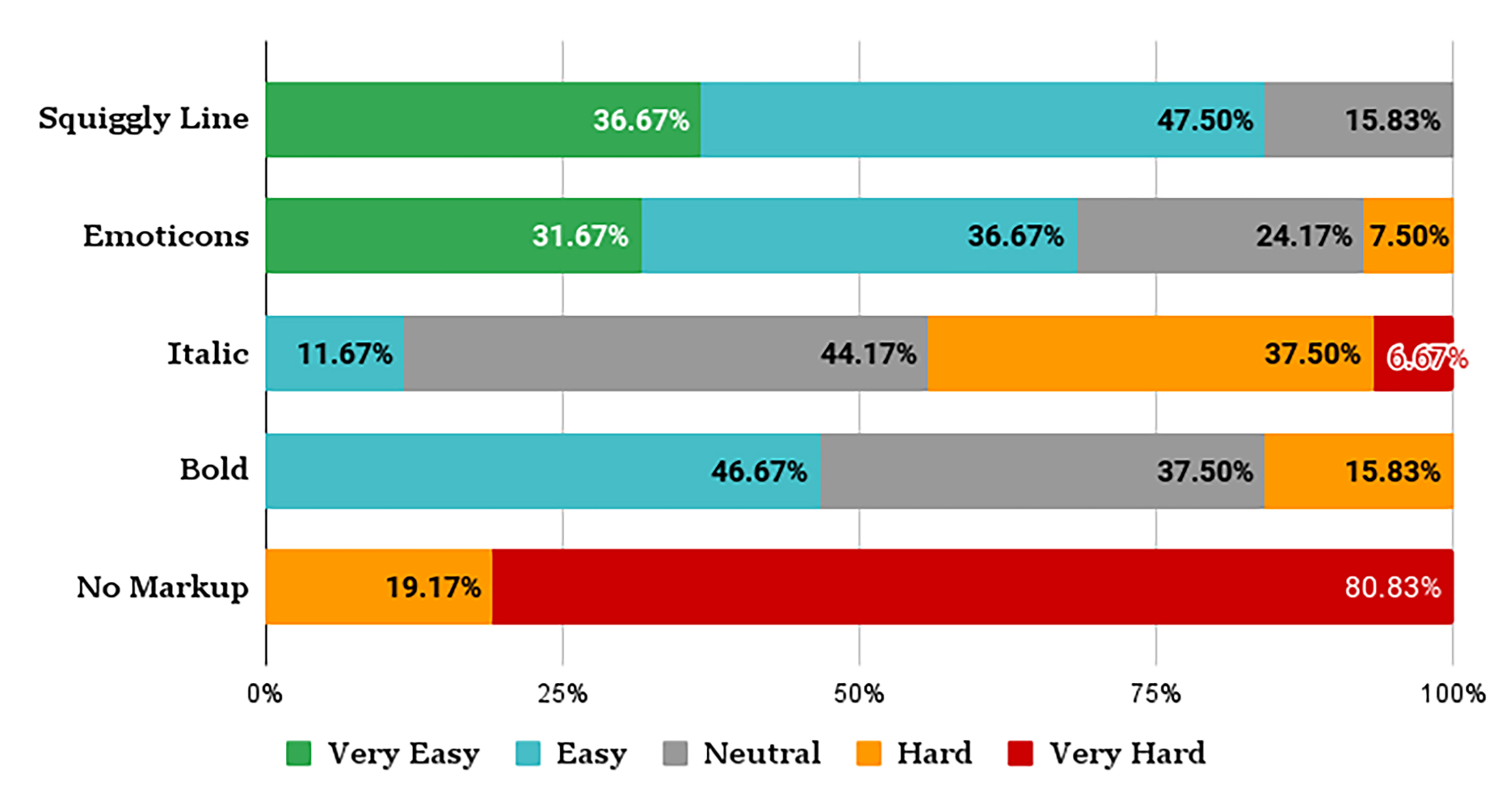}
  \label{fig:con_clear}}
  % \end{subfigure}
  % \begin{subfigure}[b]{0.49\textwidth}
  \subfloat[Easy to Notice]
   {\includegraphics[width=0.49\textwidth]{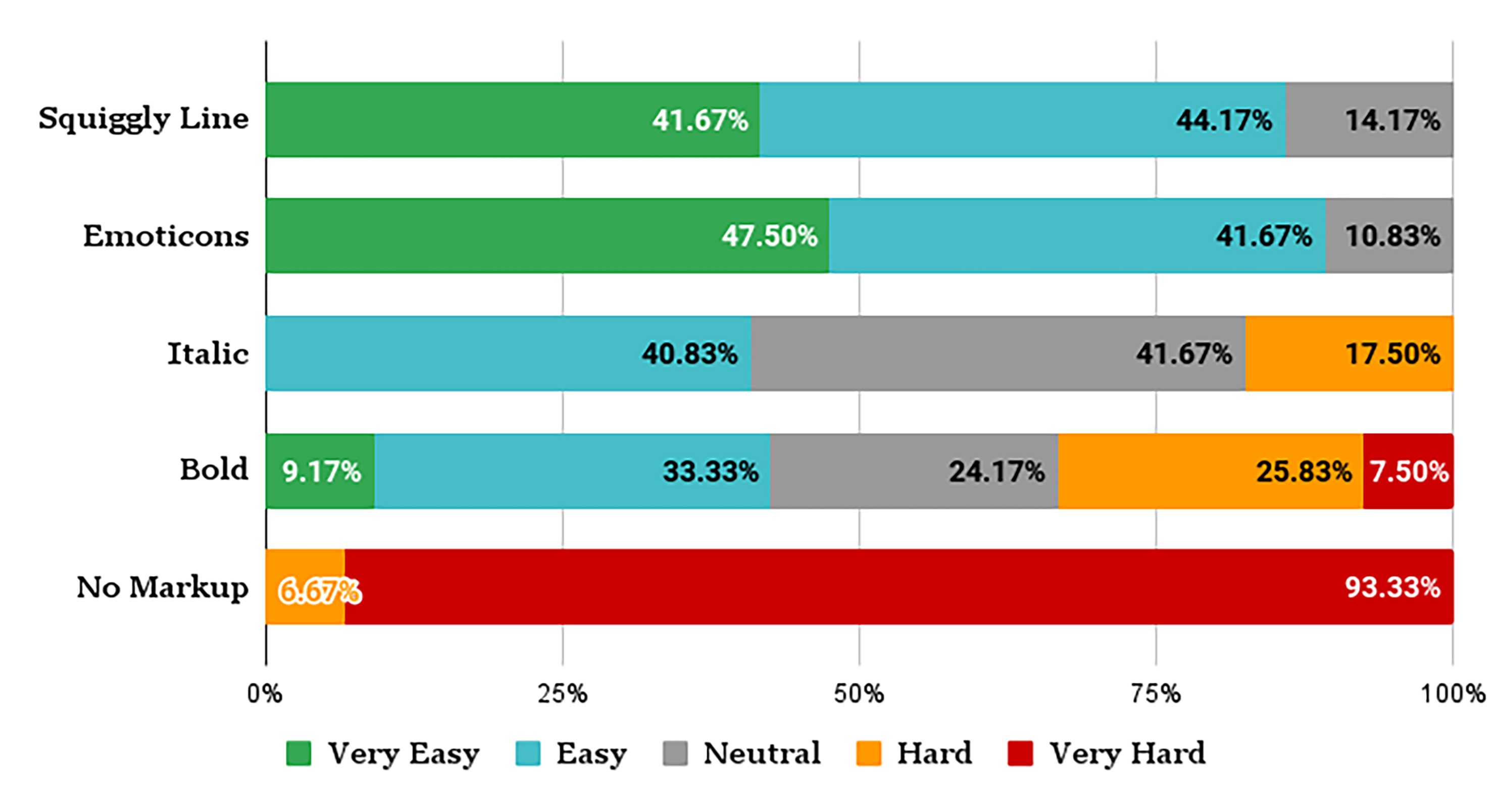}
  \label{fig:con_ez}}\par
  % \end{subfigure}
  % \begin{subfigure}[b]{0.49\textwidth}
  \subfloat[Distraction]
   {\includegraphics[width=0.49\textwidth]{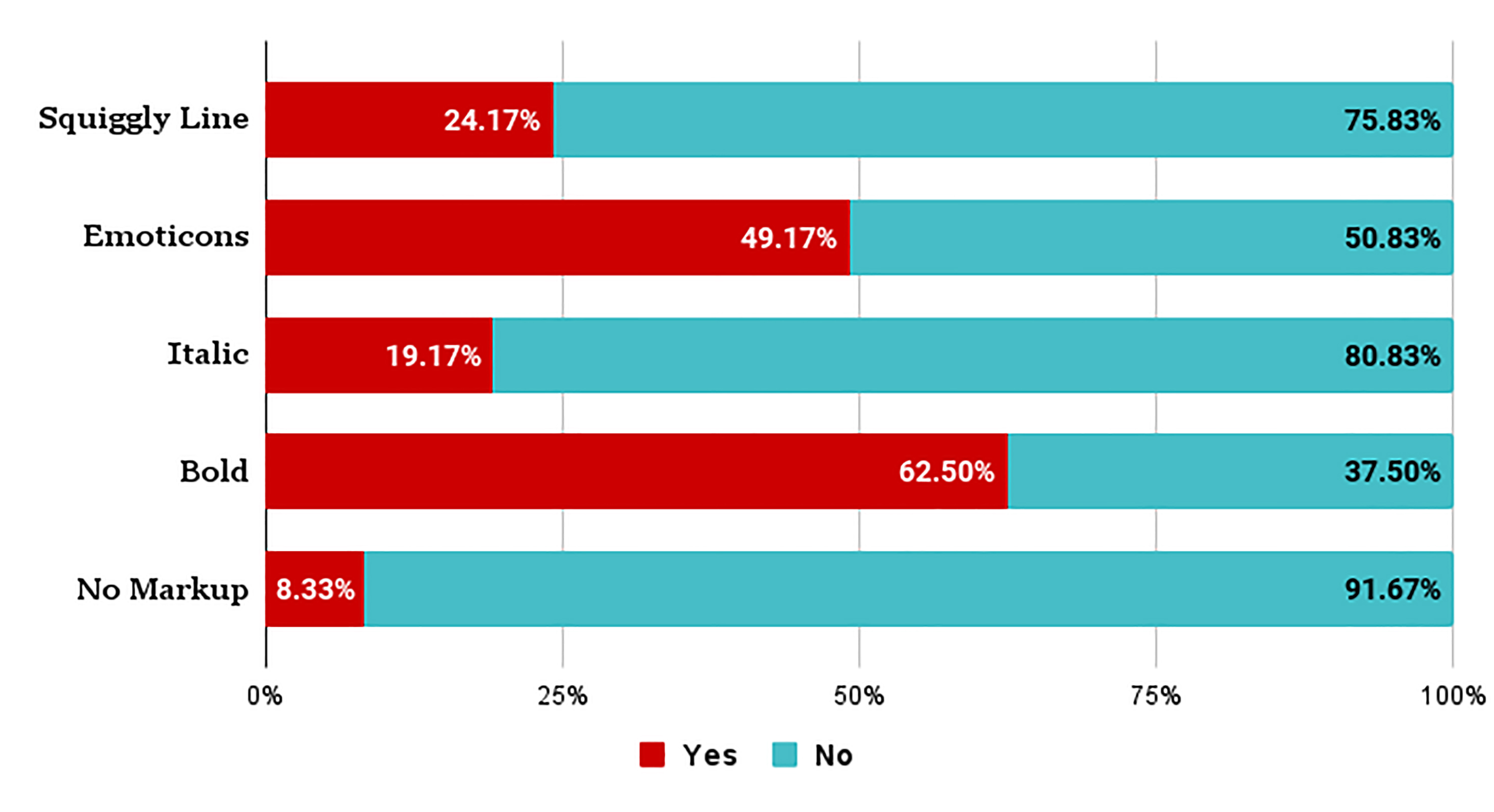}
  \label{fig:con_dis}}
  % \end{subfigure}
  % \begin{subfigure}[b]{0.49\textwidth}
  \subfloat[Preference]
   {\includegraphics[width=0.49\textwidth]{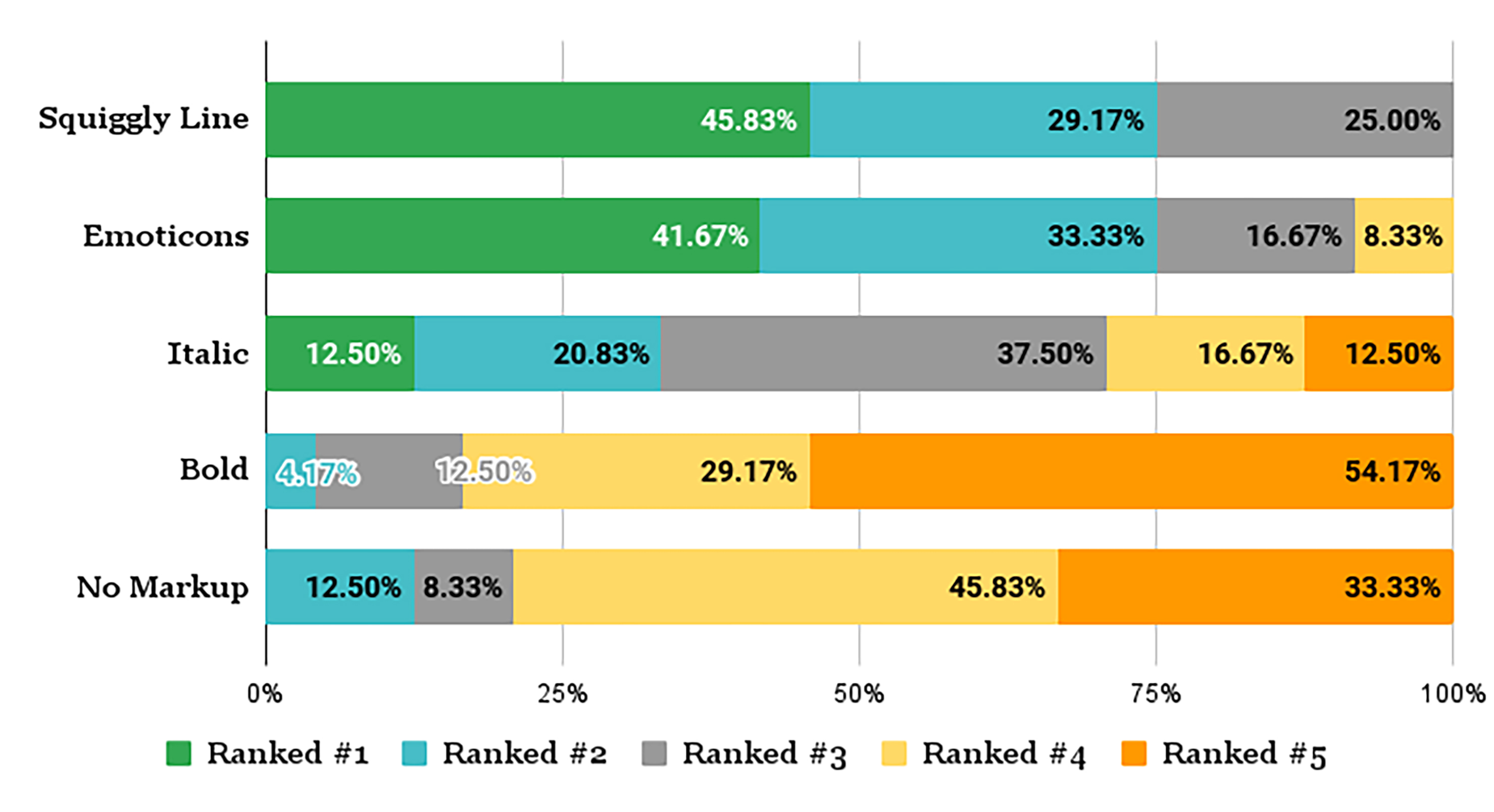}
  \label{fig:con_rank}}
  % \end{subfigure}
  \caption{Percentage Comparison of Markup Types}
  \label{Fig:markup_results}
\end{figure}

We performed an analysis based on the data we collected from total of 600 $(24 \times 5 \times 5)$ sessions. Figure \ref{Fig:markup_results} provides a summary of results of all four questions.\\
\textbf{Clear and Understandable:}
Friedman test revealed significance among markups (Friedman's \(\chi^2 = 365.017, df = 4, p < .001\)). Wilcoxon signed-rank test for for post-hoc comparisons revealed that red squiggly lines  (\(M = 4.208, SD = 0.697\)) and emoticons  (\(M = 3.925, SD = 0.927\)) are significantly clear and understandable compared to italic (\(M = 2.608, SD = 0.781\)), bold  (\(M = 3.308, SD = 0.731\)) and no markup  (\(M = 1.192, SD = 0.395\)), \(p < .001\). Red squiggly lines are significantly clear and understandable compared to emoticons, \(p < .001\). Italic and bold are significantly clear and understandable over no markup, \(p < .001\). Results suggested that red squiggly lines are clear and understandable compared to other methods and Figure \ref{fig:con_clear} displays the summary of results.\\
\textbf{Easy to Notice:}
Friedman reveled a significance among the markups (Friedman's \(\chi^2 = 363.870, df = 4, p < .001\)). Wilcoxon signed-rank test revealed that emoticons (\(M =  4.367, SD = 0.673\)) and red squiggly lines (\(M = 4.275, SD = 0.698\)) were significantly more noticeable compared to italic (\(M = 3.233, SD = 0.730\)), bold (\(M = 3.108, SD = 1.121\)) and no markups (\(M = 1.067, SD = 0.250\)), \(p < .001\). Italic and bold showed higher significance over no markups, \(p < .001\). We could not find significance among other groups, \(p = .05\). 
Results revealed that emoticons following red squiggly lines are easily noticeable compared to other methods, likely due to their frequent use in everyday educational activities. \\
\textbf{Distraction:}
Distraction levels were assessed using a binary yes/no question,  with Figure \ref{fig:con_dis} representing the responses for each markup type. Friedman test showed a significance among the markups (Friedman's \(\chi^2 = 100.444, df = 4, p < .001\)). Wilcoxon signed-rank test revealed no markups are significantly less distracting compared to emoticons and bold (\(p < .001\)). No markups were significantly less distracting compared to italic (\(p = .029\)) and red squiggly lines (\(p = .003\)). Red squiggly lines and italics were found to be significantly less distracting than bold and emoticons (\(p < .001\)). We could not find any significance among red squiggly lines and italics,  \(p = .377\). 
These results suggest that while markups can distract from reading, the impact is less pronounced with less obtrusive markups, such as red squiggly lines and italics. \\
\textbf{Preference:}
Participants were asked to rank the five markups from Most Preferred (1) to Least Preferred (5). Figure \ref{fig:con_rank} illustrates the percentage of ranks for each markup. The Friedman test revealed a statistically significant difference in preference rankings among the markups(Friedman's \(\chi^2 = 51.967, df = 4, p < .001\)). Ranking revealed that red squiggly lines was the most preferred choice among participants, followed by emoticons, italic, no markup, and Bold. Wilcoxon signed-rank tests for pairwise comparisons revealed that red squiggly lines and emoticons are significantly preferred compared to bold and no markup [\(p < 0.01\)], and with italic [\(p = 0.005\),\(p = 0.006\)], respectively. Also, we could not find any significant difference among red squiggly lines and emoticons, \(p = 0.703\). 

\subsection{Study 3: Word-by-Word Highlighting Technique}
\subsubsection{Study Design}
Primary objective of the study was to identify the most effective word-by-word highlighting strategy, specifically addressing RQ3. We evaluated six highlighting strategies (Figure \ref{fig:text-highlight}) with three questions
(Table \ref{tab:study3q}). The goal of the study is to identify highlighting strategies which are less distracting and easier to follow ultimately most effective for the karaoke-like presentation method.

\subsubsection{Apparatus}
For this study, we designed a simple mobile app which displays text with all six highlighting strategies. We chose texts from the school textbook\footnote[10]{\url{http://www.edupub.gov.lk/Administrator/Sinhala/9/sin\%20g-9/sinhala\%20g9.pdf} (last accessed January 02, 2024)} suggested by the teacher. Participants were able to guide themselves through the words with a touch on the screen. We selected ‘Red Color’ for the ‘Text Color Highlighting Strategy’, and ‘Yellow Color’ for ‘Background Color Highlighting Strategy’ which were suggested by participants. (Figure \ref{fig:text-highlight})

\subsubsection{Procedure}
Before the study, we explained to the participants the purpose of the study and introduced them to the mobile app. Participants used the mobile app to get an understanding of how it works. Each participant read text for about 20 minutes from each highlighting strategy and we counterbalanced the order of highlighting strategy with a Latin-square design. Text was displayed as shown in Figure \ref{fig:text-highlight}.

After each highlighting method, participants answered two questions (Table \ref{tab:study3q} Q1, Q2), and after completing the study they ranked the strategies according to preference. 

\begin{table} [ht]
  \caption{List of Questions Used in Study 3}
  {\begin{tabular}{p{0.5\textwidth}p{0.4\textwidth}}
    \toprule
    Questions & Scale\\
    \midrule
    Q1. I found it easy to follow the text with this highlighting strategy & 5-point Likert Scale from Strongly Disagree to Strongly Agree.\\
    Q2. It was distracting to me& Yes=1/No=0 \\
    Q3. Rank the highlighting strategies according to preference& Most Preferred to Least Preferred (1 to 6) \\
    \bottomrule
  \end{tabular}}
  \label{tab:study3q}
\end{table}

\subsubsection{Study Results}
\begin{figure}[ht]
  \centering
  % \begin{subfigure}[b]{0.49\textwidth}
  \subfloat[Easy to Follow]
   {\includegraphics[width=0.49\textwidth]{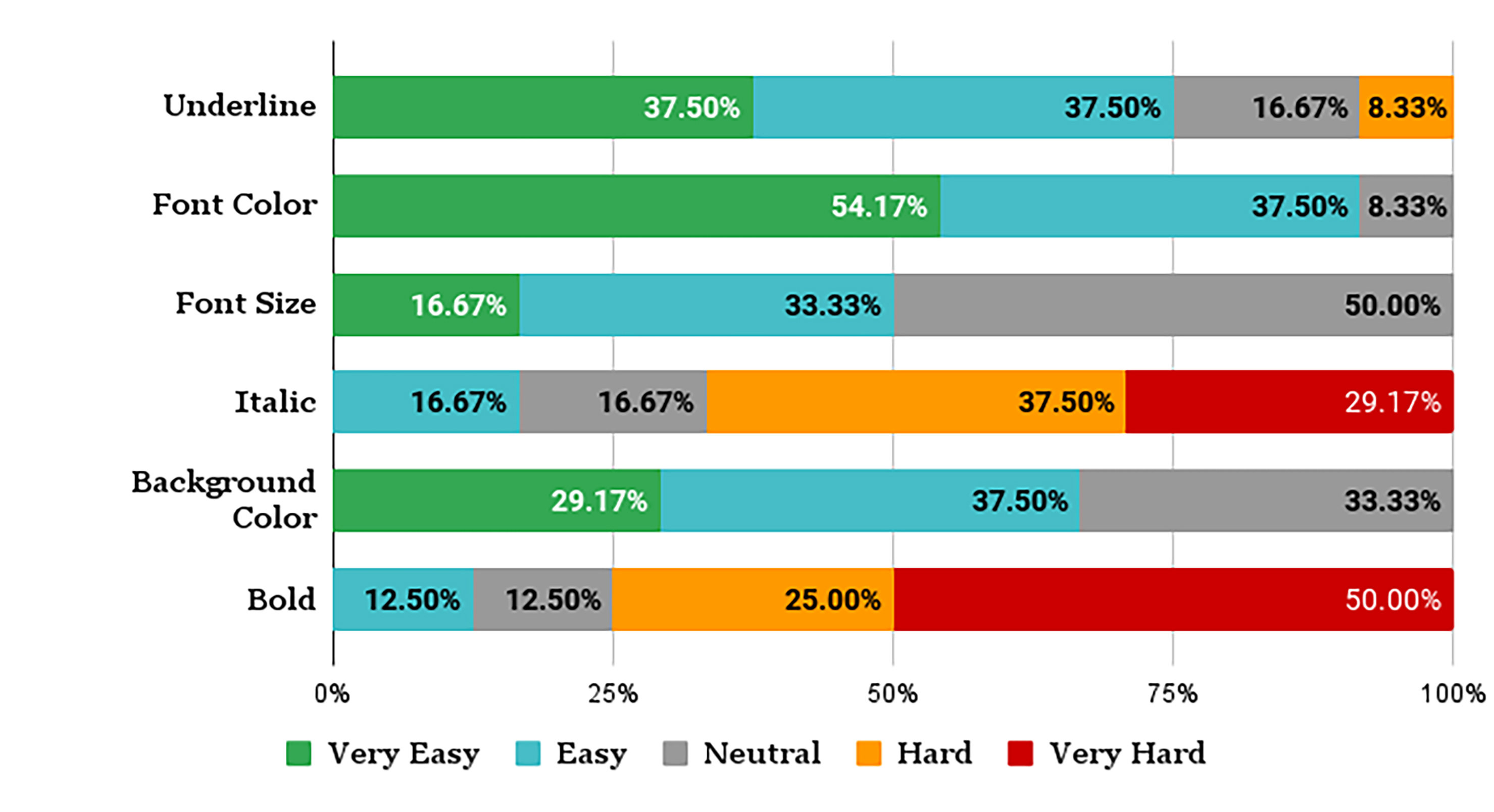}
  \label{fig:high_ez}}
  % \end{subfigure}
  % \begin{subfigure}[b]{0.49\textwidth}
  \subfloat[Distraction]
   {\includegraphics[width=0.49\textwidth]{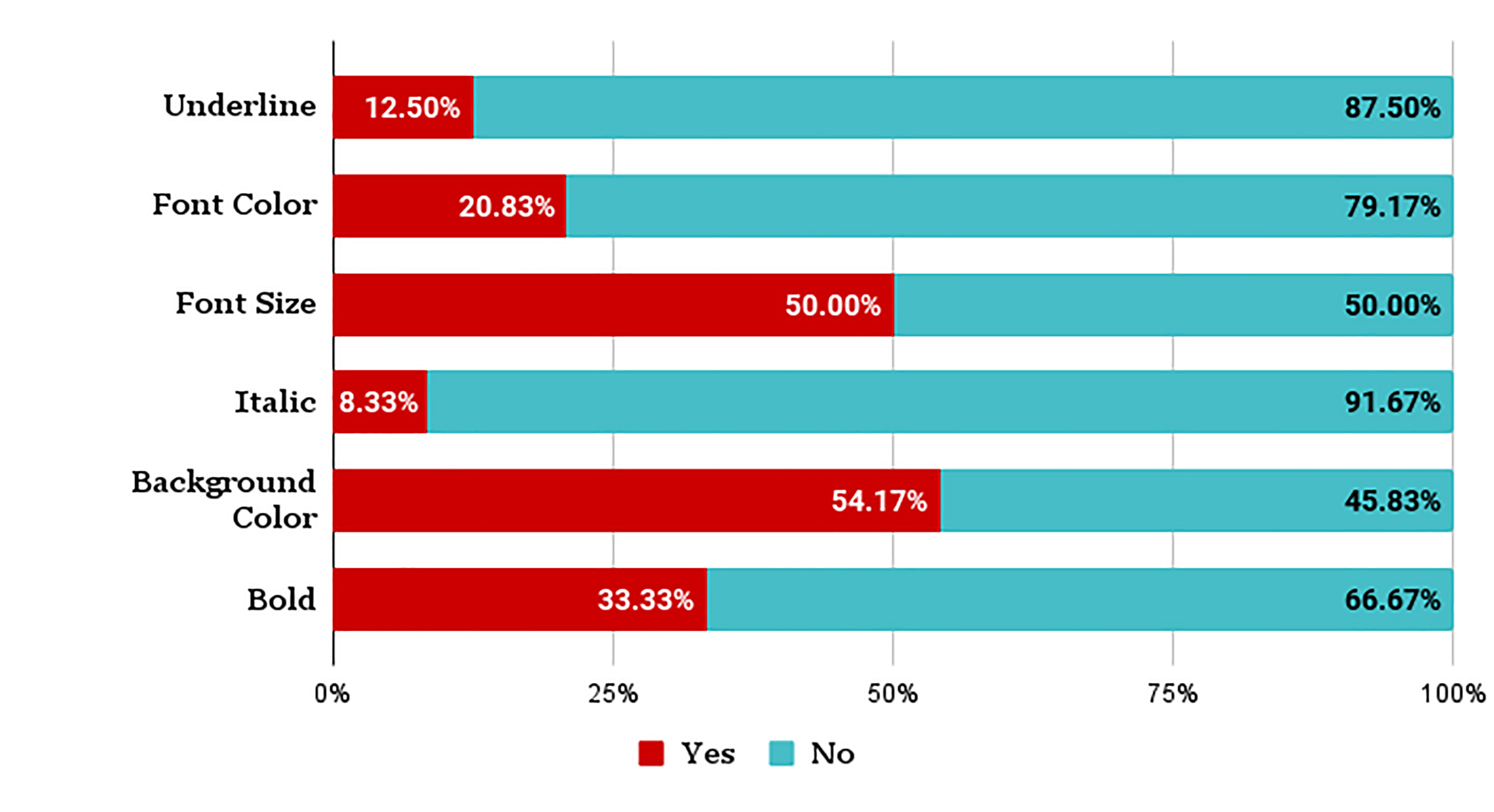}
  \label{fig:high_dis}}\par
  % \end{subfigure}
  % \begin{subfigure}[b]{0.49\textwidth}
  \subfloat[Preference]
   {\includegraphics[width=0.49\textwidth]{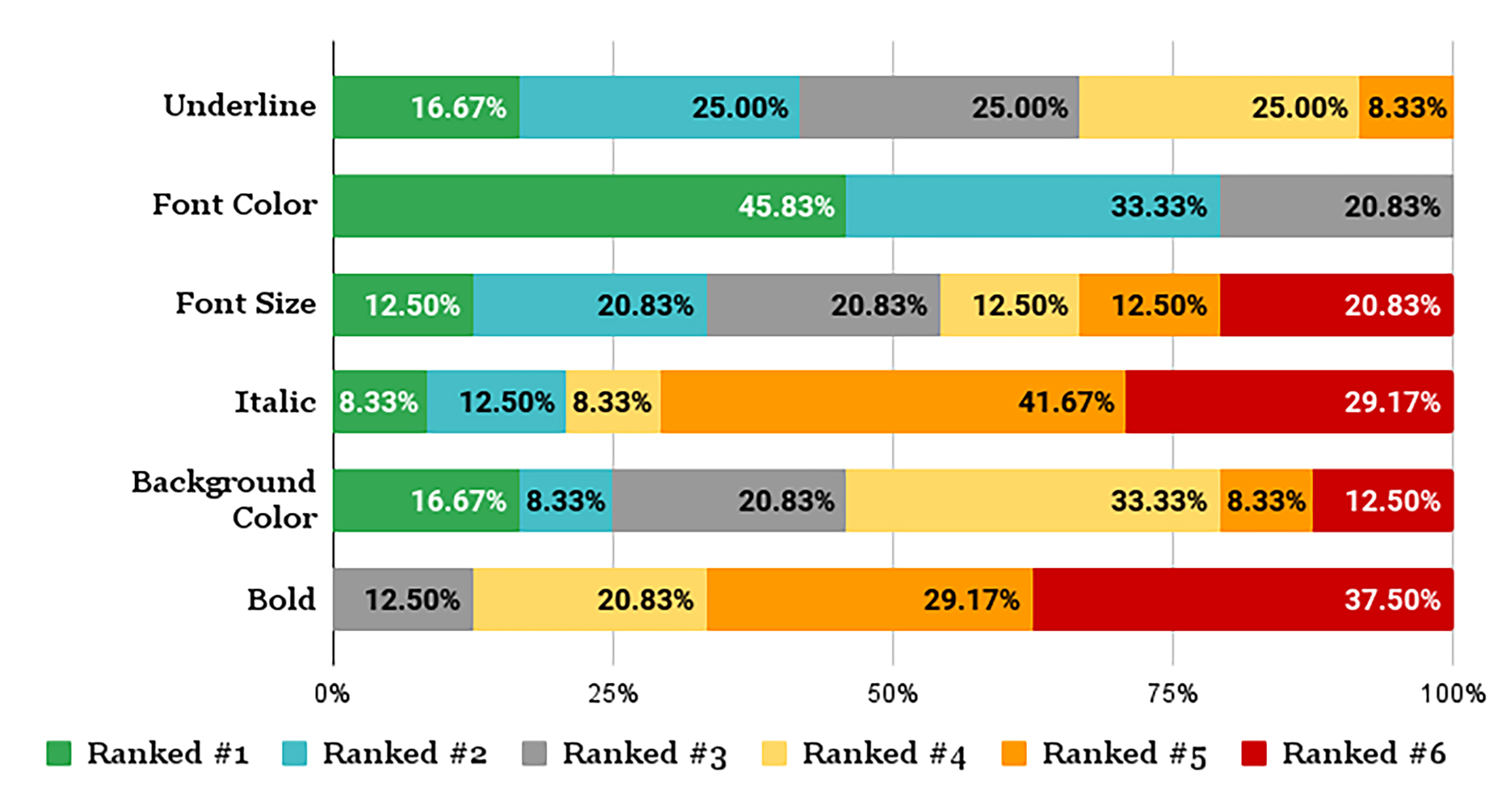}
  \label{fig:high_rank}}
  % \end{subfigure}
  \caption{Percentage Comparison of Highlighting Strategies}
  \label{Fig:Highlighting_Results}
\end{figure}

We performed an analysis based on the data we collected from all 24 participants. Figure \ref{Fig:Highlighting_Results} displays the responses for easy-to-follow, distractibility and preference rankings for each highlighting strategy.\\
\textbf{Easy to Follow Ratings:}
Friedman test revealed significance among each highlighting strategy (Friedman's \(\chi^2 = 62.258, df = 5, p < .001\)). Wilcoxon signed-rank test for post-hoc comparisons revealed  Font Color (\(M = 4.458, SD = 0.658\)) was rated significantly easy to follow compared to Bold (\(M = 1.875, SD = 1.076\)), Font Size (\(M = 3.667, SD = 0.761\)), and Italic (\(M = 2.208, SD = 1.062\)), \(p < .01\). Also, Underline (\(M= 4.042, SD=0.955\)), Background Color (\(M= 3.958, SD=0.806\)) and Font Size were significantly easier to follow compared to Italic and Bold,\(p < 0.01\). We could not find statistically significant differences in easy-to-follow ratings among other pairs, \(p > 0.05\). Figure \ref{fig:high_ez} displays the responses for each highlighting strategy. Results show that Font-Color is easier to follow compared to other highlighting strategies.\\
\textbf{Distraction:}
Distraction was tested using a binary question of yes/no, with Figure \ref{fig:high_dis} representing the responses for each highlighting strategy. Friedman test showed a significance among the highlighting strategies  (Friedman's \(\chi^2 = 19.424, df = 5, p = .001\)). Wilcoxon signed-rank test revealed underline and italic are significantly less distracting compared to Font-Size and Background Color, \(p < .05\). We could not find any significance between other highlighting strategies, \(p > .05\).\\
\textbf{Preference Rankings:}
Participants were asked to rank the six highlighting strategies from Most Preferred (1) to Least Preferred (6). Figure \ref{fig:high_rank} represents the percentage of ranks for each highlighting strategy. The Friedman test revealed a statistically significant difference in preference rankings among the highlighting strategies (Friedman's \(\chi^2 = 44.690, df = 5, p < .001\)). Ranking showed that Font Color was the most preferred choice among participants, followed by Underline, Background Color, Font Size, Italic, and Bold. Wilcoxon signed-rank tests for pairwise comparisons revealed that Font Color is significantly more preferred compared to other strategies, except Underline \(p < 0.001\) and significantly preferred over underlining, \(p = 0.002\). Underline is significantly preferred over bold and italic, \(p < 0.001\).  We could not find any statistically significance between other strategies, \(p > 0.05\). Results suggested that Font Color is significantly preferred among the highlighting strategies. 

\subsection{Study 4: Caption Presentation Method}

\subsubsection{Study Design}
The study is designed to evaluate RQ4 with four presentation methods (Figure \ref{fig:presentation}), These independent variables were tested against dependent variables. 
This study was conducted to evaluate the effect of caption presentation methods (Figure \ref{fig:presentation}) on Comprehension, Reading Speed, Subjective Task Load assessed by Raw TLX (RTLX) \citep{NasaTaskLoad2006} and perceived satisfaction using the ``overall user reactions to the system'' part of the Questionnaire for User Interaction Satisfaction (QUIS) \citep{QUIS1988} (Figure \ref{fig:appendix-quis}), specifically addressing RQ6. In addition, participants provided their qualitative feedback about how each presentation method affects the overall learning experience.

\subsubsection{Apparatus}
Study was conducted using our system (Section \ref{sec:system_implementation}). Captions placed `directly-below' and we selected `Font-Color' for the highlighting strategy for Karaoke-Like presentation method. These design decisions for the study were taken from two preliminary studies. The teacher selected materials from school textbook\footnote[11]{\url{http://www.edupub.gov.lk/Administrator/Sinhala/9/sin\%20g-9/sinhala\%20g9.pdf}(last accessed January 02, 2024)} for the study. We developed four separate prototypes with presentation types RSVP, Single-Line, Multi-Line (two lines)  and Karaoke-Like. (Figure \ref{fig:presentation}) 

\subsubsection{Procedure}
For the study, we asked each participant to engage in 4 classroom sessions. In each session, the Teacher of Deaf (ToD) verbally delivered the content and the participants' task was to read the real-time captions displayed on the AR glass. Each session was performed using all four presentation methods, and we counterbalanced the order of presentation methods with a Latin-square design. 

Before beginning the study, we assisted participants in putting on the AR glasses and provided them with a Bluetooth mouse. Utilizing the ToD (Task-oriented Dialogue) system, we conducted a training session for the participants, teaching them how to read real-time captions during a classroom session led by a teacher. Following this, we outlined the study's objectives, introduced reading strategies, and demonstrated how to use the mouse to navigate the Karaoke-Like reading method.

% \begin{figure}[ht]
%   \centering
%   \includegraphics[width=\linewidth]{figures/final-eval-01.jpg}
%   \caption{caption}
%   \Description{description}
%   \label{Fig:Classroom}
% \end{figure}

Each session had an average of 200 words (SD=0.81) of text and ToD created 5 comprehension questions for each session. With 24 participants, 4 sessions and 4 presentation methods, participants performed a total of 384 $(24 \times 4 \times 4) $ assessment sessions. 

After participants completed a single session from a single presentation method, we asked them to go through a comprehension test followed by the RTLX and the ``overall user reactions to the system'' part of the QUIS questionnaire (Figure \ref{fig:appendix-quis}). The Comprehension test was created by the expert ToD which contained both multiple choice questions and closed-ended questions with single answers. The grading scale, which ranges from 0 to 10, was also developed by the same expert ToD. The study spanned for 8 days, with an average of 6 hours per day.  After each participant completed all the sessions with all the presentation methods we asked them to provide their qualitative feedback about the presentation methods.

\subsubsection{Study Results}
For our study, each participant engaged in a total of 4 sessions, with each session being performed using all 4 presentation methods. Subsequently, a comprehension test was administered and evaluated by a teacher, who recorded the marks. Additionally, we measured reading speed by words per minute (WPM), and for subjective assessments, we conducted both the QUIS and RTLX tests. A descriptive summary of the results is presented in Table \ref{tab:main_results} to provide an overview of the findings. 

\begin{table}[ht]
  \tbl{Descriptive results of study 4}
  {\begin{tabular}{lrrrrrrrr}
    \toprule
    & \multicolumn{2}{c}{RSVP} & \multicolumn{2}{c}{Single-Line} & \multicolumn{2}{c}{Multi-Line} & \multicolumn{2}{c}{Karaoke-Like}\\
 & M& SD& M& SD& M& SD&M& SD\\
    \midrule
     Comprehension& 6.854& 1.205& 7.896& 1.310& 8.396& 0.923&9.135&0.816\\
     Reading Speed& 45.337& 7.252& 44.384& 7.640& 37.201& 4.956&43.527&7.042\\
     QUIS& 42.229&4.079& 39.198& 4.692& 40.156& 4.729&44.729&5.236\\
     RTLX& 58.281& 8.971& 57.292& 7.568& 62.639& 8.394&55.451&7.471\\
    \bottomrule
  \end{tabular}}
  \label{tab:main_results}
\end{table}

\subsubsection{Quantitative Results}
\begin{figure}[ht]
  \centering
   % \begin{subfigure}[b]{0.48\textwidth}
  \subfloat[Comprehension]
   {\includegraphics[width=0.48\textwidth]{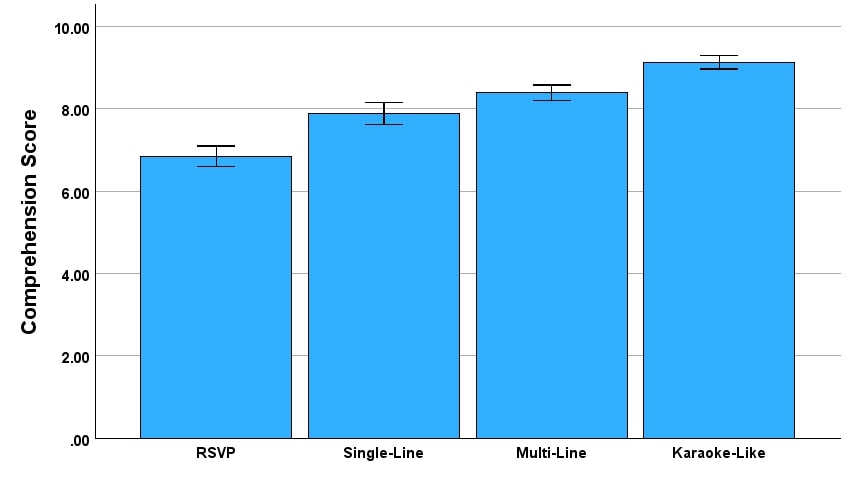}
  \label{fig:com}}
  % \end{subfigure}
  %  \begin{subfigure}[b]{0.48\textwidth}
  \subfloat[Reading Speed]
   {\includegraphics[width=0.48\textwidth]{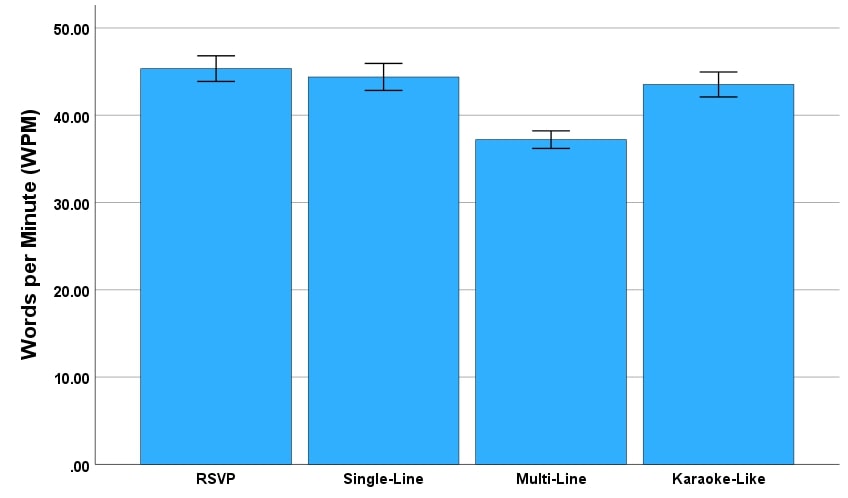}
  \label{fig:rs}}\par
  % \end{subfigure}
  %  \begin{subfigure}[b]{0.48\textwidth}
  \subfloat[Subjective Task Load]
   {\includegraphics[width=0.48\textwidth]{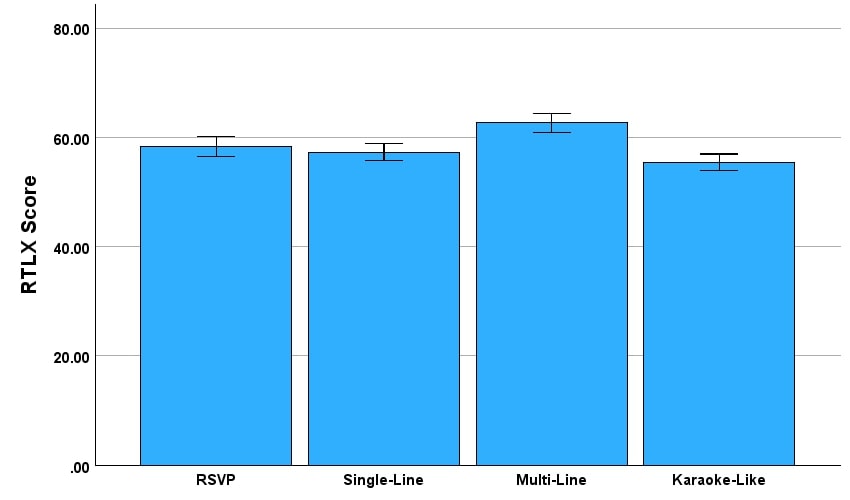}
  \label{fig:rtlx}}
  % \end{subfigure}
  %  \begin{subfigure}[b]{0.48\textwidth}
  \subfloat[Perceived Satisfaction]
   {\includegraphics[width=0.48\textwidth]{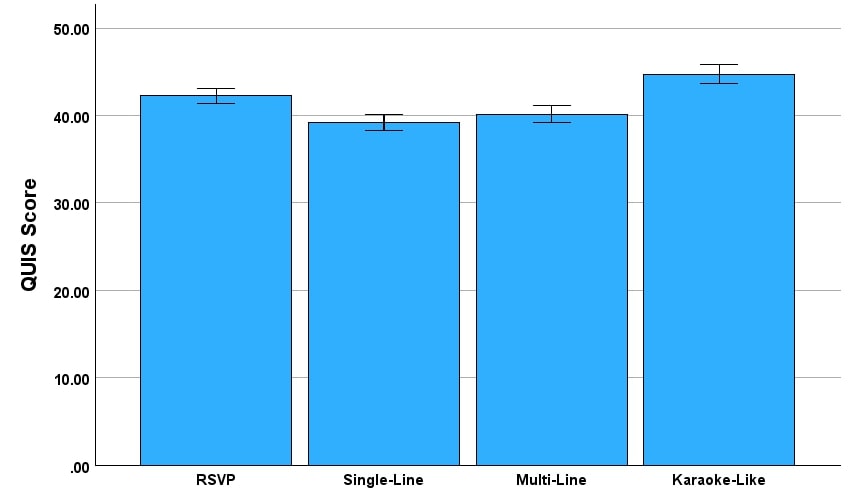}
  \label{fig:quis}}
  % \end{subfigure}
  \caption{Mean Comparison of Presentation Method Results}
  \label{Fig:main_results}
\end{figure}

\textbf{Comprehension:}
Comparing the comprehension scores (Figure \ref{fig:com}), one-way repeated measures ANOVA revealed a significant effect of presentation method (\(F (3, 285) = 105.700, p < .001, \eta_{p}^{2} = .527\)). Post-hoc pairwise comparisons using paired t-test with Bonferroni correction revealed that Karaoke-like presentations have significantly higher comprehension compared to RSVP, Single-Line, and Multi-Line, \(p < .001\). Multi-line showed significantly higher comprehension scores than RSVP and Single-Line, \(p < .001\). Single-Line showed significantly higher comprehension compared to RSVP,  \(p < .001\). Results suggested that comprehension increases significantly with the amount of content presented at a time and adding a guided reading method significantly increases comprehension.\\
\textbf{Reading Speed:}
In reading speed analysis (Figure \ref{fig:rs}), one-way repeated measures ANOVA  revealed a significant effect of the presentation method (\(F (3, 285) = 84.167, p < .001, \eta_p^2 = .470\)). Post-hoc pairwise comparisons using paired t-test with Bonferroni correction revealed that RSVP, Single Line and Karaoke-Like have significantly higher reading speed compared to Multi-Line, \(p < .001\). We could not find any significance between Karaoke-Like and Single-Line presentations, and RSVP and Single-Line \(p > 0.05\). Results suggested that reading speed is significantly higher when less amount of content is displayed compared to more amount of content. However, results showed that with the reading assistant method (Karaoke-Like) significantly increased reading speed even with more content.\\
\textbf{Reading Efficiency:}
Wilcoxon signed-rank test for reading efficiency, calculated by multiplying each reading speed with the comprehension, and results suggested that the Karaoke-Like presentation method significantly enhances reading efficiency over other presentation methods, \(p < .001\). \\
\textbf{Subjective Task Load:}
Comparison of RTLX results is displayed in (Figure \ref{fig:rtlx}), and the one-way repeated measures ANOVA test showed a significant difference among the presentation methods  (\(F (3, 285) = 18.431, p < .001, \eta_p^2 = .162\)). Pairwise comparisons revealed that Karaoke-Like presentation has significantly less subjective load compared to Multi-Line (\(p < 0.001\)) and RSVP presentations (\(p < 0.05\)). RSVP and Single-Line resulted in significantly less subjective load compared to Multi-Line, \(p < 0.001\). However, we could not find any significance between Karaoke-Like and Single-Line, and RSVP and Single-Line, \(p > 0.05\). Results suggested that multiple lines of text significantly increase the cognitive load and with a reading assistant method like Karaoke-Like it is significantly decreased.\\
\textbf{Perceived Satisfaction:}
One-way repeated measures ANOVA revealed a higher significant difference among the prototypes for QUIS scores (\(F (3, 285) = 36.996, p < .001, \eta_p^2 = .280\)). Pairwise comparisons using paired t-test with Bonferroni correction revealed a higher significance for Karaoke-Like over other methods, \(p < 0.001\). RSVP showed a higher significance over Multi-Line (\(p = 0.003\)) and Single-Line presentations (\(p < 0.001\)). We could not find any significance between Multi-line and Single-line, \(p = 0.345\). (Figure \ref{fig:quis})

\subsubsection{Qualitative Feedback}
After participants completed all the sessions we asked them about their qualitative experience of using each presentation method. Participants found the Karaoke-Like method effective for focusing on one word at a time, enhancing reading effectiveness, \textit{“With the highlighted word, I can focus solely on that, making reading easier.”} (S2), \textit{“It’s like someone assisting me to read, for me it is very effective.”} (S4). 
Additionally, participants found the Karaoke-like method to be interactive and motivating, with one participant mentioning, \textit{``I like it very much, I feel like I need to read the words.''} (S5). Participants mentioned that Karoake-Like helps them to read words at their own speed which increases their comprehension, \textit{``I am having control to read is really good, I can read at my own pace''}. In contrast, the RSVP method was deemed easy to read but challenging to comprehend, as expressed by one participant, \textit{``It’s just one word at a time, I can't get a meaning from that, this method is hard for me to understand.''} (S2). Participants highlighted that Multi-line presentation is difficult to read, \textit{``More words feels like a lot of tasks to me.''} (S7). However, participants mentioned that even with Karaoke-like presentations it feels like a lot of tasks but is better than Multi-line presentations, \textit{``Highlighting is better as I could focus only on the red color word.''} (S9).

\subsection{Study 5: Visualization of Personal Utterances}
\subsubsection{Study Design}
This study aimed to evaluate the effect of visualizing personal utterances within the user's line of sight, addressing RQ5. We developed four designs through a User-Centered Design (UCD) process: separated, separated with color, no markup, and not separated with color (refer to Figure \ref{fig:utterance}). Participants were asked to rank these designs and provide qualitative feedback on their preferences.

\subsubsection{Apparatus and Procedure}
Since the objective of our implementation was to mainly test visual designs, we developed a separate mobile app to detect teachers' voice transcriptions and our system (Section \ref{sec:system_implementation}) to detect personal transcriptions. Teachers' transcriptions were sent over an Application Programming Interface (API) to a server and retrieved by the mobile app on wearers' AR glass. We make sure teacher stand from a distance such that transcriptions don't get overlapped. (Figure \ref{fig:utterance})

Participants engaged in interactive classroom activities, responding to questions posed by the teacher. Each design was tested in a 15-minute session, with the total study duration averaging 7 hours per day over four days.

After completing all sessions, participants were asked to rank the designs and provide feedback on each.

\subsubsection{Study Results}
Even though participants suggested to use visualize the personal utterances in the UCD process, most participants (n=13) found out that it is very distracting and preferred not to visualize personal utterances, Some comments include, \textit{``Its annoying, sometimes I still can't distinguish.''} (S21), Student S15  found the simultaneous processing of visual and auditory information to be overwhelming, stating, \textit{``Trying to read my own words while listening and reading others' is overwhelming. It splits my attention too much.''}. This sentiment was echoed by S7, who noted, \textit{``With both my words and others' floating around, the screen gets too busy. Its hard to focus on what's important.''} For some, like S3, the visualization of their own speech seemed redundant: \textit{``I know what I said; I don't need to see it. I would rather the system help me with what I don't know—like what others are saying.''} S20 suggested system can be useful in noisy environments, \textit{``Maybe in a noisy environment or a group conversation, it could help. But in here, its just unnecessary visual noise.''}.

Second most preferred (n=8) design is separated with color (Figure \ref{fig:v4}). This approach received positive feedback for enhancing clarity and aiding in the distinction between the wearer's speech and that of others. Participants noted that this design helped in catching mistakes in real-time and made the conversation flow more smoothly, \textit{``Having my own words in a distinct color actually helped me catch a few mistakes in what I was saying. Its useful feedback in real-time.''} (S14), \textit{``The color differentiation helps a lot. It's easier to tell my words from theirs, which makes the conversation flow better for me.''} (S9). This feedback underscores the potential benefits of displaying personal utterances.

\section{Overall System Evaluation}
Following the completion of all the studies, all 24 students and 2 teachers were invited to share their feedback regarding the overall system.  This study is conducted to get an understanding of the effectiveness, usability of the system and preference to use it in their classroom settings. Overall usability was assessed using a System Usability Scale (SUS) \citep{SUS1995}. Finally, students and teachers provided their qualitative feedback about the system. Two researchers independently conducted the qualitative analysis using a thematic approach based on inductive themes, achieving an average ICC of 0.83, which indicates good agreement between the coders.

\subsection{Quantitative}
The average System Usability Scale (SUS) score given by the 24 participants was 78.64 (\(SD=4.77\)) and detailed analysis is displayed in Figure \ref{fig:appendix-susfigure}. Twenty-one participants (87.50\%) showed their preference to use the current system for their classroom activities while two participants were uncertain and one rejected mentioning that the system causes extra tiredness for the eyes.

\subsection{Qualitative}
\textbf{Literacy Development and Educational Activity Support:} Students and teachers recognized the system's potential to enhance literacy skills. Teacher T1 highlighted, \textit{``With long-term use of the system students get used to read and comprehend quickly.''}. Student S7 mentioned, \textit{``I usually remember words by writing, now I can write when the teacher speaks.''}, S14 added, \textit{``Now I can write short notes of the ongoing classroom session.''}. T2 highlighted that students' curiosity of learning words has improved with the system, \textit{``After a session student was curious about new words encountered.''} Teachers noted that the system can be used in speech therapy activities, \textit{ ``Students can pronounce displayed word following my lips, which is ideal as they get to see the actual word as well.''} (T1). \\
\textbf{Engaging Learning Experience:} Teachers noted students' increased engagement in classroom sessions, \textit{``System is interactive for them, especially Karaoke-Like presentations, they seem very engaged during classroom sessions.''} (T2). Students mentioned that the system motivated them to engage in classroom sessions, \textit{``This is a new experience for me, I really like this, Where can I buy this system?''} (S3)\\
\textbf{Enhanced Learning Speed and Effectiveness:} Students and teachers noted the system's ability to improve their learning speed, helping them follow classroom sessions more effectively: \textit{``Now, I don't have to look at the book to read the words.''} (S17), T2 mentioned, \textit{``I can verbally deliver the classroom session, we don't have to write everything on a whiteboard.''}\\
\textbf{Challenges and Concerns:} Students reported challenges in the system, including the need for assistance with the device setup, discomfort due to its weight, eye strain, and dizziness after prolonged use.

\section{Discussion and Limitations}
\textbf{Placement of Captions\\}
Similar to findings reported by Jain et al. \citep{ExploringAugmentedReality2018}, participants indicated a preference to place captions directly below the user. Captions placed near the speaker left, right, or directly below showed a significant increase in comprehension compared to traditional bottom-centered captions. This could imply that keeping captions near the speaker helps viewers integrate visual cues (eg. lip movements and facial expressions) with the text, enhancing understanding. Reading speed shows a significance increase when captions are directly below the speaker facilitating synchronization with lip movements more effectively. However, comfort levels were found to be higher when captions are positioned away from the speaker's body suggesting that captions overlaying the speaker's body can become distracting, particularly could be due to the reduced readability (e.g., clash colors with speakers' clothing) and increased cognitive load when maintained captions in the central field of view for extended periods. Overall findings suggest the importance of customized caption placement in enhancing the educational experience. The diverse preferences underscore the need for adaptable captioning systems that cater to individual requirements, thereby optimizing comprehension, comfort, and reading speed in learning environments.\\ \\ 
\textbf{Markup for Uncertain Words\\}
We discovered that participants showed a preference for using emoticons, particularly 'angry' emoticons, and red squiggly lines to indicate uncertain words. This finding contrasts with Berke et al. \citep{confidence} research, which revealed that participants generally did not favor the use of markups, attributing this to the potential distraction caused by altered text appearance, or to the specific accuracy levels of ASR for English captions and the literacy levels of the participants involved. Unlike Berke et al.'s study, our research utilized 'Sinhala' captions and incorporated design input directly from the participants, who expressed a strong desire for markups. Our study, conducted with students using complex educational materials, found that although markups were perceived as distracting, they were significantly preferred for their role in enhancing understanding of the content. Moreover, markups encouraged students to pay closer attention to lip reading when encountering uncertain words, thereby reducing overall distraction. Given these findings, it's evident that the context in which captions and markups are used plays a crucial role in their effectiveness and acceptance among users. The preference for specific types of markups, like emoticons and visual indicators, underscores the need for customizable captioning solutions that cater to the diverse needs and preferences of users, especially in educational settings.\\ \\
\textbf{Word-by-Word Highlighting Strategy\\}
Participants showed a strong preference for text color highlighting as the most effective word-by-word highlighting strategy, citing its ability to guide the reader through the text seamlessly and maintain focus without causing significant distractions. This method was notably rated higher on the 'easy to follow' scale compared to alternatives like italicizing or underlining. Research supports this preference, as studies have shown that color-coded text can effectively direct attention, improve comprehension, and enhance learning outcomes in reading tasks \citep{OZCELIK2010110}.\\ \\
\textbf{Caption Presentation Method\\} 
Presentation methods revealed a marked preference for karaoke-style presentations, which was found to significantly improve reading efficiency over other methods. This method outperformed others in terms of comprehension, particularly when compared to Rapid Serial Visual Presentation (RSVP), Single-Line, and Multi-Line methods. Similar to the findings in \cite{SpeechBubbles2018}, we found that presenting more content simultaneously enhances comprehension. The Karaoke-Like method, which capitalizes on this principle by allowing for the simultaneous presentation of more content while also enabling readers to progress at their own pace, emerged as the most effective approach, outperforming all other methods in our comparison. This approach also addresses issues identified in initial prototypes, where participants struggled with word focus, slowing down their reading speed. Despite presenting more content, the karaoke method facilitated faster reading speeds by enabling focus on individual words. User satisfaction with the karaoke method scored significantly higher on the Questionnaire for User Interaction Satisfaction (QUIS) than all other methods. Participants described this method as ``very interactive,'' highlighting its engagement factor. For subjective task load, the RSVP method showed the lowest values and the multi-line approach was found to be more demanding. Notably, the karaoke method was effective in reducing cognitive load, as participants appreciated focusing on one word at a time, which simplified content processing. These insights underscore the importance of adaptive and user-friendly presentation methods in enhancing reading comprehension and user experience, particularly in educational and accessibility-focused applications.\\ \\
\textbf{Visualization of Personal Utterances\\}
Participants reported finding the display of personal utterances distracting for several reasons, including difficulties in distinguishing between different sources of speech, a sense of being overwhelmed, and visual clutter. However, they also acknowledged the potential utility of such displays in noisy environments and group conversations, where distinguishing speech is inherently challenging. This ambivalence seems rooted in the participants' primary focus on the teacher's captions during classroom sessions. Interestingly, some participants expressed a preference to visualize personal utterances in specific scenarios where verbatim accuracy of personal utterances was crucial (eg. practicing speech). This aspect underscores the need for further investigation through more comprehensive studies involving a larger participant pool and diverse settings beyond the classroom environment to fully understand the nuances of user preferences and requirements.\\ \\
\textbf{Overall System\\}
In our system evaluation, we found that students increasingly preferred using our system for their learning interactions. Initially uncertain, students' preferences shifted positively as we addressed various design challenges. Both students and teachers provided qualitative feedback, highlighting that the system not only improved the overall learning experience but also offered significant benefits in areas such as reading, writing, vocabulary, and more which could be tested by conducting long-term studies.\\ \\
\textbf{Limitations\\}
Our studies were conducted in a controlled classroom environment characterized by low noise levels and consistent lighting conditions, but in real-world classrooms with diverse and dynamic settings, there will be challenges with the readability and accuracy of ASR. Our study sessions involved one-on-one interactions between a student and a teacher but in real-world scenarios, multiple students will be engaging in the sessions which might bring new challenges from factors like size of the classroom and students positioning. We conducted studies in a classroom of a selected deaf school but in real-world scenarios, the classroom environments can be changed. Considering the financial needs of most deaf students, we employed a low-cost AR device for the study, which had limitations such as font size and text-to-eye distance. Additionally, the hardware form factors we selected may have introduced field-of-view limitations and affected the user experience during prolonged use due to the device's weight. The selection of study materials was guided by teachers to minimize confounding factors, such as the presence of overly unfamiliar vocabulary. The potential influence of specific material choices on the results was not explicitly addressed in this study. Also, we used the same set of participants throughout the studies which might have developed a learning effect as they progressed through the prototypes. It is important to note that our present study did not specifically address or account for these potential factors.

\section{Conclusion and Future Work}
In this paper, we have demonstrated the design, development, and evaluation of our low-cost real-time AR captioning system to support the unique educational needs of DHH students in a specialized school of deaf. Our comprehensive need analysis identified the significant challenges faced by DHH students in specialized educational settings due to limited exposure to written and spoken language and a lack of tailored educational tools. Through an iterative user-centered design process we developed an AR interface that introduces caption presentation methods, placement strategies, and visualizations to represent uncertain transcriptions and personal utterances, addressing the limitations in existing designs that affect the overall learning experience in an educational setting. Our comprehensive user studies identified that our prototypes significantly enhanced the learning experience, notably improving reading efficiency, comprehension, and speed. Additionally, our solutions were found to reduce cognitive load and minimize distractions, thereby improving overall usability and engagement in the learning process. The findings and methodologies presented in this study are poised to guide future developments in tailored AR-based educational tools, offering valuable insights for researchers, designers, and educators aiming to create more engaging and accessible learning experiences for DHH students. 

As a next step in our research, we intend to conduct a longitudinal study to rigorously assess the enduring effects of our intervention on the comprehensive linguistic development of DHH students. This extended research will encompass vocabulary, speaking, reading, and writing skills, aiming to provide a comprehensive understanding of how our solution influences DHH students' language proficiency over an extended period. By evaluating the sustained benefits, we aim to further scale our solution such that DHH students can engage in mainstream school sessions, thus significantly enhancing their accessibility to a wide array of learning resources. We also plan to conduct user studies to evaluate the effectiveness of the system to wide range of age groups such as primary school and undergraduate students. We will further validate the potential of our AR-based real-time captions system as a transformative tool in DHH education, ultimately improving their academic performance and language proficiency, offering them improved opportunities for success and inclusion in educational and professional settings.

\section*{Acknowledgement(s)}
We would like to acknowledge the teachers and students who participated in this project. Finally, we acknowledge that all authors made critical contributions at various junctures of this long complex project.
% An unnumbered section, e.g.\ \verb"\section*{Acknowledgements}", may be used for thanks, etc.\ if required and included \emph{in the non-anonymous version} before any Notes or References.

\section*{Disclosure Statement}
We have no conflicts of interest to declare.

\section*{Ethics Consideration}
The studies were conducted in accordance with the ethical research guidelines provided by the Ethics Review Committee (ERC) of the University of Colombo School of Computing.

\bibliographystyle{apacite}

\begin{thebibliography}{}

\bibitem [\protect \citeauthoryear {%
Adamo-Villani%
\ \BBA {} Anasingaraju%
}{%
Adamo-Villani%
\ \BBA {} Anasingaraju%
}{%
{\protect \APACyear {2017}}%
}]{%
Adamo-Villani2016Holographic}
\APACinsertmetastar {%
Adamo-Villani2016Holographic}%
\begin{APACrefauthors}%
Adamo-Villani, N.%
\BCBT {}\ \BBA {} Anasingaraju, S.%
\end{APACrefauthors}%
\unskip\
\newblock
\APACrefYearMonthDay{2017}{}{}.
\newblock
{\BBOQ}\APACrefatitle {Holographic Signing Avatars for Deaf Education} {Holographic signing avatars for deaf education}.{\BBCQ}
\newblock
\BIn{} \APACrefbtitle {E-Learning, E-Education, and Online Training} {E-learning, e-education, and online training}\ (\BPGS\ 54--61).
\newblock
\APACaddressPublisher{Cham}{Springer International Publishing}.
\newblock
\begin{APACrefDOI} \doi{10.1007/978-3-319-49625-2_7} \end{APACrefDOI}
\PrintBackRefs{\CurrentBib}

\bibitem [\protect \citeauthoryear {%
Akçayır%
\ \BBA {} Akçayır%
}{%
Akçayır%
\ \BBA {} Akçayır%
}{%
{\protect \APACyear {2017}}%
}]{%
akcayir2017advantages}
\APACinsertmetastar {%
akcayir2017advantages}%
\begin{APACrefauthors}%
Akçayır, M.%
\BCBT {}\ \BBA {} Akçayır, G.%
\end{APACrefauthors}%
\unskip\
\newblock
\APACrefYearMonthDay{2017}{}{}.
\newblock
{\BBOQ}\APACrefatitle {Advantages and challenges associated with augmented reality for education: A systematic review of the literature} {Advantages and challenges associated with augmented reality for education: A systematic review of the literature}.{\BBCQ}
\newblock
\APACjournalVolNumPages{Educational Research Review}{20}{}{1-11}.
\newblock
\begin{APACrefURL} \url{https://www.sciencedirect.com/science/article/pii/S1747938X16300616} \end{APACrefURL}
\newblock
\begin{APACrefDOI} \doi{10.1016/j.edurev.2016.11.002} \end{APACrefDOI}
\PrintBackRefs{\CurrentBib}

\bibitem [\protect \citeauthoryear {%
Al-Megren%
\ \BBA {} Almutairi%
}{%
Al-Megren%
\ \BBA {} Almutairi%
}{%
{\protect \APACyear {2018}}%
{\protect \APACexlab {{\protect \BCnt {1}}}}}]{%
Al-Megren_Almutairi_2018}
\APACinsertmetastar {%
Al-Megren_Almutairi_2018}%
\begin{APACrefauthors}%
Al-Megren, S.%
\BCBT {}\ \BBA {} Almutairi, A.%
\end{APACrefauthors}%
\unskip\
\newblock
\APACrefYearMonthDay{2018{\protect \BCnt {1}}}{Dec.}{}.
\newblock
{\BBOQ}\APACrefatitle {ANALYSIS OF USER REQUIREMENTS FOR A MOBILE AUGMENTED REALITY APPLICATION TO SUPPORT LITERACY DEVELOPMENT AMONGST HEARING-IMPAIRED CHILDREN} {Analysis of user requirements for a mobile augmented reality application to support literacy development amongst hearing-impaired children}.{\BBCQ}
\newblock
\APACjournalVolNumPages{Journal of Information and Communication Technology}{18}{1}{97–121}.
\newblock
\begin{APACrefURL} \url{https://e-journal.uum.edu.my/index.php/jict/article/view/jict2019.18.1.6} \end{APACrefURL}
\newblock
\begin{APACrefDOI} \doi{10.32890/jict2019.18.1.6} \end{APACrefDOI}
\PrintBackRefs{\CurrentBib}

\bibitem [\protect \citeauthoryear {%
Al-Megren%
\ \BBA {} Almutairi%
}{%
Al-Megren%
\ \BBA {} Almutairi%
}{%
{\protect \APACyear {2018}}%
{\protect \APACexlab {{\protect \BCnt {2}}}}}]{%
AssessingTheEffectiveness2018}
\APACinsertmetastar {%
AssessingTheEffectiveness2018}%
\begin{APACrefauthors}%
Al-Megren, S.%
\BCBT {}\ \BBA {} Almutairi, A.%
\end{APACrefauthors}%
\unskip\
\newblock
\APACrefYearMonthDay{2018{\protect \BCnt {2}}}{}{}.
\newblock
{\BBOQ}\APACrefatitle {Assessing the Effectiveness of an Augmented Reality Application for the Literacy Development of Arabic Children with Hearing Impairments} {Assessing the effectiveness of an augmented reality application for the literacy development of arabic children with hearing impairments}.{\BBCQ}
\newblock
\BIn{} P\BHBI L\BPBI P.~Rau\ (\BED), \APACrefbtitle {Cross-Cultural Design. Applications in Cultural Heritage, Creativity and Social Development} {Cross-cultural design. applications in cultural heritage, creativity and social development}\ (\BPGS\ 3--18).
\newblock
\APACaddressPublisher{Cham}{Springer International Publishing}.
\PrintBackRefs{\CurrentBib}

\bibitem [\protect \citeauthoryear {%
Arici%
, Yildirim%
, Caliklar%
\BCBL {}\ \BBA {} Yilmaz%
}{%
Arici%
\ \protect \BOthers {.}}{%
{\protect \APACyear {2019}}%
{\protect \APACexlab {{\protect \BCnt {2}}}}}]{%
researchtrends}
\APACinsertmetastar {%
researchtrends}%
\begin{APACrefauthors}%
Arici, F.%
, Yildirim, P.%
, Caliklar, c.%
\BCBL {}\ \BBA {} Yilmaz, R\BPBI M.%
\end{APACrefauthors}%
\unskip\
\newblock
\APACrefYearMonthDay{2019{\protect \BCnt {2}}}{dec}{}.
\newblock
{\BBOQ}\APACrefatitle {Research trends in the use of augmented reality in science education: Content and bibliometric mapping analysis} {Research trends in the use of augmented reality in science education: Content and bibliometric mapping analysis}.{\BBCQ}
\newblock
\APACjournalVolNumPages{Comput. Educ.}{142}{C}{}.
\newblock
\begin{APACrefURL} \url{https://doi.org/10.1016/j.compedu.2019.103647} \end{APACrefURL}
\newblock
\begin{APACrefDOI} \doi{10.1016/j.compedu.2019.103647} \end{APACrefDOI}
\PrintBackRefs{\CurrentBib}

\bibitem [\protect \citeauthoryear {%
Arici%
, Yildirim%
, Caliklar%
\BCBL {}\ \BBA {} Yilmaz%
}{%
Arici%
\ \protect \BOthers {.}}{%
{\protect \APACyear {2019}}%
{\protect \APACexlab {{\protect \BCnt {1}}}}}]{%
arici2019research}
\APACinsertmetastar {%
arici2019research}%
\begin{APACrefauthors}%
Arici, F.%
, Yildirim, P.%
, Caliklar, S.%
\BCBL {}\ \BBA {} Yilmaz, R\BPBI M.%
\end{APACrefauthors}%
\unskip\
\newblock
\APACrefYearMonthDay{2019{\protect \BCnt {1}}}{}{}.
\newblock
{\BBOQ}\APACrefatitle {Research trends in the use of augmented reality in science education: Content and bibliometric mapping analysis} {Research trends in the use of augmented reality in science education: Content and bibliometric mapping analysis}.{\BBCQ}
\newblock
\APACjournalVolNumPages{Computers \& Education}{142}{}{103647}.
\PrintBackRefs{\CurrentBib}

\bibitem [\protect \citeauthoryear {%
Baker%
\ \BBA {} Lambourne%
}{%
Baker%
\ \BBA {} Lambourne%
}{%
{\protect \APACyear {1984}}%
}]{%
baker1982handbook}
\APACinsertmetastar {%
baker1982handbook}%
\begin{APACrefauthors}%
Baker, R.%
\BCBT {}\ \BBA {} Lambourne, A.%
\end{APACrefauthors}%
\unskip\
\newblock
\APACrefYear{1984}.
\newblock
\APACrefbtitle {Handbook for Television Subtitlers} {Handbook for television subtitlers}.
\newblock
\APACaddressPublisher{Winchester, Hampshire}{Independent Broadcasting Authority}.
\PrintBackRefs{\CurrentBib}

\bibitem [\protect \citeauthoryear {%
Berger%
, Niebuhr%
\BCBL {}\ \BBA {} Fischer%
}{%
Berger%
\ \protect \BOthers {.}}{%
{\protect \APACyear {2018}}%
}]{%
ElicitingExtraProminence2018}
\APACinsertmetastar {%
ElicitingExtraProminence2018}%
\begin{APACrefauthors}%
Berger, S.%
, Niebuhr, O.%
\BCBL {}\ \BBA {} Fischer, K.%
\end{APACrefauthors}%
\unskip\
\newblock
\APACrefYearMonthDay{2018}{}{}.
\newblock
{\BBOQ}\APACrefatitle {{Eliciting extra prominence in read-speech tasks: The effects of different text-highlighting methods on acoustic cues to perceived prominence}} {{Eliciting extra prominence in read-speech tasks: The effects of different text-highlighting methods on acoustic cues to perceived prominence}}.{\BBCQ}
\newblock
\BIn{} \APACrefbtitle {Proc. Speech Prosody 2018} {Proc. speech prosody 2018}\ (\BPGS\ 75--79).
\newblock
\APACaddressPublisher{Poznań, Poland}{International Speech Communication Association (ISCA)}.
\newblock
\begin{APACrefDOI} \doi{10.21437/SpeechProsody.2018-15} \end{APACrefDOI}
\PrintBackRefs{\CurrentBib}

\bibitem [\protect \citeauthoryear {%
Berke%
, Caulfield%
\BCBL {}\ \BBA {} Huenerfauth%
}{%
Berke%
\ \protect \BOthers {.}}{%
{\protect \APACyear {2017}}%
}]{%
confidence}
\APACinsertmetastar {%
confidence}%
\begin{APACrefauthors}%
Berke, L.%
, Caulfield, C.%
\BCBL {}\ \BBA {} Huenerfauth, M.%
\end{APACrefauthors}%
\unskip\
\newblock
\APACrefYearMonthDay{2017}{}{}.
\newblock
{\BBOQ}\APACrefatitle {Deaf and Hard-of-Hearing Perspectives on Imperfect Automatic Speech Recognition for Captioning One-on-One Meetings} {Deaf and hard-of-hearing perspectives on imperfect automatic speech recognition for captioning one-on-one meetings}.{\BBCQ}
\newblock
\BIn{} \APACrefbtitle {Proceedings of the 19th International ACM SIGACCESS Conference on Computers and Accessibility} {Proceedings of the 19th international acm sigaccess conference on computers and accessibility}\ (\BPG~155–164).
\newblock
\APACaddressPublisher{New York, NY, USA}{Association for Computing Machinery}.
\newblock
\begin{APACrefURL} \url{https://doi.org/10.1145/3132525.3132541} \end{APACrefURL}
\newblock
\begin{APACrefDOI} \doi{10.1145/3132525.3132541} \end{APACrefDOI}
\PrintBackRefs{\CurrentBib}

\bibitem [\protect \citeauthoryear {%
Birinci%
\ \BBA {} Sari\c{c}oban%
}{%
Birinci%
\ \BBA {} Sari\c{c}oban%
}{%
{\protect \APACyear {2021}}%
}]{%
JOURNALOF2021}
\APACinsertmetastar {%
JOURNALOF2021}%
\begin{APACrefauthors}%
Birinci, F.%
\BCBT {}\ \BBA {} Sari\c{c}oban, A.%
\end{APACrefauthors}%
\unskip\
\newblock
\APACrefYearMonthDay{2021}{03}{}.
\newblock
{\BBOQ}\APACrefatitle {JOURNAL OF LANGUAGE AND LINGUISTIC STUDIES The effectiveness of visual materials in teaching vocabulary to deaf students of EFL} {Journal of language and linguistic studies the effectiveness of visual materials in teaching vocabulary to deaf students of efl}.{\BBCQ}
\newblock
\APACjournalVolNumPages{Journal of Language and Linguistic Studies}{17}{}{}.
\newblock
\begin{APACrefDOI} \doi{10.52462/jlls.43} \end{APACrefDOI}
\PrintBackRefs{\CurrentBib}

\bibitem [\protect \citeauthoryear {%
Braun%
\ \BBA {} Clarke%
}{%
Braun%
\ \BBA {} Clarke%
}{%
{\protect \APACyear {2006}}%
}]{%
BraunClarke2006}
\APACinsertmetastar {%
BraunClarke2006}%
\begin{APACrefauthors}%
Braun, V.%
\BCBT {}\ \BBA {} Clarke, V.%
\end{APACrefauthors}%
\unskip\
\newblock
\APACrefYearMonthDay{2006}{}{}.
\newblock
{\BBOQ}\APACrefatitle {Using thematic analysis in psychology} {Using thematic analysis in psychology}.{\BBCQ}
\newblock
\APACjournalVolNumPages{Qualitative Research in Psychology}{3}{2}{77--101}.
\newblock
\begin{APACrefURL} \url{http://eprints.uwe.ac.uk/11735} \end{APACrefURL}
\PrintBackRefs{\CurrentBib}

\bibitem [\protect \citeauthoryear {%
Brooke%
}{%
Brooke%
}{%
{\protect \APACyear {1995}}%
}]{%
SUS1995}
\APACinsertmetastar {%
SUS1995}%
\begin{APACrefauthors}%
Brooke, J.%
\end{APACrefauthors}%
\unskip\
\newblock
\APACrefYearMonthDay{1995}{11}{}.
\newblock
{\BBOQ}\APACrefatitle {SUS: A quick and dirty usability scale} {Sus: A quick and dirty usability scale}.{\BBCQ}
\newblock
\APACjournalVolNumPages{Usability Eval. Ind.}{189}{}{}.
\PrintBackRefs{\CurrentBib}

\bibitem [\protect \citeauthoryear {%
Brown%
\ \protect \BOthers {.}}{%
Brown%
\ \protect \BOthers {.}}{%
{\protect \APACyear {2015}}%
}]{%
DynamicSubtitles2015}
\APACinsertmetastar {%
DynamicSubtitles2015}%
\begin{APACrefauthors}%
Brown, A.%
, Jones, R.%
, Crabb, M.%
, Sandford, J.%
, Brooks, M.%
, Armstrong, M.%
\BCBL {}\ \BBA {} Jay, C.%
\end{APACrefauthors}%
\unskip\
\newblock
\APACrefYearMonthDay{2015}{}{}.
\newblock
{\BBOQ}\APACrefatitle {Dynamic Subtitles: The User Experience} {Dynamic subtitles: The user experience}.{\BBCQ}
\newblock
\BIn{} \APACrefbtitle {Proceedings of the ACM International Conference on Interactive Experiences for TV and Online Video} {Proceedings of the acm international conference on interactive experiences for tv and online video}\ (\BPG~103–112).
\newblock
\APACaddressPublisher{New York, NY, USA}{Association for Computing Machinery}.
\newblock
\begin{APACrefURL} \url{https://doi.org/10.1145/2745197.2745204} \end{APACrefURL}
\newblock
\begin{APACrefDOI} \doi{10.1145/2745197.2745204} \end{APACrefDOI}
\PrintBackRefs{\CurrentBib}

\bibitem [\protect \citeauthoryear {%
Chi%
, Hong%
, Gumbrecht%
\BCBL {}\ \BBA {} Card%
}{%
Chi%
\ \protect \BOthers {.}}{%
{\protect \APACyear {2005}}%
}]{%
ScentHighlights2005}
\APACinsertmetastar {%
ScentHighlights2005}%
\begin{APACrefauthors}%
Chi, E\BPBI H.%
, Hong, L.%
, Gumbrecht, M.%
\BCBL {}\ \BBA {} Card, S\BPBI K.%
\end{APACrefauthors}%
\unskip\
\newblock
\APACrefYearMonthDay{2005}{}{}.
\newblock
{\BBOQ}\APACrefatitle {ScentHighlights: Highlighting Conceptually-Related Sentences during Reading} {Scenthighlights: Highlighting conceptually-related sentences during reading}.{\BBCQ}
\newblock
\BIn{} \APACrefbtitle {Proceedings of the 10th International Conference on Intelligent User Interfaces} {Proceedings of the 10th international conference on intelligent user interfaces}\ (\BPG~272–274).
\newblock
\APACaddressPublisher{New York, NY, USA}{Association for Computing Machinery}.
\newblock
\begin{APACrefURL} \url{https://doi.org/10.1145/1040830.1040895} \end{APACrefURL}
\newblock
\begin{APACrefDOI} \doi{10.1145/1040830.1040895} \end{APACrefDOI}
\PrintBackRefs{\CurrentBib}

\bibitem [\protect \citeauthoryear {%
Chiaro%
, Heiss%
\BCBL {}\ \BBA {} Bucaria%
}{%
Chiaro%
\ \protect \BOthers {.}}{%
{\protect \APACyear {2008}}%
}]{%
BetweenText2008}
\APACinsertmetastar {%
BetweenText2008}%
\begin{APACrefauthors}%
Chiaro, D.%
, Heiss, C.%
\BCBL {}\ \BBA {} Bucaria, C.%
\end{APACrefauthors}%
\unskip\
\newblock
\APACrefYear{2008}.
\newblock
\APACrefbtitle {Between Text and Image: Updating Research in Screen Translation} {Between text and image: Updating research in screen translation}.
\newblock
\APACaddressPublisher{Amsterdam, Netherlands}{John Benjamins Publishing Company}.
\PrintBackRefs{\CurrentBib}

\bibitem [\protect \citeauthoryear {%
Chin%
, Diehl%
\BCBL {}\ \BBA {} Norman%
}{%
Chin%
\ \protect \BOthers {.}}{%
{\protect \APACyear {1988}}%
}]{%
QUIS1988}
\APACinsertmetastar {%
QUIS1988}%
\begin{APACrefauthors}%
Chin, J\BPBI P.%
, Diehl, V\BPBI A.%
\BCBL {}\ \BBA {} Norman, K\BPBI L.%
\end{APACrefauthors}%
\unskip\
\newblock
\APACrefYearMonthDay{1988}{}{}.
\newblock
{\BBOQ}\APACrefatitle {Development of an Instrument Measuring User Satisfaction of the Human-Computer Interface} {Development of an instrument measuring user satisfaction of the human-computer interface}.{\BBCQ}
\newblock
\BIn{} \APACrefbtitle {Proceedings of the SIGCHI Conference on Human Factors in Computing Systems} {Proceedings of the sigchi conference on human factors in computing systems}\ (\BPG~213–218).
\newblock
\APACaddressPublisher{New York, NY, USA}{Association for Computing Machinery}.
\newblock
\begin{APACrefURL} \url{https://doi.org/10.1145/57167.57203} \end{APACrefURL}
\newblock
\begin{APACrefDOI} \doi{10.1145/57167.57203} \end{APACrefDOI}
\PrintBackRefs{\CurrentBib}

\bibitem [\protect \citeauthoryear {%
Clark%
}{%
Clark%
}{%
{\protect \APACyear {1981}}%
}]{%
Clark1981-ca}
\APACinsertmetastar {%
Clark1981-ca}%
\begin{APACrefauthors}%
Clark, J\BPBI G.%
\end{APACrefauthors}%
\unskip\
\newblock
\APACrefYearMonthDay{1981}{07}{}.
\newblock
{\BBOQ}\APACrefatitle {Uses and abuses of hearing loss classification} {Uses and abuses of hearing loss classification}.{\BBCQ}
\newblock
\APACjournalVolNumPages{ASHA}{23}{7}{493-500}.
\PrintBackRefs{\CurrentBib}

\bibitem [\protect \citeauthoryear {%
Dostal%
\ \BBA {} Wolbers%
}{%
Dostal%
\ \BBA {} Wolbers%
}{%
{\protect \APACyear {2014}}%
}]{%
DevelopingLanguage2014}
\APACinsertmetastar {%
DevelopingLanguage2014}%
\begin{APACrefauthors}%
Dostal, H\BPBI M.%
\BCBT {}\ \BBA {} Wolbers, K\BPBI A.%
\end{APACrefauthors}%
\unskip\
\newblock
\APACrefYearMonthDay{2014}{}{}.
\newblock
{\BBOQ}\APACrefatitle {Developing language and writing skills of deaf and hard of hearing students: A simultaneous approach} {Developing language and writing skills of deaf and hard of hearing students: A simultaneous approach}.{\BBCQ}
\newblock
\APACjournalVolNumPages{Literacy Research and Instruction}{53}{3}{245–268}.
\newblock
\begin{APACrefDOI} \doi{10.1080/19388071.2014.907382} \end{APACrefDOI}
\PrintBackRefs{\CurrentBib}

\bibitem [\protect \citeauthoryear {%
Duchnicky%
\ \BBA {} Kolers%
}{%
Duchnicky%
\ \BBA {} Kolers%
}{%
{\protect \APACyear {1983}}%
}]{%
duchnicky1983readability}
\APACinsertmetastar {%
duchnicky1983readability}%
\begin{APACrefauthors}%
Duchnicky, R\BPBI L.%
\BCBT {}\ \BBA {} Kolers, P\BPBI A.%
\end{APACrefauthors}%
\unskip\
\newblock
\APACrefYearMonthDay{1983}{}{}.
\newblock
{\BBOQ}\APACrefatitle {Readability of Text Scrolled on Visual Display Terminals as a Function of Window Size} {Readability of text scrolled on visual display terminals as a function of window size}.{\BBCQ}
\newblock
\APACjournalVolNumPages{Human Factors}{25}{6}{683–692}.
\newblock
\begin{APACrefDOI} \doi{10.1177/001872088302500605} \end{APACrefDOI}
\PrintBackRefs{\CurrentBib}

\bibitem [\protect \citeauthoryear {%
Dye%
, Baril%
\BCBL {}\ \BBA {} Bavelier%
}{%
Dye%
\ \protect \BOthers {.}}{%
{\protect \APACyear {2007}}%
}]{%
Dye2007Which}
\APACinsertmetastar {%
Dye2007Which}%
\begin{APACrefauthors}%
Dye, M\BPBI W\BPBI G.%
, Baril, D\BPBI E.%
\BCBL {}\ \BBA {} Bavelier, D.%
\end{APACrefauthors}%
\unskip\
\newblock
\APACrefYearMonthDay{2007}{}{}.
\newblock
{\BBOQ}\APACrefatitle {Which aspects of visual attention are changed by deafness? The case of the Attentional Network Test} {Which aspects of visual attention are changed by deafness? the case of the attentional network test}.{\BBCQ}
\newblock
\APACjournalVolNumPages{Neuropsychologia}{45}{}{1801-1811}.
\newblock
\begin{APACrefDOI} \doi{10.1016/j.neuropsychologia.2006.12.019} \end{APACrefDOI}
\PrintBackRefs{\CurrentBib}

\bibitem [\protect \citeauthoryear {%
Forster%
}{%
Forster%
}{%
{\protect \APACyear {1970}}%
}]{%
VisualPerception1970}
\APACinsertmetastar {%
VisualPerception1970}%
\begin{APACrefauthors}%
Forster, K.%
\end{APACrefauthors}%
\unskip\
\newblock
\APACrefYearMonthDay{1970}{07}{}.
\newblock
{\BBOQ}\APACrefatitle {Visual perception of rapidly presented word sequences of varying complexity} {Visual perception of rapidly presented word sequences of varying complexity}.{\BBCQ}
\newblock
\APACjournalVolNumPages{Perception \& Psychophysics}{8}{}{215-221}.
\newblock
\begin{APACrefDOI} \doi{10.3758/BF03210208} \end{APACrefDOI}
\PrintBackRefs{\CurrentBib}

\bibitem [\protect \citeauthoryear {%
Garberoglio%
, Johnson%
, Sales%
\BCBL {}\ \BBA {} Cawthon%
}{%
Garberoglio%
\ \protect \BOthers {.}}{%
{\protect \APACyear {2021}}%
}]{%
Garberoglio2021}
\APACinsertmetastar {%
Garberoglio2021}%
\begin{APACrefauthors}%
Garberoglio, C\BPBI L.%
, Johnson, P\BPBI M.%
, Sales, A.%
\BCBL {}\ \BBA {} Cawthon, S\BPBI W.%
\end{APACrefauthors}%
\unskip\
\newblock
\APACrefYearMonthDay{2021}{}{}.
\newblock
{\BBOQ}\APACrefatitle {Change Over Time in Educational Attainment for Deaf Individuals from 2008-2018} {Change over time in educational attainment for deaf individuals from 2008-2018}.{\BBCQ}
\newblock
\APACjournalVolNumPages{Journal of Postsecondary Education and Disability}{34}{3}{253--272}.
\PrintBackRefs{\CurrentBib}

\bibitem [\protect \citeauthoryear {%
Garz{\'o}n%
, Pav{\'o}n%
\BCBL {}\ \BBA {} Baldiris%
}{%
Garz{\'o}n%
\ \protect \BOthers {.}}{%
{\protect \APACyear {2019}}%
}]{%
garzon2019systematic}
\APACinsertmetastar {%
garzon2019systematic}%
\begin{APACrefauthors}%
Garz{\'o}n, J.%
, Pav{\'o}n, J.%
\BCBL {}\ \BBA {} Baldiris, S.%
\end{APACrefauthors}%
\unskip\
\newblock
\APACrefYearMonthDay{2019}{}{}.
\newblock
{\BBOQ}\APACrefatitle {Systematic review and meta-analysis of augmented reality in educational settings} {Systematic review and meta-analysis of augmented reality in educational settings}.{\BBCQ}
\newblock
\APACjournalVolNumPages{Virtual Reality}{23}{}{447--459}.
\newblock
\begin{APACrefDOI} \doi{10.1007/s10055-019-00379-9} \end{APACrefDOI}
\PrintBackRefs{\CurrentBib}

\bibitem [\protect \citeauthoryear {%
Hart%
}{%
Hart%
}{%
{\protect \APACyear {2006}}%
}]{%
NasaTaskLoad2006}
\APACinsertmetastar {%
NasaTaskLoad2006}%
\begin{APACrefauthors}%
Hart, S\BPBI G.%
\end{APACrefauthors}%
\unskip\
\newblock
\APACrefYearMonthDay{2006}{}{}.
\newblock
{\BBOQ}\APACrefatitle {Nasa-Task Load Index (NASA-TLX); 20 Years Later} {Nasa-task load index (nasa-tlx); 20 years later}.{\BBCQ}
\newblock
\APACjournalVolNumPages{Proceedings of the Human Factors and Ergonomics Society Annual Meeting}{50}{9}{904-908}.
\newblock
\begin{APACrefDOI} \doi{10.1177/154193120605000909} \end{APACrefDOI}
\PrintBackRefs{\CurrentBib}

\bibitem [\protect \citeauthoryear {%
Hollander%
, Wolfe%
\BCBL {}\ \BBA {} Chicken%
}{%
Hollander%
\ \protect \BOthers {.}}{%
{\protect \APACyear {1999}}%
}]{%
Friedman1999}
\APACinsertmetastar {%
Friedman1999}%
\begin{APACrefauthors}%
Hollander, M.%
, Wolfe, D\BPBI A.%
\BCBL {}\ \BBA {} Chicken, E.%
\end{APACrefauthors}%
\unskip\
\newblock
\APACrefYear{1999}.
\newblock
\APACrefbtitle {Nonparametric Statistical Methods} {Nonparametric statistical methods}.
\newblock
\APACaddressPublisher{New York, NY, USA}{Wiley}.
\PrintBackRefs{\CurrentBib}

\bibitem [\protect \citeauthoryear {%
Hong%
, Wang%
, Xu%
, Yan%
\BCBL {}\ \BBA {} Chua%
}{%
Hong%
\ \protect \BOthers {.}}{%
{\protect \APACyear {2010}}%
}]{%
DynamicCaptioning2010}
\APACinsertmetastar {%
DynamicCaptioning2010}%
\begin{APACrefauthors}%
Hong, R.%
, Wang, M.%
, Xu, M.%
, Yan, S.%
\BCBL {}\ \BBA {} Chua, T\BHBI S.%
\end{APACrefauthors}%
\unskip\
\newblock
\APACrefYearMonthDay{2010}{}{}.
\newblock
{\BBOQ}\APACrefatitle {Dynamic captioning: video accessibility enhancement for hearing impairment} {Dynamic captioning: video accessibility enhancement for hearing impairment}.{\BBCQ}
\newblock
\BIn{} \APACrefbtitle {Proceedings of the 18th ACM International Conference on Multimedia} {Proceedings of the 18th acm international conference on multimedia}\ (\BPG~421–430).
\newblock
\APACaddressPublisher{New York, NY, USA}{Association for Computing Machinery}.
\newblock
\begin{APACrefURL} \url{https://doi.org/10.1145/1873951.1874013} \end{APACrefURL}
\newblock
\begin{APACrefDOI} \doi{10.1145/1873951.1874013} \end{APACrefDOI}
\PrintBackRefs{\CurrentBib}

\bibitem [\protect \citeauthoryear {%
Ibáñez%
\ \BBA {} Delgado-Kloos%
}{%
Ibáñez%
\ \BBA {} Delgado-Kloos%
}{%
{\protect \APACyear {2018}}%
}]{%
ibanez2018augmented}
\APACinsertmetastar {%
ibanez2018augmented}%
\begin{APACrefauthors}%
Ibáñez, M\BHBI B.%
\BCBT {}\ \BBA {} Delgado-Kloos, C.%
\end{APACrefauthors}%
\unskip\
\newblock
\APACrefYearMonthDay{2018}{}{}.
\newblock
{\BBOQ}\APACrefatitle {Augmented reality for STEM learning: A systematic review} {Augmented reality for stem learning: A systematic review}.{\BBCQ}
\newblock
\APACjournalVolNumPages{Computers \& Education}{123}{}{109--123}.
\PrintBackRefs{\CurrentBib}

\bibitem [\protect \citeauthoryear {%
Ioannou%
\ \BBA {} Constantinou%
}{%
Ioannou%
\ \BBA {} Constantinou%
}{%
{\protect \APACyear {2018}}%
}]{%
Ioannou2017Augmented}
\APACinsertmetastar {%
Ioannou2017Augmented}%
\begin{APACrefauthors}%
Ioannou, A.%
\BCBT {}\ \BBA {} Constantinou, V.%
\end{APACrefauthors}%
\unskip\
\newblock
\APACrefYearMonthDay{2018}{}{}.
\newblock
{\BBOQ}\APACrefatitle {Augmented Reality Supporting Deaf Students in Mainstream Schools: Two Case Studies of Practical Utility of the Technology} {Augmented reality supporting deaf students in mainstream schools: Two case studies of practical utility of the technology}.{\BBCQ}
\newblock
\BIn{} \APACrefbtitle {Interactive Mobile Communication Technologies and Learning} {Interactive mobile communication technologies and learning}\ (\BPG~387-396).
\newblock
\APACaddressPublisher{Cham}{Springer International Publishing}.
\newblock
\begin{APACrefDOI} \doi{10.1007/978-3-319-75175-7_39} \end{APACrefDOI}
\PrintBackRefs{\CurrentBib}

\bibitem [\protect \citeauthoryear {%
Izaguirre%
, Abásolo%
\BCBL {}\ \BBA {} Collazos%
}{%
Izaguirre%
\ \protect \BOthers {.}}{%
{\protect \APACyear {2020}}%
}]{%
izaguirre2020mobile}
\APACinsertmetastar {%
izaguirre2020mobile}%
\begin{APACrefauthors}%
Izaguirre, E\BPBI D\BPBI P.%
, Abásolo, M\BPBI J.%
\BCBL {}\ \BBA {} Collazos, C\BPBI A.%
\end{APACrefauthors}%
\unskip\
\newblock
\APACrefYearMonthDay{2020}{}{}.
\newblock
{\BBOQ}\APACrefatitle {Mobile technology and extended reality for deaf people: A systematic review of the open access literature} {Mobile technology and extended reality for deaf people: A systematic review of the open access literature}.{\BBCQ}
\newblock
\BIn{} \APACrefbtitle {2020 XV Conferencia Latinoamericana de Tecnologias de Aprendizaje (LACLO)} {2020 xv conferencia latinoamericana de tecnologias de aprendizaje (laclo)}\ (\BPGS\ 1--8).
\PrintBackRefs{\CurrentBib}

\bibitem [\protect \citeauthoryear {%
Jachova%
, Kovacevic%
, Ristovska%
\BCBL {}\ \BBA {} Radovanovic%
}{%
Jachova%
\ \protect \BOthers {.}}{%
{\protect \APACyear {2023}}%
}]{%
jachova2023deaf}
\APACinsertmetastar {%
jachova2023deaf}%
\begin{APACrefauthors}%
Jachova, Z.%
, Kovacevic, J.%
, Ristovska, L.%
\BCBL {}\ \BBA {} Radovanovic, V.%
\end{APACrefauthors}%
\unskip\
\newblock
\APACrefYearMonthDay{2023}{09}{}.
\newblock
{\BBOQ}\APACrefatitle {Deaf and hard of hearing students in inclusive education classrooms} {Deaf and hard of hearing students in inclusive education classrooms}.{\BBCQ}
\newblock
\BIn{} \APACrefbtitle {Proceedings of the 6th International Scientific Conference 30 Years of Studies in Special Education and Rehabilitation} {Proceedings of the 6th international scientific conference 30 years of studies in special education and rehabilitation}\ (\BPGS\ 532--546).
\newblock
\APACaddressPublisher{Skopje}{Faculty of Philosophy, Skopje}.
\PrintBackRefs{\CurrentBib}

\bibitem [\protect \citeauthoryear {%
Jain%
, Chinh%
, Findlater%
, Kushalnagar%
\BCBL {}\ \BBA {} Froehlich%
}{%
Jain%
\ \protect \BOthers {.}}{%
{\protect \APACyear {2018}}%
}]{%
ExploringAugmentedReality2018}
\APACinsertmetastar {%
ExploringAugmentedReality2018}%
\begin{APACrefauthors}%
Jain, D.%
, Chinh, B.%
, Findlater, L.%
, Kushalnagar, R.%
\BCBL {}\ \BBA {} Froehlich, J.%
\end{APACrefauthors}%
\unskip\
\newblock
\APACrefYearMonthDay{2018}{}{}.
\newblock
{\BBOQ}\APACrefatitle {Exploring Augmented Reality Approaches to Real-Time Captioning: A Preliminary Autoethnographic Study} {Exploring augmented reality approaches to real-time captioning: A preliminary autoethnographic study}.{\BBCQ}
\newblock
\BIn{} \APACrefbtitle {Proceedings of the 2018 ACM Conference Companion Publication on Designing Interactive Systems} {Proceedings of the 2018 acm conference companion publication on designing interactive systems}\ (\BPG~7–11).
\newblock
\APACaddressPublisher{New York, NY, USA}{Association for Computing Machinery}.
\newblock
\begin{APACrefURL} \url{https://doi.org/10.1145/3197391.3205404} \end{APACrefURL}
\newblock
\begin{APACrefDOI} \doi{10.1145/3197391.3205404} \end{APACrefDOI}
\PrintBackRefs{\CurrentBib}

\bibitem [\protect \citeauthoryear {%
Kafle%
\ \BBA {} Huenerfauth%
}{%
Kafle%
\ \BBA {} Huenerfauth%
}{%
{\protect \APACyear {2016}}%
}]{%
Kafle2016Effect}
\APACinsertmetastar {%
Kafle2016Effect}%
\begin{APACrefauthors}%
Kafle, S.%
\BCBT {}\ \BBA {} Huenerfauth, M.%
\end{APACrefauthors}%
\unskip\
\newblock
\APACrefYearMonthDay{2016}{}{}.
\newblock
{\BBOQ}\APACrefatitle {{Effect of Speech Recognition Errors on Text Understandability for People who are Deaf or Hard of Hearing}} {{Effect of Speech Recognition Errors on Text Understandability for People who are Deaf or Hard of Hearing}}.{\BBCQ}
\newblock
\BIn{} \APACrefbtitle {Proc. 7th Workshop on Speech and Language Processing for Assistive Technologies (SLPAT 2016)} {Proc. 7th workshop on speech and language processing for assistive technologies (slpat 2016)}\ (\BPGS\ 20--25).
\newblock
\APACaddressPublisher{San Francisco, USA}{International Speech Communication Association (ISCA)}.
\newblock
\begin{APACrefDOI} \doi{10.21437/SLPAT.2016-4} \end{APACrefDOI}
\PrintBackRefs{\CurrentBib}

\bibitem [\protect \citeauthoryear {%
Karamitroglou%
}{%
Karamitroglou%
}{%
{\protect \APACyear {1998}}%
}]{%
AProposedSet1998}
\APACinsertmetastar {%
AProposedSet1998}%
\begin{APACrefauthors}%
Karamitroglou, F.%
\end{APACrefauthors}%
\unskip\
\newblock
\APACrefYearMonthDay{1998}{}{}.
\newblock
{\BBOQ}\APACrefatitle {A Proposed Set of Subtitling Standards in Europe} {A proposed set of subtitling standards in europe}.{\BBCQ}
\newblock
\APACjournalVolNumPages{Translation Journal}{2}{2}{1-15}.
\PrintBackRefs{\CurrentBib}

\bibitem [\protect \citeauthoryear {%
Kercher%
\ \BBA {} Rowe%
}{%
Kercher%
\ \BBA {} Rowe%
}{%
{\protect \APACyear {2012}}%
}]{%
Kercher2012Improving}
\APACinsertmetastar {%
Kercher2012Improving}%
\begin{APACrefauthors}%
Kercher, K.%
\BCBT {}\ \BBA {} Rowe, D\BPBI C.%
\end{APACrefauthors}%
\unskip\
\newblock
\APACrefYearMonthDay{2012}{}{}.
\newblock
{\BBOQ}\APACrefatitle {Improving the learning experience for the deaf through augment reality innovations} {Improving the learning experience for the deaf through augment reality innovations}.{\BBCQ}
\newblock
\BIn{} \APACrefbtitle {2012 18th International ICE Conference on Engineering, Technology and Innovation} {2012 18th international ice conference on engineering, technology and innovation}\ (\BPG~1-11).
\newblock
\APACaddressPublisher{Munich, Germany}{IEEE}.
\newblock
\begin{APACrefDOI} \doi{10.1109/ICE.2012.6297673} \end{APACrefDOI}
\PrintBackRefs{\CurrentBib}

\bibitem [\protect \citeauthoryear {%
Koolstra%
, Peeters%
\BCBL {}\ \BBA {} Spinhof%
}{%
Koolstra%
\ \protect \BOthers {.}}{%
{\protect \APACyear {2002}}%
}]{%
TheProsAndCons2002}
\APACinsertmetastar {%
TheProsAndCons2002}%
\begin{APACrefauthors}%
Koolstra, C.%
, Peeters, A.%
\BCBL {}\ \BBA {} Spinhof, H.%
\end{APACrefauthors}%
\unskip\
\newblock
\APACrefYearMonthDay{2002}{09}{}.
\newblock
{\BBOQ}\APACrefatitle {The Pros and Cons of Dubbing and Subtitling} {The pros and cons of dubbing and subtitling}.{\BBCQ}
\newblock
\APACjournalVolNumPages{European Journal of Communication - EUR J COMMUN}{17}{}{325-354}.
\newblock
\begin{APACrefDOI} \doi{10.1177/0267323102017003694} \end{APACrefDOI}
\PrintBackRefs{\CurrentBib}

\bibitem [\protect \citeauthoryear {%
Kurzhals%
, Cetinkaya%
, Hu%
, Wang%
\BCBL {}\ \BBA {} Weiskopf%
}{%
Kurzhals%
\ \protect \BOthers {.}}{%
{\protect \APACyear {2017}}%
}]{%
CloseToTheAction2017}
\APACinsertmetastar {%
CloseToTheAction2017}%
\begin{APACrefauthors}%
Kurzhals, K.%
, Cetinkaya, E.%
, Hu, Y.%
, Wang, W.%
\BCBL {}\ \BBA {} Weiskopf, D.%
\end{APACrefauthors}%
\unskip\
\newblock
\APACrefYearMonthDay{2017}{}{}.
\newblock
{\BBOQ}\APACrefatitle {Close to the Action: Eye-Tracking Evaluation of Speaker-Following Subtitles} {Close to the action: Eye-tracking evaluation of speaker-following subtitles}.{\BBCQ}
\newblock
\BIn{} \APACrefbtitle {Proceedings of the 2017 CHI Conference on Human Factors in Computing Systems} {Proceedings of the 2017 chi conference on human factors in computing systems}\ (\BPG~6559–6568).
\newblock
\APACaddressPublisher{New York, NY, USA}{Association for Computing Machinery}.
\newblock
\begin{APACrefURL} \url{https://doi.org/10.1145/3025453.3025772} \end{APACrefURL}
\newblock
\begin{APACrefDOI} \doi{10.1145/3025453.3025772} \end{APACrefDOI}
\PrintBackRefs{\CurrentBib}

\bibitem [\protect \citeauthoryear {%
Kushalnagar%
\ \BBA {} Kushalnagar%
}{%
Kushalnagar%
\ \BBA {} Kushalnagar%
}{%
{\protect \APACyear {2014}}%
}]{%
CollaborativeGaze2014}
\APACinsertmetastar {%
CollaborativeGaze2014}%
\begin{APACrefauthors}%
Kushalnagar, R.%
\BCBT {}\ \BBA {} Kushalnagar, P.%
\end{APACrefauthors}%
\unskip\
\newblock
\APACrefYearMonthDay{2014}{}{}.
\newblock
{\BBOQ}\APACrefatitle {Collaborative Gaze Cues and Replay for Deaf and Hard of Hearing Students} {Collaborative gaze cues and replay for deaf and hard of hearing students}.{\BBCQ}
\newblock
\BIn{} K.~Miesenberger, D.~Fels, D.~Archambault, P.~Pe{\v{n}}{\'a}z\BCBL {}\ \BBA {} W.~Zagler\ (\BEDS), \APACrefbtitle {Computers Helping People with Special Needs} {Computers helping people with special needs}\ (\BPGS\ 415--422).
\newblock
\APACaddressPublisher{Cham}{Springer International Publishing}.
\PrintBackRefs{\CurrentBib}

\bibitem [\protect \citeauthoryear {%
Lederberg%
, Schick%
\BCBL {}\ \BBA {} Spencer%
}{%
Lederberg%
\ \protect \BOthers {.}}{%
{\protect \APACyear {2013}}%
}]{%
Lederberg2013}
\APACinsertmetastar {%
Lederberg2013}%
\begin{APACrefauthors}%
Lederberg, A\BPBI R.%
, Schick, B.%
\BCBL {}\ \BBA {} Spencer, P\BPBI E.%
\end{APACrefauthors}%
\unskip\
\newblock
\APACrefYearMonthDay{2013}{}{}.
\newblock
{\BBOQ}\APACrefatitle {Language and literacy development of deaf and hard-of-hearing children: successes and challenges.} {Language and literacy development of deaf and hard-of-hearing children: successes and challenges.}{\BBCQ}
\newblock
\APACjournalVolNumPages{Developmental Psychology}{49}{}{15-30}.
\newblock
\begin{APACrefDOI} \doi{10.1037/a0029558} \end{APACrefDOI}
\PrintBackRefs{\CurrentBib}

\bibitem [\protect \citeauthoryear {%
Leigh%
\ \BBA {} Crowe%
}{%
Leigh%
\ \BBA {} Crowe%
}{%
{\protect \APACyear {2020}}%
}]{%
EvidenceBasedPractices2020}
\APACinsertmetastar {%
EvidenceBasedPractices2020}%
\begin{APACrefauthors}%
Leigh, G.%
\BCBT {}\ \BBA {} Crowe, K.%
\end{APACrefauthors}%
\unskip\
\newblock
\APACrefYearMonthDay{2020}{09}{}.
\newblock
\APACrefbtitle {Evidence-Based Practices for Teaching Learners who are Deaf or Hard of Hearing in Regular Classrooms.} {Evidence-based practices for teaching learners who are deaf or hard of hearing in regular classrooms.}
\newblock
\APACaddressPublisher{}{Oxford University Press}.
\newblock
\begin{APACrefURL} \url{https://oxfordre.com/education/view/10.1093/acrefore/9780190264093.001.0001/acrefore-9780190264093-e-1258} \end{APACrefURL}
\newblock
\begin{APACrefDOI} \doi{10.1093/acrefore/9780190264093.013.1258} \end{APACrefDOI}
\PrintBackRefs{\CurrentBib}

\bibitem [\protect \citeauthoryear {%
Luckner%
\ \BBA {} Pierce%
}{%
Luckner%
\ \BBA {} Pierce%
}{%
{\protect \APACyear {2013}}%
}]{%
Luckner2013Response}
\APACinsertmetastar {%
Luckner2013Response}%
\begin{APACrefauthors}%
Luckner, J\BPBI L.%
\BCBT {}\ \BBA {} Pierce, C\BPBI D.%
\end{APACrefauthors}%
\unskip\
\newblock
\APACrefYearMonthDay{2013}{}{}.
\newblock
{\BBOQ}\APACrefatitle {Response to Intervention and Students Who are Deaf or Hard of Hearing} {Response to intervention and students who are deaf or hard of hearing}.{\BBCQ}
\newblock
\APACjournalVolNumPages{Deafness \& Education International}{15}{}{222 - 240}.
\newblock
\begin{APACrefDOI} \doi{10.1179/1557069X13Y.0000000027} \end{APACrefDOI}
\PrintBackRefs{\CurrentBib}

\bibitem [\protect \citeauthoryear {%
Luo%
, Weng%
, Songrui%
, Hao%
\BCBL {}\ \BBA {} Tu%
}{%
Luo%
\ \protect \BOthers {.}}{%
{\protect \APACyear {2022}}%
}]{%
Luo2022Avatar}
\APACinsertmetastar {%
Luo2022Avatar}%
\begin{APACrefauthors}%
Luo, L.%
, Weng, D.%
, Songrui, G.%
, Hao, J.%
\BCBL {}\ \BBA {} Tu, Z.%
\end{APACrefauthors}%
\unskip\
\newblock
\APACrefYearMonthDay{2022}{}{}.
\newblock
{\BBOQ}\APACrefatitle {Avatar Interpreter: Improving Classroom Experiences for Deaf and Hard-of-Hearing People Based on Augmented Reality} {Avatar interpreter: Improving classroom experiences for deaf and hard-of-hearing people based on augmented reality}.{\BBCQ}
\newblock
\BIn{} \APACrefbtitle {Extended Abstracts of the 2022 CHI Conference on Human Factors in Computing Systems.} {Extended abstracts of the 2022 chi conference on human factors in computing systems.}
\newblock
\APACaddressPublisher{New York, NY, USA}{Association for Computing Machinery}.
\newblock
\begin{APACrefURL} \url{https://doi.org/10.1145/3491101.3519799} \end{APACrefURL}
\newblock
\begin{APACrefDOI} \doi{10.1145/3491101.3519799} \end{APACrefDOI}
\PrintBackRefs{\CurrentBib}

\bibitem [\protect \citeauthoryear {%
Marschark%
\ \protect \BOthers {.}}{%
Marschark%
\ \protect \BOthers {.}}{%
{\protect \APACyear {2006}}%
}]{%
marschark2006benefits}
\APACinsertmetastar {%
marschark2006benefits}%
\begin{APACrefauthors}%
Marschark, M.%
, Leigh, G.%
, Sapere, P.%
, Burnham, D.%
, Convertino, C.%
, Stinson, M.%
\BDBL {}Noble, W.%
\end{APACrefauthors}%
\unskip\
\newblock
\APACrefYearMonthDay{2006}{08}{}.
\newblock
{\BBOQ}\APACrefatitle {{Benefits of Sign Language Interpreting and Text Alternatives for Deaf Students' Classroom Learning}} {{Benefits of Sign Language Interpreting and Text Alternatives for Deaf Students' Classroom Learning}}.{\BBCQ}
\newblock
\APACjournalVolNumPages{The Journal of Deaf Studies and Deaf Education}{11}{4}{421-437}.
\newblock
\begin{APACrefURL} \url{https://doi.org/10.1093/deafed/enl013} \end{APACrefURL}
\newblock
\begin{APACrefDOI} \doi{10.1093/deafed/enl013} \end{APACrefDOI}
\PrintBackRefs{\CurrentBib}

\bibitem [\protect \citeauthoryear {%
Masson%
}{%
Masson%
}{%
{\protect \APACyear {1983}}%
}]{%
masson1983conceptual}
\APACinsertmetastar {%
masson1983conceptual}%
\begin{APACrefauthors}%
Masson, M\BPBI E\BPBI J.%
\end{APACrefauthors}%
\unskip\
\newblock
\APACrefYearMonthDay{1983}{May}{}.
\newblock
{\BBOQ}\APACrefatitle {Conceptual processing of text during skimming and rapid sequential reading} {Conceptual processing of text during skimming and rapid sequential reading}.{\BBCQ}
\newblock
\APACjournalVolNumPages{Memory \& Cognition}{11}{3}{262--274}.
\newblock
\begin{APACrefDOI} \doi{10.3758/BF03196973} \end{APACrefDOI}
\PrintBackRefs{\CurrentBib}

\bibitem [\protect \citeauthoryear {%
Ozcelik%
, Arslan-Ari%
\BCBL {}\ \BBA {} Cagiltay%
}{%
Ozcelik%
\ \protect \BOthers {.}}{%
{\protect \APACyear {2010}}%
}]{%
OZCELIK2010110}
\APACinsertmetastar {%
OZCELIK2010110}%
\begin{APACrefauthors}%
Ozcelik, E.%
, Arslan-Ari, I.%
\BCBL {}\ \BBA {} Cagiltay, K.%
\end{APACrefauthors}%
\unskip\
\newblock
\APACrefYearMonthDay{2010}{}{}.
\newblock
{\BBOQ}\APACrefatitle {Why does signaling enhance multimedia learning? Evidence from eye movements} {Why does signaling enhance multimedia learning? evidence from eye movements}.{\BBCQ}
\newblock
\APACjournalVolNumPages{Computers in Human Behavior}{26}{1}{110-117}.
\newblock
\begin{APACrefURL} \url{https://doi.org/10.1016/j.chb.2009.09.001} \end{APACrefURL}
\newblock
\begin{APACrefDOI} \doi{10.1016/j.chb.2009.09.001} \end{APACrefDOI}
\PrintBackRefs{\CurrentBib}

\bibitem [\protect \citeauthoryear {%
Paul%
\ \protect \BOthers {.}}{%
Paul%
\ \protect \BOthers {.}}{%
{\protect \APACyear {2020}}%
}]{%
paul2020current}
\APACinsertmetastar {%
paul2020current}%
\begin{APACrefauthors}%
Paul, R.%
, Paatsch, L.%
, Caselli, N.%
, Garberoglio, C\BPBI L.%
, Goldin-Meadow, S.%
\BCBL {}\ \BBA {} Lederberg, A.%
\end{APACrefauthors}%
\unskip\
\newblock
\APACrefYearMonthDay{2020}{}{}.
\newblock
{\BBOQ}\APACrefatitle {Current research in pragmatic language use among deaf and hard of hearing children} {Current research in pragmatic language use among deaf and hard of hearing children}.{\BBCQ}
\newblock
\APACjournalVolNumPages{Pediatrics}{146}{Supplement\_3}{S237--S245}.
\PrintBackRefs{\CurrentBib}

\bibitem [\protect \citeauthoryear {%
Peng%
\ \protect \BOthers {.}}{%
Peng%
\ \protect \BOthers {.}}{%
{\protect \APACyear {2018}}%
}]{%
SpeechBubbles2018}
\APACinsertmetastar {%
SpeechBubbles2018}%
\begin{APACrefauthors}%
Peng, Y\BHBI H.%
, Hsi, M\BHBI W.%
, Taele, P.%
, Lin, T\BHBI Y.%
, Lai, P\BHBI E.%
, Hsu, L.%
\BDBL {}Chen, M\BPBI Y.%
\end{APACrefauthors}%
\unskip\
\newblock
\APACrefYearMonthDay{2018}{}{}.
\newblock
{\BBOQ}\APACrefatitle {SpeechBubbles: Enhancing Captioning Experiences for Deaf and Hard-of-Hearing People in Group Conversations} {Speechbubbles: Enhancing captioning experiences for deaf and hard-of-hearing people in group conversations}.{\BBCQ}
\newblock
\BIn{} \APACrefbtitle {Proceedings of the 2018 CHI Conference on Human Factors in Computing Systems} {Proceedings of the 2018 chi conference on human factors in computing systems}\ (\BPG~1–10).
\newblock
\APACaddressPublisher{New York, NY, USA}{Association for Computing Machinery}.
\newblock
\begin{APACrefURL} \url{https://doi.org/10.1145/3173574.3173867} \end{APACrefURL}
\newblock
\begin{APACrefDOI} \doi{10.1145/3173574.3173867} \end{APACrefDOI}
\PrintBackRefs{\CurrentBib}

\bibitem [\protect \citeauthoryear {%
Ponce%
\ \BBA {} Mayer%
}{%
Ponce%
\ \BBA {} Mayer%
}{%
{\protect \APACyear {2014}}%
}]{%
AnEyeMovement2014}
\APACinsertmetastar {%
AnEyeMovement2014}%
\begin{APACrefauthors}%
Ponce, H.%
\BCBT {}\ \BBA {} Mayer, R.%
\end{APACrefauthors}%
\unskip\
\newblock
\APACrefYearMonthDay{2014}{12}{}.
\newblock
{\BBOQ}\APACrefatitle {An eye movement analysis of highlighting and graphic organizer study aids for learning from expository text} {An eye movement analysis of highlighting and graphic organizer study aids for learning from expository text}.{\BBCQ}
\newblock
\APACjournalVolNumPages{Computers in Human Behavior}{41}{}{21–32}.
\newblock
\begin{APACrefDOI} \doi{10.1016/j.chb.2014.09.010} \end{APACrefDOI}
\PrintBackRefs{\CurrentBib}

\bibitem [\protect \citeauthoryear {%
Proksch%
\ \BBA {} Bavelier%
}{%
Proksch%
\ \BBA {} Bavelier%
}{%
{\protect \APACyear {2002}}%
}]{%
Proksch2002Changes}
\APACinsertmetastar {%
Proksch2002Changes}%
\begin{APACrefauthors}%
Proksch, J.%
\BCBT {}\ \BBA {} Bavelier, D.%
\end{APACrefauthors}%
\unskip\
\newblock
\APACrefYearMonthDay{2002}{}{}.
\newblock
{\BBOQ}\APACrefatitle {Changes in the Spatial Distribution of Visual Attention after Early Deafness} {Changes in the spatial distribution of visual attention after early deafness}.{\BBCQ}
\newblock
\APACjournalVolNumPages{Journal of Cognitive Neuroscience}{14}{}{687-701}.
\newblock
\begin{APACrefDOI} \doi{10.1162/08989290260138591} \end{APACrefDOI}
\PrintBackRefs{\CurrentBib}

\bibitem [\protect \citeauthoryear {%
Qi%
\ \BBA {} Mitchell%
}{%
Qi%
\ \BBA {} Mitchell%
}{%
{\protect \APACyear {2012}}%
}]{%
QiMitchell2012}
\APACinsertmetastar {%
QiMitchell2012}%
\begin{APACrefauthors}%
Qi, S.%
\BCBT {}\ \BBA {} Mitchell, R\BPBI E.%
\end{APACrefauthors}%
\unskip\
\newblock
\APACrefYearMonthDay{2012}{}{}.
\newblock
{\BBOQ}\APACrefatitle {Large-Scale Academic Achievement Testing of Deaf and Hard-of-Hearing Students: Past, Present, and Future} {Large-scale academic achievement testing of deaf and hard-of-hearing students: Past, present, and future}.{\BBCQ}
\newblock
\APACjournalVolNumPages{The Journal of Deaf Studies and Deaf Education}{17}{1}{1--18}.
\PrintBackRefs{\CurrentBib}

\bibitem [\protect \citeauthoryear {%
Razalli%
, Hashim%
\BCBL {}\ \BBA {} Mamat%
}{%
Razalli%
\ \protect \BOthers {.}}{%
{\protect \APACyear {2018}}%
}]{%
EffectsOfBilingual2018}
\APACinsertmetastar {%
EffectsOfBilingual2018}%
\begin{APACrefauthors}%
Razalli, A.%
, Hashim, T.%
\BCBL {}\ \BBA {} Mamat, N.%
\end{APACrefauthors}%
\unskip\
\newblock
\APACrefYearMonthDay{2018}{11}{}.
\newblock
{\BBOQ}\APACrefatitle {Effects of Bilingual Approach in Malay Language Teaching for Hearing Impaired Students} {Effects of bilingual approach in malay language teaching for hearing impaired students}.{\BBCQ}
\newblock
\APACjournalVolNumPages{International Journal of Academic Research in Progressive Education and Development}{7}{}{109-121}.
\newblock
\begin{APACrefDOI} \doi{10.6007/IJARPED/v7-i4/4799} \end{APACrefDOI}
\PrintBackRefs{\CurrentBib}

\bibitem [\protect \citeauthoryear {%
Rello%
, Saggion%
\BCBL {}\ \BBA {} Baeza-Yates%
}{%
Rello%
\ \protect \BOthers {.}}{%
{\protect \APACyear {2014}}%
}]{%
KeywordHighlighting2014}
\APACinsertmetastar {%
KeywordHighlighting2014}%
\begin{APACrefauthors}%
Rello, L.%
, Saggion, H.%
\BCBL {}\ \BBA {} Baeza-Yates, R.%
\end{APACrefauthors}%
\unskip\
\newblock
\APACrefYearMonthDay{2014}{{\APACmonth{04}}}{}.
\newblock
{\BBOQ}\APACrefatitle {Keyword Highlighting Improves Comprehension for People with Dyslexia} {Keyword highlighting improves comprehension for people with dyslexia}.{\BBCQ}
\newblock
\BIn{} S.~Williams, A.~Siddharthan\BCBL {}\ \BBA {} A.~Nenkova\ (\BEDS), \APACrefbtitle {Proceedings of the 3rd Workshop on Predicting and Improving Text Readability for Target Reader Populations ({PITR})} {Proceedings of the 3rd workshop on predicting and improving text readability for target reader populations ({PITR})}\ (\BPGS\ 30--37).
\newblock
\APACaddressPublisher{Gothenburg, Sweden}{Association for Computational Linguistics}.
\newblock
\begin{APACrefURL} \url{https://aclanthology.org/W14-1204} \end{APACrefURL}
\newblock
\begin{APACrefDOI} \doi{10.3115/v1/W14-1204} \end{APACrefDOI}
\PrintBackRefs{\CurrentBib}

\bibitem [\protect \citeauthoryear {%
Rzayev%
, Wo\'{z}niak%
, Dingler%
\BCBL {}\ \BBA {} Henze%
}{%
Rzayev%
\ \protect \BOthers {.}}{%
{\protect \APACyear {2018}}%
}]{%
ReadingOnSmartGlasses2018}
\APACinsertmetastar {%
ReadingOnSmartGlasses2018}%
\begin{APACrefauthors}%
Rzayev, R.%
, Wo\'{z}niak, P\BPBI W.%
, Dingler, T.%
\BCBL {}\ \BBA {} Henze, N.%
\end{APACrefauthors}%
\unskip\
\newblock
\APACrefYearMonthDay{2018}{}{}.
\newblock
{\BBOQ}\APACrefatitle {Reading on Smart Glasses: The Effect of Text Position, Presentation Type and Walking} {Reading on smart glasses: The effect of text position, presentation type and walking}.{\BBCQ}
\newblock
\BIn{} \APACrefbtitle {Proceedings of the 2018 CHI Conference on Human Factors in Computing Systems} {Proceedings of the 2018 chi conference on human factors in computing systems}\ (\BPG~1–9).
\newblock
\APACaddressPublisher{New York, NY, USA}{Association for Computing Machinery}.
\newblock
\begin{APACrefURL} \url{https://doi.org/10.1145/3173574.3173619} \end{APACrefURL}
\newblock
\begin{APACrefDOI} \doi{10.1145/3173574.3173619} \end{APACrefDOI}
\PrintBackRefs{\CurrentBib}

\bibitem [\protect \citeauthoryear {%
Seita%
, Lee%
, Andrew%
, Shinohara%
\BCBL {}\ \BBA {} Huenerfauth%
}{%
Seita%
\ \protect \BOthers {.}}{%
{\protect \APACyear {2022}}%
}]{%
remote}
\APACinsertmetastar {%
remote}%
\begin{APACrefauthors}%
Seita, M.%
, Lee, S.%
, Andrew, S.%
, Shinohara, K.%
\BCBL {}\ \BBA {} Huenerfauth, M.%
\end{APACrefauthors}%
\unskip\
\newblock
\APACrefYearMonthDay{2022}{}{}.
\newblock
{\BBOQ}\APACrefatitle {Remotely Co-Designing Features for Communication Applications using Automatic Captioning with Deaf and Hearing Pairs} {Remotely co-designing features for communication applications using automatic captioning with deaf and hearing pairs}.{\BBCQ}
\newblock
\BIn{} \APACrefbtitle {Proceedings of the 2022 CHI Conference on Human Factors in Computing Systems.} {Proceedings of the 2022 chi conference on human factors in computing systems.}
\newblock
\APACaddressPublisher{New York, NY, USA}{Association for Computing Machinery}.
\newblock
\begin{APACrefURL} \url{https://doi.org/10.1145/3491102.3501843} \end{APACrefURL}
\newblock
\begin{APACrefDOI} \doi{10.1145/3491102.3501843} \end{APACrefDOI}
\PrintBackRefs{\CurrentBib}

\bibitem [\protect \citeauthoryear {%
Spencer%
\ \BBA {} Marschark%
}{%
Spencer%
\ \BBA {} Marschark%
}{%
{\protect \APACyear {2010}}%
}]{%
spencer2010evidence}
\APACinsertmetastar {%
spencer2010evidence}%
\begin{APACrefauthors}%
Spencer, P\BPBI E.%
\BCBT {}\ \BBA {} Marschark, M.%
\end{APACrefauthors}%
\unskip\
\newblock
\APACrefYear{2010}.
\newblock
\APACrefbtitle {Evidence-based practice in educating deaf and hard-of-hearing students} {Evidence-based practice in educating deaf and hard-of-hearing students}.
\newblock
\APACaddressPublisher{Oxford, England}{Oxford University Press}.
\PrintBackRefs{\CurrentBib}

\bibitem [\protect \citeauthoryear {%
Vertanen%
\ \BBA {} Kristensson%
}{%
Vertanen%
\ \BBA {} Kristensson%
}{%
{\protect \APACyear {2008}}%
}]{%
OnTheBenefits2008}
\APACinsertmetastar {%
OnTheBenefits2008}%
\begin{APACrefauthors}%
Vertanen, K.%
\BCBT {}\ \BBA {} Kristensson, P\BPBI O.%
\end{APACrefauthors}%
\unskip\
\newblock
\APACrefYearMonthDay{2008}{}{}.
\newblock
{\BBOQ}\APACrefatitle {On the Benefits of Confidence Visualization in Speech Recognition} {On the benefits of confidence visualization in speech recognition}.{\BBCQ}
\newblock
\BIn{} \APACrefbtitle {Proceedings of the SIGCHI Conference on Human Factors in Computing Systems} {Proceedings of the sigchi conference on human factors in computing systems}\ (\BPG~1497–1500).
\newblock
\APACaddressPublisher{New York, NY, USA}{Association for Computing Machinery}.
\newblock
\begin{APACrefURL} \url{https://doi.org/10.1145/1357054.1357288} \end{APACrefURL}
\newblock
\begin{APACrefDOI} \doi{10.1145/1357054.1357288} \end{APACrefDOI}
\PrintBackRefs{\CurrentBib}

\bibitem [\protect \citeauthoryear {%
Wang%
, Nagano%
, Kashino%
\BCBL {}\ \BBA {} Igarashi%
}{%
Wang%
\ \protect \BOthers {.}}{%
{\protect \APACyear {2016}}%
}]{%
VisualizingVideoSounds2016}
\APACinsertmetastar {%
VisualizingVideoSounds2016}%
\begin{APACrefauthors}%
Wang, F.%
, Nagano, H.%
, Kashino, K.%
\BCBL {}\ \BBA {} Igarashi, T.%
\end{APACrefauthors}%
\unskip\
\newblock
\APACrefYearMonthDay{2016}{09}{}.
\newblock
{\BBOQ}\APACrefatitle {Visualizing Video Sounds With Sound Word Animation to Enrich User Experience} {Visualizing video sounds with sound word animation to enrich user experience}.{\BBCQ}
\newblock
\APACjournalVolNumPages{IEEE Transactions on Multimedia}{PP}{}{1-1}.
\newblock
\begin{APACrefDOI} \doi{10.1109/TMM.2016.2613641} \end{APACrefDOI}
\PrintBackRefs{\CurrentBib}

\bibitem [\protect \citeauthoryear {%
Ziadat%
\ \BBA {} Al~Rahmneh%
}{%
Ziadat%
\ \BBA {} Al~Rahmneh%
}{%
{\protect \APACyear {2020}}%
}]{%
ziadat2020learning}
\APACinsertmetastar {%
ziadat2020learning}%
\begin{APACrefauthors}%
Ziadat, A.%
\BCBT {}\ \BBA {} Al~Rahmneh, A.%
\end{APACrefauthors}%
\unskip\
\newblock
\APACrefYearMonthDay{2020}{Oct.}{}.
\newblock
{\BBOQ}\APACrefatitle {The learning, social, and economic challenges facing the deaf and hearing-impaired individuals} {The learning, social, and economic challenges facing the deaf and hearing-impaired individuals}.{\BBCQ}
\newblock
\APACjournalVolNumPages{Cypriot Journal of Educational Sciences}{15}{5}{976--988}.
\newblock
\begin{APACrefURL} \url{https://doi.org/10.18844/cjes.v15i5.5130} \end{APACrefURL}
\newblock
\begin{APACrefDOI} \doi{10.18844/cjes.v15i5.5130} \end{APACrefDOI}
\PrintBackRefs{\CurrentBib}

\end{thebibliography}

% -----------------
\section{Appendices}
\appendix

\section{Mean Comparison of SUS Questions}
\label{appendix-sus}

\begin{figure}[H]
    \centering  \includegraphics[width=1\linewidth]{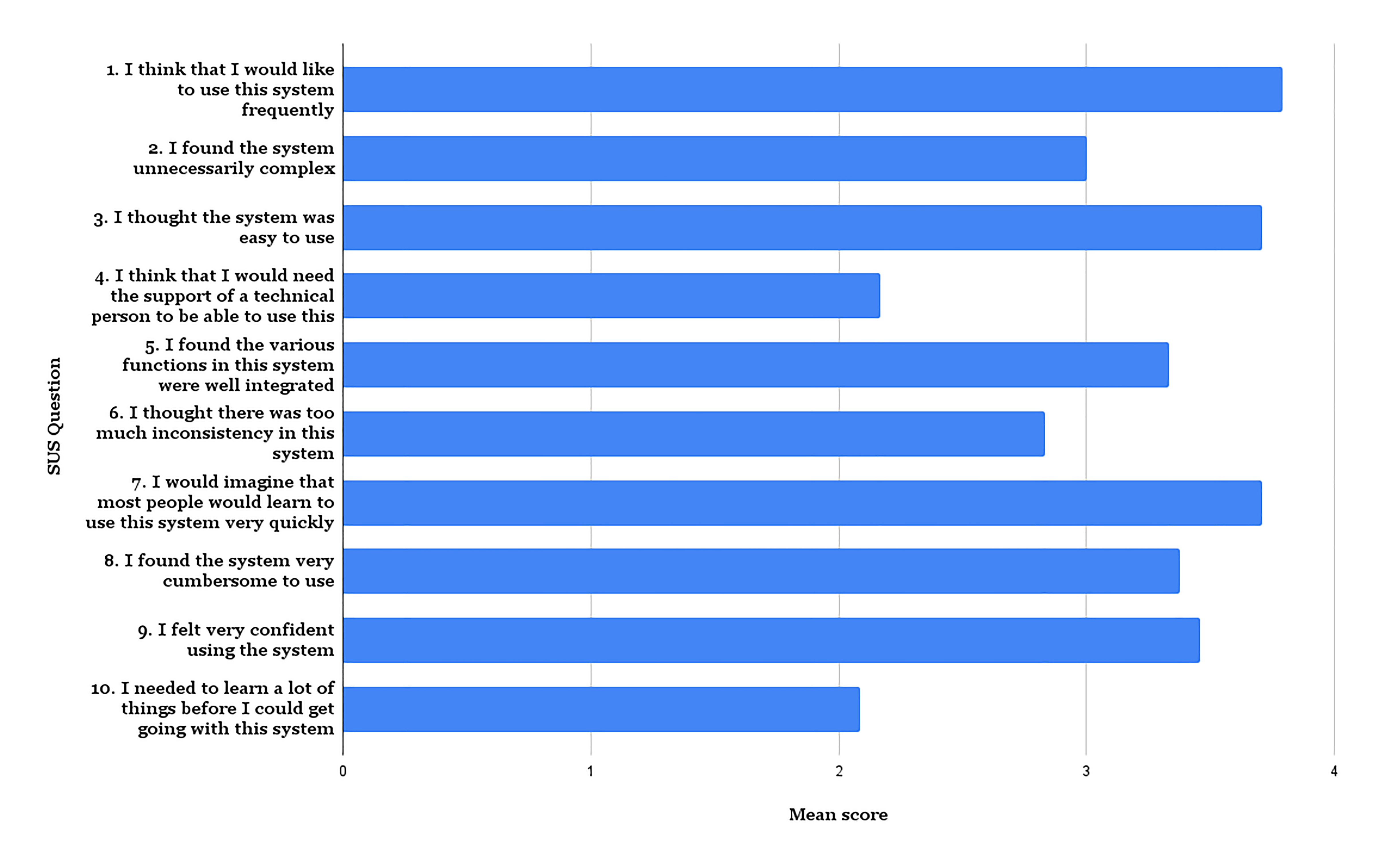}
    \caption{Mean comparison of SUS questions}
    \label{fig:appendix-susfigure}
\end{figure}

\section{“Overall user reactions to the system”
part of the Questionnaire for User Interaction Satisfaction (QUIS)}
\label{appendix-quis}

\begin{figure}[H]
    \centering  \includegraphics[width=1\linewidth]{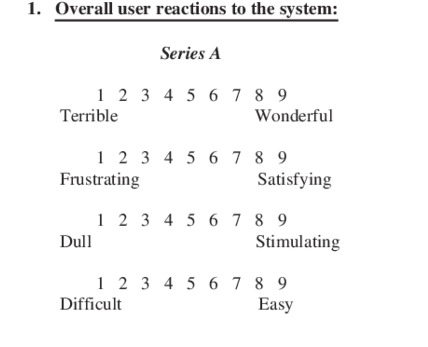}
    \caption{“Overall user reactions to the system”
part of the Questionnaire for User Interaction Satisfaction (QUIS)}
    \label{fig:appendix-quis}
\end{figure}

\section{System SDK Documentation}
Upon request, we are open to providing our open-source codebase, Unity SDKs, and datasets.

\end{document}